\documentclass[fleqn,usenatbib,useAMS]{mnras}

\usepackage{graphicx}	
\usepackage{amsmath}	
\usepackage{amssymb}	
\usepackage{multicol}        
\usepackage{bm}		
\usepackage{pdflscape}	
\usepackage{float} 
\usepackage{enumitem} 
\usepackage{caption} 

\usepackage[T1]{fontenc}
\usepackage{ae,aecompl}

\usepackage{newtxtext,newtxmath}

\usepackage{lastpage} 
\usepackage{xcolor}

\title[Offset stream self-crossing]{\textit{Spin-induced offset stream self-crossing shocks in tidal disruption events}}

\author[T. Jankovič et al.]{T. Jankovič$^{1}${\thanks{Contact e-mail: \href{mailto:taj.jankovic@ung.si}{taj.jankovic@ung.si}}, C. Bonnerot$^{2,\, 3}$, A. Gomboc$^{1}$
}%
\\
$^{1}$ Center for Astrophysics and Cosmology, School of Science, University of Nova Gorica, Vipavska 13, SI-5000 Nova Gorica, Slovenia\\
$^{2}$  Niels Bohr International Academy, Niels Bohr Institute, Blegdamsvej 17, DK-2100 Copenhagen Ø, Denmark\\
$^{3}$  School of Physics and Astronomy \& Institute for Gravitational Wave Astronomy, University of Birmingham, Birmingham B15 2TT, UK
}

\date{Last updated 2024 February 19; in original form 2023 March 31}

\pubyear{2024}

\begin{document}
\label{firstpage}
\pagerange{\pageref{firstpage}--\pageref{lastpage}}
\maketitle

\begin{abstract}
Tidal disruption events occur when a star is disrupted by a supermassive black hole, resulting in an elongated stream of gas that partly falls back to the pericenter. Due to apsidal precession, the returning stream may collide with itself, leading to a self-crossing shock that launches an outflow. If the black hole spins, this collision may additionally be affected by Lense-Thirring precession that can cause an offset between the two stream components. We study the impact of this effect on the outflow properties by carrying out local simulations of collisions between offset streams. As the offset increases, we find that the geometry of the outflow becomes less spherical and more collimated along the directions of the incoming streams, with less gas getting unbound by the interaction. However, even the most grazing collisions we consider significantly affect the trajectories of the colliding gas, likely promoting subsequent strong interactions near the black hole and rapid disc formation. We analytically compute the dependence of the offset to stream width ratio, finding that even slowly spinning black holes can cause both strong and grazing collisions. {We estimate that the self-crossing shock luminosity is lower for an offset collision than an
aligned one since radiation energy injected by the shock is significantly lower for more offset collisions. We find that the deviation from outflow sphericity may cause significant variations in the efficiency at which X-ray radiation from the disc is reprocessed to the optical band, depending on the viewing angle, and increase the degree of the observed polarization}. These potentially observable features hold the promise of constraining the black hole spin from tidal disruption events.

\end{abstract}




\begingroup
\let\clearpage\relax
\endgroup
\newpage

 \section{Introduction}

\noindent A stellar tidal disruption event (TDE) occurs at the center of a galaxy when a star is scattered in close proximity of a supermassive black hole. This highly energetic event, which can outshine the host galaxy, is a consequence of the star being disrupted due to the black hole's tidal field. After the disruption, approximately half of the debris escapes from the gravitational potential of the black hole, while the other half returns to the black hole's vicinity. There it can form an accretion disc, which may emit radiation for months to years \citep{rees}.  

By observing the light curve of a TDE it is possible to learn more about the properties of the disrupted star and its initial orbit, and also about the properties of the supermassive black hole and accretion physics (\citealt{komossa}; \citealt{van_Velzen_2019}; {\citealt{Yao_2023arXiv230306523Y}}).  The information about the orbit of the disrupted star offers a way to study the stellar dynamics in galactic centers. The mass and size of the star would provide a way to determine which stars are more prone to be disrupted to get insight into the initial mass function in the proximity of black holes. The emitted signal also depends on the black hole's mass and spin, whose determination is crucial for the understanding of the demographic of black holes as well as their growth and evolution. To date, around $100$ events of this type have been observed, in different ranges of the electromagnetic spectrum (e.g. \citealt{Alexander_2020,Saxton_2021, Petrushevska_2023}). This field will be revolutionized starting 2025 with the advent of the Vera  Rubin Observatory, which will detect thousands of new TDEs, thus increasing the sample size by one or two orders of magnitude \citep{Bricman_2020}.

The evolution of a TDE starts with the stellar disruption when the star is deformed into an elongated stream of debris, due to the black hole's tidal field. Bound parts of the debris move on a range of elliptic orbits and return to the proximity of the black hole with a fallback rate that depends on the properties of the system (e.g. \citealt{Lodato_2009MNRAS.392..332L,Guillochon_2013, Bonnerot_2022,Jankovic_2023https://doi.org/10.48550/arxiv.2302.00607}). Debris with more negative energy returns sooner and collides with the still in-falling stream due to the relativistic apsidal precession. This collision leads to a self-crossing shock that dissipates the kinetic energy of the debris, causing the formation of an outflow, part of which can later circularize into an accretion disc \citep{Sadowski_2016,liptai2019disc, Bonnerot_2020, Bonnerot_2021}.

A global numerical simulation of the entire TDE process requires a resolution higher than what is computationally feasible to date, mostly due to the significantly larger longitudinal extent of the debris stream than its transverse width \citep{bonnerot_2020_book}. Therefore, several numerical studies focused on disruptions of stars on elliptic orbits or by less massive black holes, in which the difference between transverse and longitudinal extents is smaller \citep{Hayasaki_2013,Shiokawa_2015, Bonnerot_2016eccentric_disks,Clerici_Gomboc_2020}. This numerical limitation prompted local studies of the self-crossing shock considering more physically realistic parameters, such as those conducted by \citet{Jiang_2016}, \citet{Lu_2019}, and {\citet{huang_2023arXiv230317443H}}, which demonstrated that the shock launches a large-scale outflow. These studies assumed that the colliding streams are similar, an assumption that depends on the effect of the nozzle shock, caused by a strong vertical compression of the in-falling stream component near the pericenter. The validity of this assumption has been later confirmed by \citet{bonnerot2021nozzle}, who found that the nozzle shock leads to minimal net expansion of the receding stream's height and width. However, \citet{Jiang_2016}, \citet{Lu_2019}, and {\citet{huang_2023arXiv230317443H}} mostly focused on stream collisions near non-rotating black holes and did not study the impact of the black hole’s spin in a systematic way.

In general, the black hole's spin and stellar angular momentum are not aligned, and, when the tip of the stream returns to the black hole's vicinity, its orbit is inclined with respect to the black hole's rotational plane. In this configuration, a relativistic effect known as Lense-Thirring precession makes the gas angular momentum vector precess around the black hole spin, potentially resulting in a misalignment between the colliding streams in the self-crossing region \citep{bonnerot_2020_book}. In most extreme cases the streams can even miss each other entirely and several additional revolutions may be necessary for them to successfully collide \citep{Batra_2022}. \citet{bonnerot2021nozzle} found that the width of the stream component receding from the black hole evolves at a slower rate than it was previously thought \citep{Guillochon_2015}, which can make this effect significant even for slowly spinning black holes. 

In this paper, we carry out the first systematic study of the effect of the black hole's rotation on the self-crossing shock. We perform local smoothed-particle hydrodynamics (SPH) simulations of collisions between two streams offset in the direction perpendicular to the orbital plane of the debris. We analyze the properties of the outflow in the local co-moving frame and also study the subsequent evolution of the outflow in the black hole's reference frame. We find that the outflow becomes less spherical as the black hole's spin increases, which also reduces the gas density along certain directions, leading to potentially observable features that could be used to constrain the black hole's spin.

This paper is organized as follows. In Section \ref{sec2}, we explain the numerical setup, initial conditions and the numerical procedure. The results are presented in Section \ref{sec3} and discussed in Section \ref{sec4}. We summarize our main conclusions in Section \ref{sec5}.

\section{SPH simulations} \label{sec2}

\noindent We simulate the collision between the two stream components involved in the self-crossing shock. The colliding streams are constructed by continuously injecting particles close to the intersection point. We consider the effect of the black hole's rotation by off-setting the streams such that only a fraction of the gas directly collides.

\subsection{Initial conditions}

\noindent Lense-Thirring precession due to the black hole's spin induces a vertical offset $\Delta z$ between the two incoming streams with cylindrical radii $H_1$ and $H_2$. By comparing $\Delta z$ to the sum of cylindrical radii of streams $H_1+H_2$, it is possible to characterize the different types of interactions. $\Delta z>H_1 + H_2$  corresponds to a situation where streams do not interact with one another and self-crossing is avoided. A complete collision would happen for $\Delta z=0$. For $0<\Delta z \leq H_1 + H_2$ only parts of the streams collide. 

We study the effect of the black hole's rotation by simulating collisions of two streams offset by different values of $\Delta z$ with the smoothed-particle hydrodynamics code \textsc{phantom} \citep{price}. Following \citet{Lu_2019}, this simulation is carried out in the frame co-moving with the common tangential velocity of the two incoming streams. In this frame, the streams collide head-on with a velocity $v$ equal to their original radial velocity. This setup is valid both if the collision is prompt or if it happens after many windings \citep{Batra_2022}. 

Our initial setup consisting of two offset incoming streams is illustrated in Figure \ref{s1}. The coordinate system is defined with unit vectors $\mathbfit{e}_{\rm x}$, $\mathbfit{e}_{\rm y}$, $\mathbfit{e}_{\rm z}$ and the origin is set to the intersection point. Therefore, gas particles of one stream have velocities $\mathbfit{v}_{\rm 1}=(0,v,0)$, while the velocity of particles in the other stream is $\mathbfit{v}_{\rm 2}=(0,-v,0)$. We use the same value $\dot{M}$ for the mass inflow rate of the two streams, $\dot{M}_1 = \dot{M}_2 = \dot{M}$. We consider a collision of two streams with circular cross-sections and equal widths $H_1 = H_2 = H$, where we define width as the cylindrical radius. This assumption is justified by the recent work of \citet{bonnerot2021nozzle}, who find that the nozzle shock leads to almost no net expansion, such that the stream size remains the same to order unity, i.e. $H_1 \approx H_2$. Values of parameters $v$, $\dot{M}$, and $H$ are set to one, which also defines our code units. Streams are also cold due to adiabatic expansion after the disruption. Therefore, we choose a value of the specific internal energy $u=10^{-5}$ (expressed in code units), such that the collision is highly supersonic.  

In our simulations, we use an adiabatic equation of state with ${\Gamma}=4/3$. This is motivated by a large trapping radius of the shocked gas and the fact, that the gas is radiation pressure dominated. The trapping radius $R_\mathrm{tr}$ defines the distance from the intersection point, beyond which the photons generated by shocks diffuse away from the outflowing gas. \citet{Lu_2019} showed that for a non-spinning black hole $ R_{\rm tr} / H \approx 100 \gg 1$, implying that the outflow evolves adiabatically until large distances. However, the situation might be different for a rotating black hole where the collision is offset. In this case, may be possible for photons to rapidly diffuse away along specific directions where the density is lower. This possibility is further discussed in Section \ref{subsec:observ_features}. The importance of the radiation pressure can be determined by comparing the radiation pressure  $P_{\gamma}$ to the gas pressure $P_\mathrm{g}$. For typical conditions in the self-crossing region, the radiation pressure dominates with a high ratio $P_{\gamma}/P_\mathrm{g} \approx 1000 \gg 1$ (\citealt{Jiang_2016}; \citealt{Lu_2019}; {\citealt{huang_2023arXiv230317443H}}).

\subsection{Numerical procedure}\label{sec:num_proc}
\begin{figure}
\centering
\includegraphics[width=0.49\textwidth]{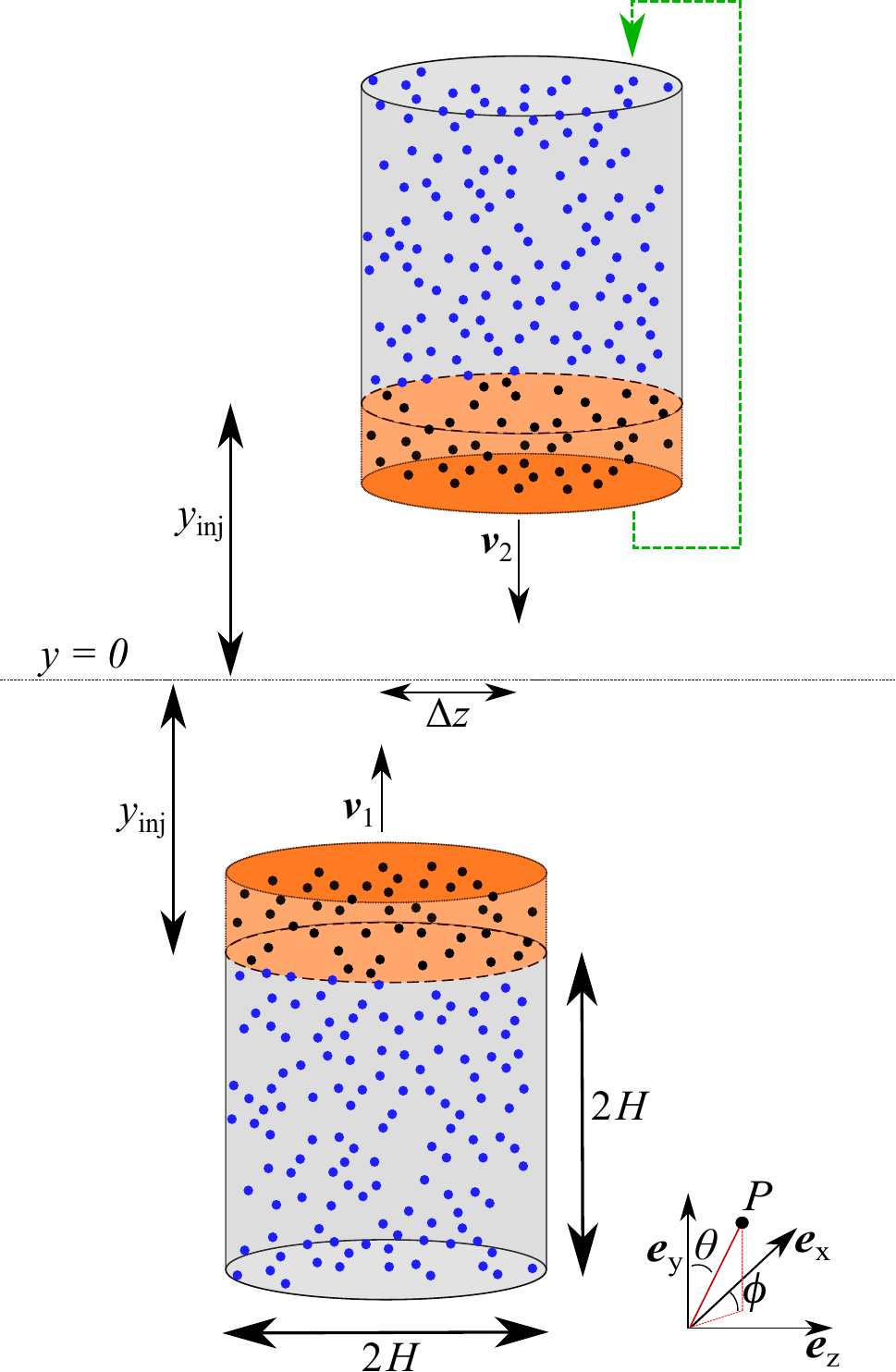}
\caption{Sketch showing our setup for the local simulations of collisions between offset streams. The streams are created by injecting SPH particles with opposite velocities $\mathbfit{v}_1$ and $\mathbfit{v}_2$ from two 3D cylinders of radius $H$ and length $2H$ that are offset by $\Delta z$. This is done by defining ghost particles inside these cylinders (blue points in the grey regions) that are cyclically added to the computational domain, where these particles (black points in the orange regions) are then hydrodynamically evolved by the code.}
\label{s1}
\end{figure}

\noindent As illustrated in Figure \ref{s1}, the streams are created by cyclically injecting SPH particles from a cylinder, which is constructed in the following way. We derive gas density in the stream from $\rho =\dot{M} /(\pi H^2 v)$, which we use to create a 3D cube with size $2H$ and total mass $M_\mathrm{cube}=\rho (2H)^3$. The cube is populated by $N$ particles with mass $m=10^{-3}$, where $N=M_\mathrm{cube}/m$, which also determines the resolution of our simulations. For this choice of $m$, the smoothing length of SPH particles is $h_\mathrm{sl}\approx (m/\rho)^{1/3}\approx 0.1H<H$, meaning that the resolution is sufficiently high.\footnote{To assess the accuracy of our study more thoroughly, we simulated stream collisions also for particle masses $m=10^{-4}$ and $10^{-2}$. We found no significant differences when decreasing the mass to less than $m=10^{-3}$, indicating that the resolution used in our simulations is sufficiently high.} We use a glass distribution, a distribution where the distances between particles are equal, to avoid the noise resulting from a random particle distribution. The glass distribution is reached by relaxing the particle distribution inside the cube using periodic boundary conditions, from which we then select a cylinder with radius $H$ and height $2H$.

The particles inside this cylinder are not all evolved initially, but instead, act as ``ghost'' particles (blue particles inside the grey regions) that are progressively injected inside the computational domain from the surfaces at $y_\mathrm{inj}=\pm 3H$ (black particles inside the orange regions in Figure \ref{s1}). At each timestep $\Delta t$, this is done by shifting their position by $\Delta y=\pm v \Delta t$, and injecting the ghost particles that have left the cylinder from the top or bottom. The number of injected particles is $\Delta N= \dot{M}\Delta t /m$. While particles initially possess mostly kinetic energy, dissipation happens through a shock when the streams collide, leading to the formation of an outflow, which we now investigate in detail.

\section{Results} \label{sec3}

\noindent In this section we present the properties of the outflow from the self-crossing shock produced by colliding streams. The collision is simulated in a local co-moving frame  according to the procedure described in Section \ref{sec2}, considering different offsets between the streams.\footnote{Movies made from the simulations are available online at \href{https://www.youtube.com/playlist?list=PLH8qhWjKWQ91adY0ae1lD2vRVbaerSWmb}{https://www.youtube.com/playlist?list=PLH8qhWjKWQ91adY0ae1lD2vRV\\baerSWmb}.} We determine various properties of the outflow in the co-moving and in the black hole's reference frames.

 \subsection{Collision dynamics and shock heating rate} \label{sec:shock}

\begin{figure*}
	\centering
	\textbf{$\Delta z/H= 0.0$}
		    \hspace{.26\linewidth}
		   \textbf{$\Delta z/H= 0.3$}
		   	    \hspace{.26\linewidth}
	    \textbf{$\Delta z/H= 1.2$}\par\medskip
	        \vspace{-0.2cm}
	\hfill
	    	\begin{minipage}[b]{0.33\linewidth}
		\includegraphics[width=\textwidth]{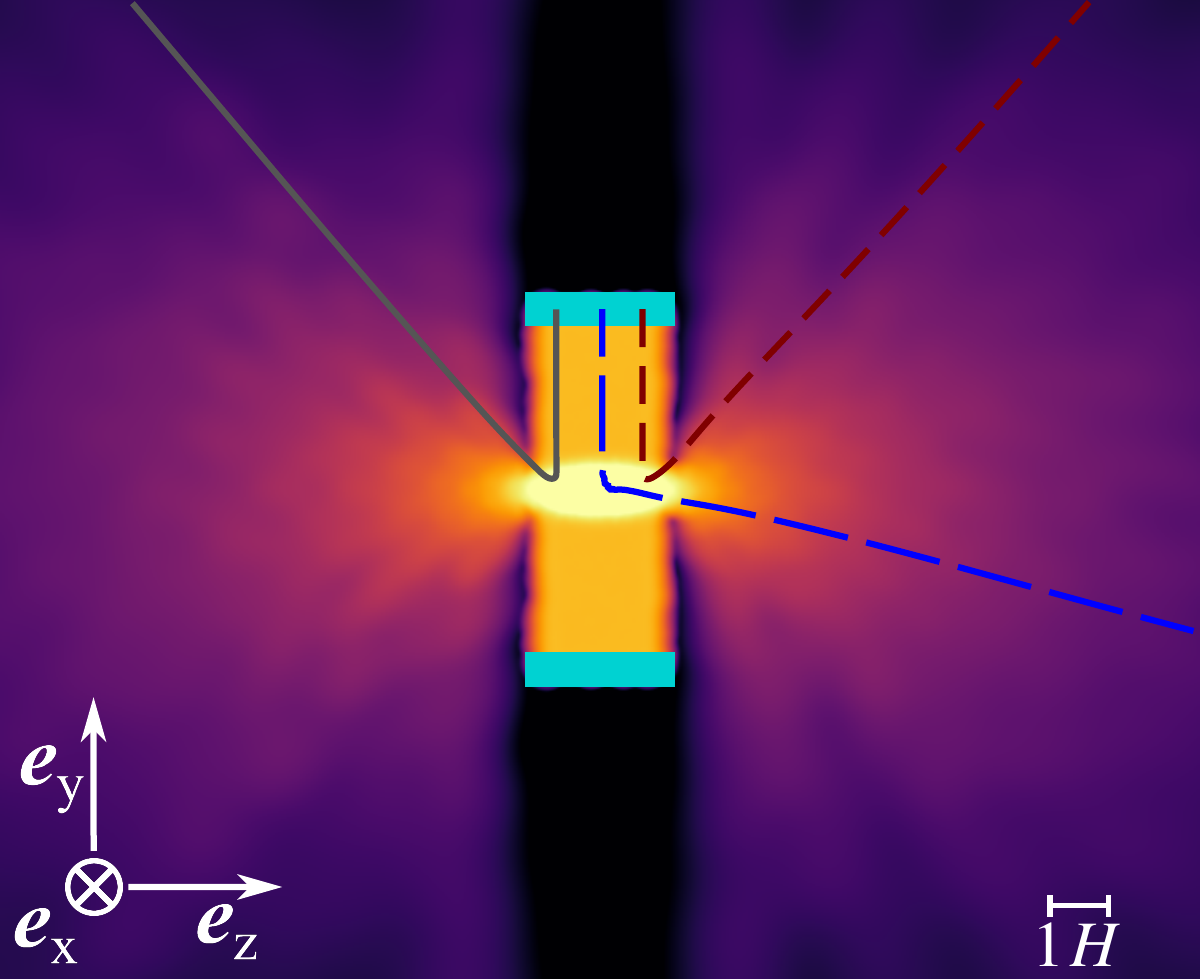}
	\end{minipage}
		\hfill
\begin{minipage}[b]{0.33\linewidth}
		\includegraphics[width=\textwidth]{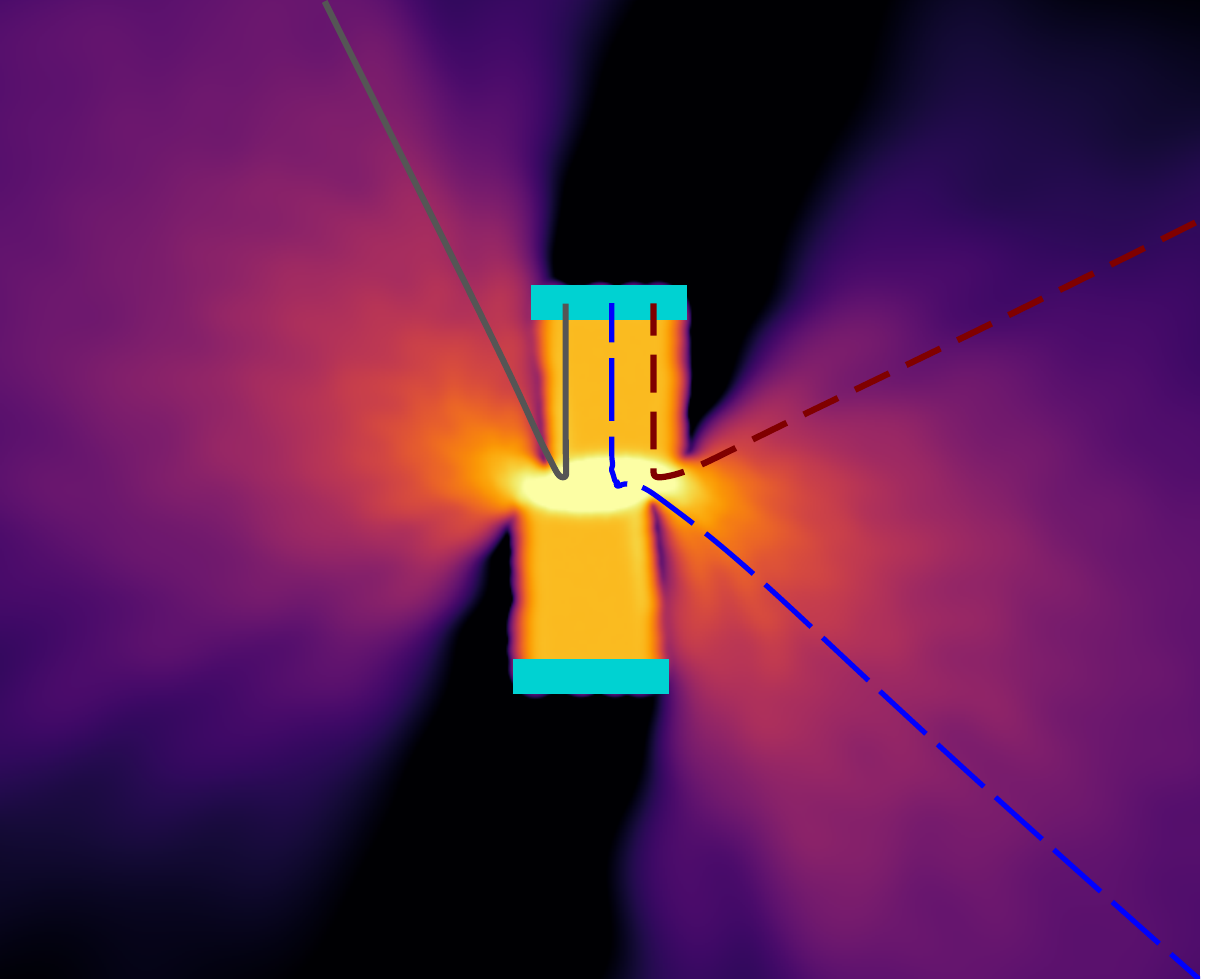}
	\end{minipage}
	\hfill
	\begin{minipage}[b]{0.33\linewidth}
		\includegraphics[width=\textwidth]{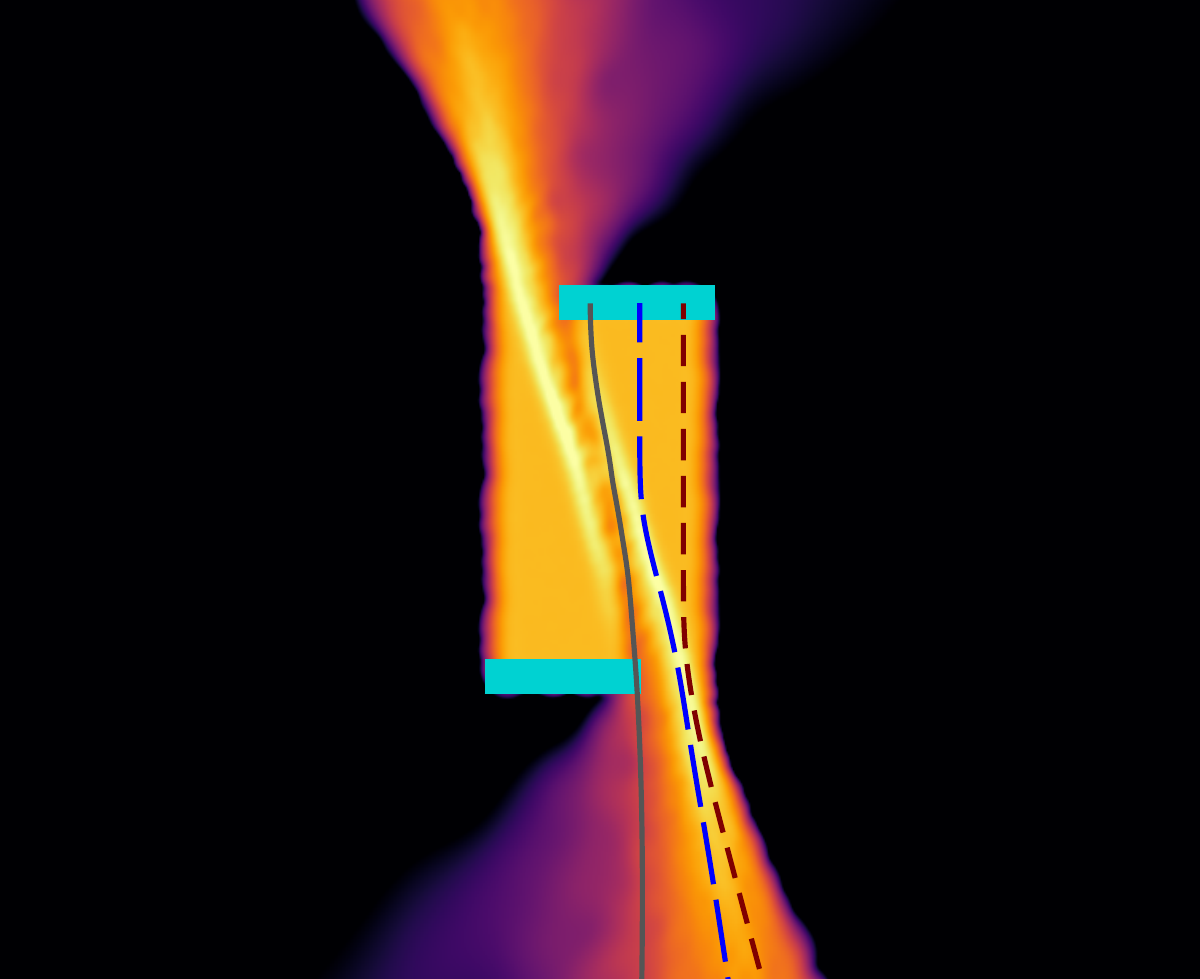}
	\end{minipage}
	\hfill
		\begin{minipage}[b]{\linewidth}
			\centering
			\includegraphics[width=\textwidth]{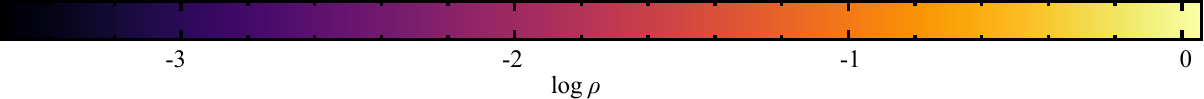}
		\end{minipage}
		\vspace*{-0.3cm}
		
	\caption{Density slices in the $zy$ plane at $x=0$ and time $t=100$ for values of the offset $\Delta z=0$ (left panel), $\Delta z=0.3H$ (middle panel) and $\Delta z=1.2H$ (right panel). Coloured lines show the trajectories of three particles, using the same colour for particles injected at the same position and time, but varying the vertical offset.}
	\label{fig:zoom}
\end{figure*}

\noindent Figure \ref{fig:zoom} shows the 3D density of gas $\rho$ contained in slices in the $zy$ plane at $x=0$ for stream offsets $\Delta z=0$, $0.3H$ and $1.2H$. Gas particles are injected at a time $t=0$ from the surfaces at $y_\mathrm{inj}$ (indicated with thick light blue segments), and the streams collide at $t=3$. Trajectories of three different particles are shown as coloured lines, where we use the same colour for particles injected at the same position and time, but varying the vertical offset. In the $\Delta z=0$ case (left panel) all the incoming gas is directly involved in the collision, which leads to the formation of a shock that dissipates kinetic energy. Consequently, pressure forces sharply increase and an outflow is launched from the collision region that is symmetric both azimuthally and with respect to the equatorial plane at $y=0$. After the gas has been forced to move from the collision region along the equatorial plane at $\theta = {\pi}/{2}$, its trajectories get further deflected away from this plane by vertical pressure forces (see particles trajectories in Figure \ref{fig:zoom}), such that the outflowing gas moves along a wide range of polar angles.
The path of a particle becomes straight after it reaches a radial distance $r\approx 3H$ from the intersection point, due to a sharp reduction of pressure forces. Inside the outflow, the internal energy of the gas gets converted back to kinetic energy until it reaches a terminal velocity equal to the initial (pre-collision) velocity, as expected from energy conservation. With time the outflow keeps expanding to reach a steady-state, a state where the properties of the outflow do not change with time. 

For $\Delta z > 0$, only a fraction of the incoming streams directly collides. However, the shock from this directly colliding gas affects also the non-directly colliding parts of the stream. The reason is that the shock wave launched by the direct collision propagates outward to affect even the gas furthest away from the intersection point. This can be seen from the highest density parts in Figure \ref{fig:zoom} (middle panel) for the $\Delta z = 0.3H$ case, which indicates the presence of shocked gas outside the directly colliding gas region. The momentum of the non-directly colliding gas does not get entirely cancelled in the $\mathbfit{e}_{\rm y}$ direction, causing the outflow to be deflected away from the equatorial plane. This effect is the strongest for $\Delta z=1.2H$, where we see that the trajectories of all three particles remain almost aligned with the incoming stream direction (right panel of Figure \ref{fig:zoom}). Distinctions between the different panels of Figure \ref{fig:zoom} suggest that stream collisions can be classified into two regimes, which we will refer to as ``strong collision'' and ``grazing collision''. In the strong collision regime, the outflow is close to spherical, being only slightly deflected from the equatorial plane. In the grazing collision regime, the outflow is primarily collimated along the trajectories of the incoming streams, which only display a low level of expansion.

\begin{figure}
 	\centering
		\begin{minipage}[b]{.99\linewidth}
		\includegraphics[width=\textwidth]{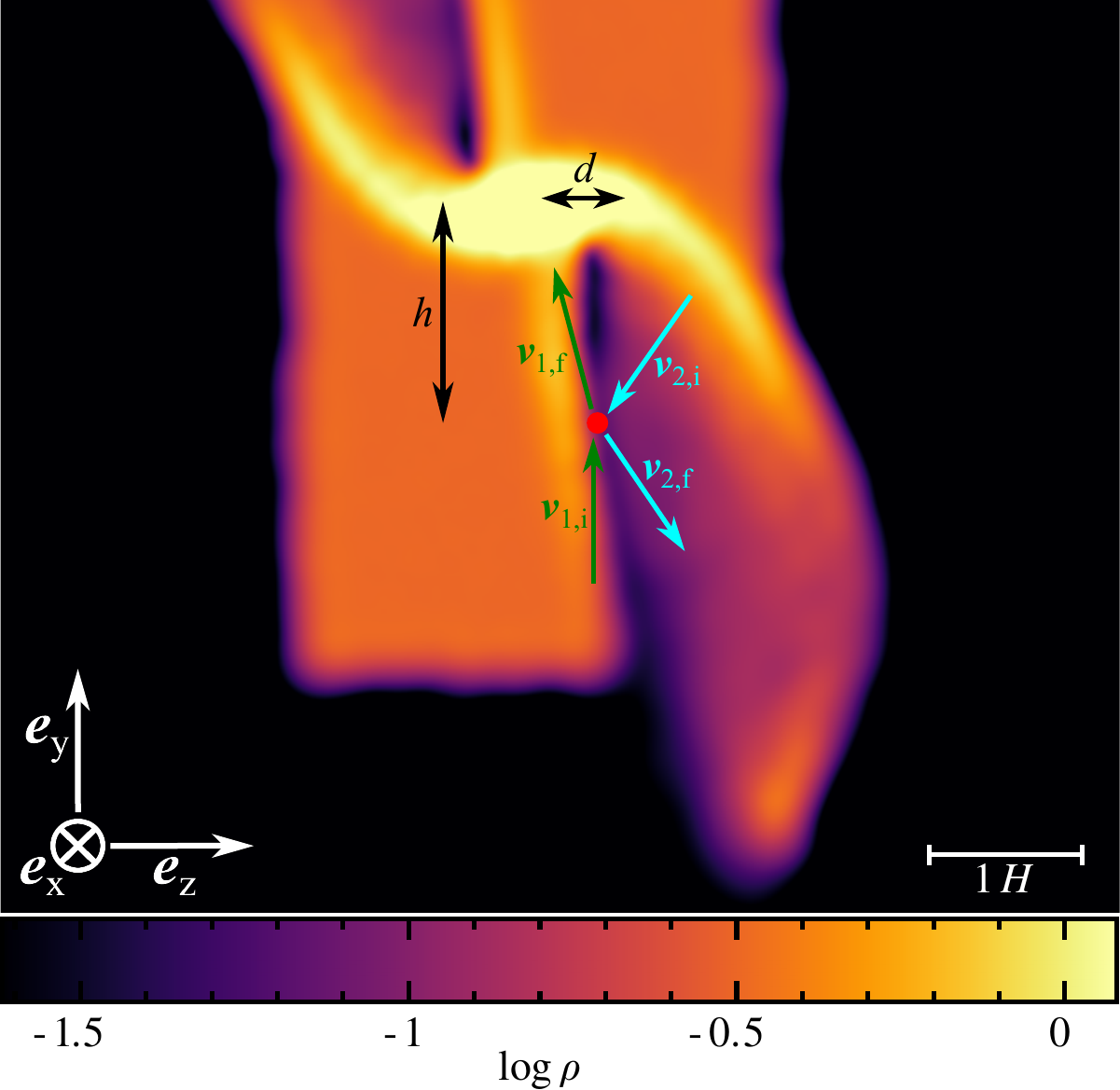}
	\end{minipage}\hfill

 	\caption{Density slice in the $zy$ plane at $x=0$ and time $t=7$, which illustrates the interaction between the gas in the incoming (upward-moving) streams (indices ``1'') and the outgoing (downward-moving) matter (indices ``2'') that has already passed through the self-crossing shock, for a value of the offset $\Delta z=1H$. The interaction (denoted by a red point) occurs at a distance $h$ from the equatorial plane and causes a deflection of the incoming stream by a distance $d$. Vector arrows represent gas velocities, where we use index ``i'' and ``f'' for components before and after the deflection, respectively.}
 	\label{fig:deflection}
 \end{figure}

The transition between the two regimes can be determined quantitatively and it is a consequence of the interaction between the gas in the incoming streams and the outgoing matter that has already passed through the self-crossing shock. This interaction results in a deflection of the gas trajectories inside the incoming streams, which reduces the amount of the incoming gas that directly collides. This mechanism can be understood more precisely from Figure \ref{fig:deflection}, where the index ``1'' refers to the gas from the incoming (upward-moving) stream involved in the collision, and the index ``2'' refers to the shocked outgoing (downward-moving) gas that causes the aforementioned deflection. Indices ``i'' and ``f'' correspond to stream velocities before and after the interaction between these two gas components, respectively. We estimate the distance $d$ by which the incoming stream gets deflected (towards the left), by modelling the interaction as an elastic collision confined to a single deflection point (red point in Figure \ref{fig:deflection}). This distance is obtained from $d\approx h\,  |v_\mathrm{1z}|/v_\mathrm{1y}$, where $h$ is the distance between the deflection point and the equatorial plane, and the two speeds are defined as $v_\mathrm{1z}\equiv \mathbfit{v}_\mathrm{1,f}\cdot \mathbfit{e}_\mathrm{z}$ and $v_\mathrm{1y}\equiv \mathbfit{v}_\mathrm{1,f}\cdot \mathbfit{e}_\mathrm{y}$. Momentum conservation imposes that $v_\mathrm{1z}=2\dot{M}_\mathrm{2,d} v_\mathrm{2z}/\dot{M}_\mathrm{1,d}$, where $\dot{M}_\mathrm{1,d}$ and $\dot{M}_\mathrm{2,d}$ denote the mass inflow rates of the two gas components involved in the deflection, and $v_\mathrm{2z}\equiv \mathbfit{v}_\mathrm{2,i}\cdot \mathbfit{e}_\mathrm{z}$. We note that not all the gas from the incoming stream is deflected, but only that contained inside the circular segment at a distance $d$ from the edge of the stream. Evaluating the area of the circular segment $A_\mathrm{d}\approx 2^{5/2}3^{-1}(d/H)^{3/2}H^2 $ at the lowest order in $d/H\ll 1$,\footnote{We note that this assumption is not entirely accurate, but we make it in order to find a simple analytical equation for $d/H$. Furthermore, we have verified our analytic approach by evaluating $A_\mathrm{d}$ without approximations and calculating $d/H$ numerically. We found only minor differences in comparison to the analytic estimate.} and making use of flux conservation $\dot{M}_\mathrm{1,d} =\dot{M}   A_\mathrm{d} / (\pi H^2) $, we find
\begin{align}
    \label{eq:d}
    \nonumber
    \frac{d}{H}&\approx \left( \frac{3\pi}{2^{3/2}}  \frac{\dot{M}_\mathrm{2,d}}{\dot{M}}   \frac{|v_\mathrm{2z}|}{v_\mathrm{1y}}   \frac{h}{H} \right)^{2/5},\\
    &\approx 0.64 \left( \frac{\dot{M}_\mathrm{2,d}}{0.5\dot{M}} \right)^{2/5}\left (  \frac{|v_\mathrm{2z}|}{0.1 v_\mathrm{1y}} \right)^{2/5} \left ( \frac{h}{2 H} \right)^{2/5}.
\end{align}
The numerical values we used here are obtained from the simulations, which lead to an evaluated distance $d\approx 0.64H$. This deflection causes an effective decrease of the stream width, such that a direct collision is eventually avoided if $\Delta z > 2(H - d) \approx 0.7H$. This analytical condition matches that found from the simulations to be in the grazing collision regime defined above. In that case, even though some of the incoming gas directly collides early on, at later times the streams end up passing above and below each other without experiencing significant dissipation (see the rightmost panel of Figure \ref{fig:zoom}). After the collision, the streams therefore, display only a low level of expansion while the trajectory of their center of mass is largely unaffected.

When the two streams collide a shock is formed and the kinetic energy of the gas is dissipated due to shock heating. We obtain an analytic estimate of the shock heating rate $\dot{E}$ from 

\begin{equation} \label{eq:dotE}
    \dot{E} \approx \xi(\Delta z) \dot{M}\Delta u,
\end{equation}
where we estimate the jump in specific internal energy $\Delta u$ from the Rankine-Hugoniot conditions as $\Delta u=  {2} v^2 \left({\Gamma} + 1\right)^{-2}= 0.37 v^2$, using ${\Gamma} = 4/3$. In Equation (\ref{eq:dotE}) we introduce a factor $\xi(\Delta z)={A_\mathrm{coll}(\Delta z)}/{A_\mathrm{0}}\in[0,1]$, which represents the fraction of the stream surfaces that directly collides initially.  $A_\mathrm{coll}$ is calculated as the intersection between the surfaces of two circular streams with a radius equal to $H$ that are offset by $\Delta z$, whereas $A_\mathrm{0}=\pi H^2$ is the cross-section of a stream. When $\Delta z=0$, the streams overlap and $A_\mathrm{coll}=A_\mathrm{0}$. However, as $\Delta z$ increases, the intersection between the streams surfaces decreases, and is reduced to zero when $\Delta z = 2H$.

\begin{figure}
 	\centering
		\begin{minipage}[b]{.99\linewidth}
		\includegraphics[width=\textwidth]{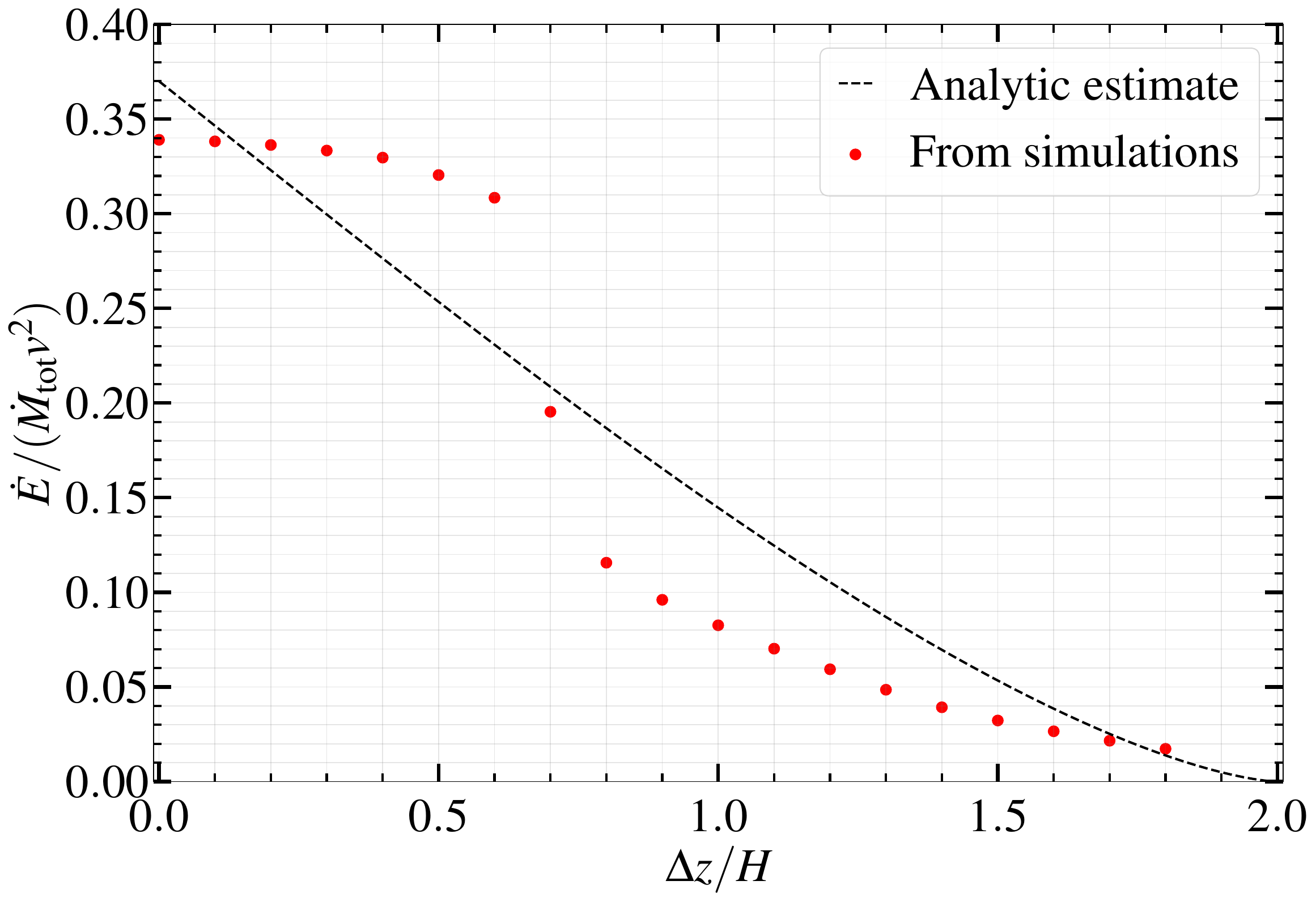}
	\end{minipage}\hfill

 	\caption{Shock heating rate $\dot{E}$ as a function of the offset $\Delta z$ expressed in units of the stream width $H$.  $\dot{E}$ is normalized by $\dot{M}_\mathrm{tot}v^2$, where $\dot{M}_\mathrm{tot}=2\dot{M}$ is the total inflow rate in the collision region and $v$ is the incoming stream velocity. The dashed line is calculated from Equation (\ref{eq:dotE}), while the red dots represent values obtained directly from simulations.}
 	\label{fig:dotE}
 \end{figure}

We have simulated 19 stream collisions corresponding to values of the offset $\Delta z \in [0, 1.8H]$,\footnote{For larger values of $\Delta z$, stream interactions are no longer physical due to the smoothing length of SPH particles, which extends outside of the stream boundaries. This has the effect of artificially enhancing the level of interaction between the streams, while for such large offsets we would physically expect the streams to be largely unaffected.} with an increment of $0.1H$. The comparison between values of $\dot{E}$ from simulations (red points) and values calculated from Equation (\ref{eq:dotE}) (black dashed line) is shown in Figure \ref{fig:dotE}. For $\Delta z \approx 0$ the values of $\dot{E}$ from simulations are lower than the analytic estimate. We attribute this to boundary effects --- values of the energy generated in shocks from simulations are lower because there is no external pressure outside of streams, leading to an expansion that adiabatically reduces the internal energy of the outflow. As the offset increases the shock heating rate becomes larger than the analytic estimate. This is because the gas that does not directly collide is still affected by the self-crossing shock due to shockwave propagation (e.g. see middle panel of Figure \ref{fig:zoom}). At $\Delta z \approx 0.7H$ we notice a sharp decline in $\dot{E}$, which marks the transition from the strong collision to the grazing collision regime. For $\Delta z \gtrsim 0.7H$, energy dissipation is lower than analytically predicted due to the effect described above (see Equation (\ref{eq:d})) that causes an effective decrease of the streams widths.

\subsection{Angular dependence of the outflow rate}\label{subsec:angular_dep}

\noindent We now calculate the dependence of the outflow rate $\dot{M}_{\rm out}$ on the direction, specified by the polar and azimuthal angles $\theta$ and $\phi$. To this aim, we first construct a spherical map, where each pixel covers the same surface area $A$, using the pixelation scheme of the  \textsc{HEALPix}\footnote{Current link to the \textsc{HEALPix} website: \href{https://healpix.sourceforge.io/}{https://healpix.sourceforge.io/}.} software \citep{2005ApJ...622..759G, Zonca2019}. We calculate the outflow rate $\dot{M}_{\rm out}=\mathrm{d} m/\mathrm{d} t $,  where $\mathrm{d} m$ is the mass of gas that crossed a given pixel during a time interval $\mathrm{d} t$. We determine $\mathrm{d} m$ and $\mathrm{d} t$ by comparing two simulation snapshots at different times, $t_1=100$ and $t_2=150$, at which the steady-state is reached. With this procedure, we are able to construct \textsc{HEALPix} maps of the outflow rate. We use these maps to calculate the normalized mass flux 
\begin{equation}\label{eq:F}
F=(\dot{M}_{\rm out}/\dot{M}_\mathrm{tot})/A_\mathrm{},
\end{equation}
through individual pixels, where the total outflow rate through the whole sphere is $\dot{M}_\mathrm{tot}=2\dot{M}=2$ due to mass conservation.

\begin{figure*}
	\centering
	\begin{minipage}[b]{.49\linewidth}
		\includegraphics[width=\textwidth]{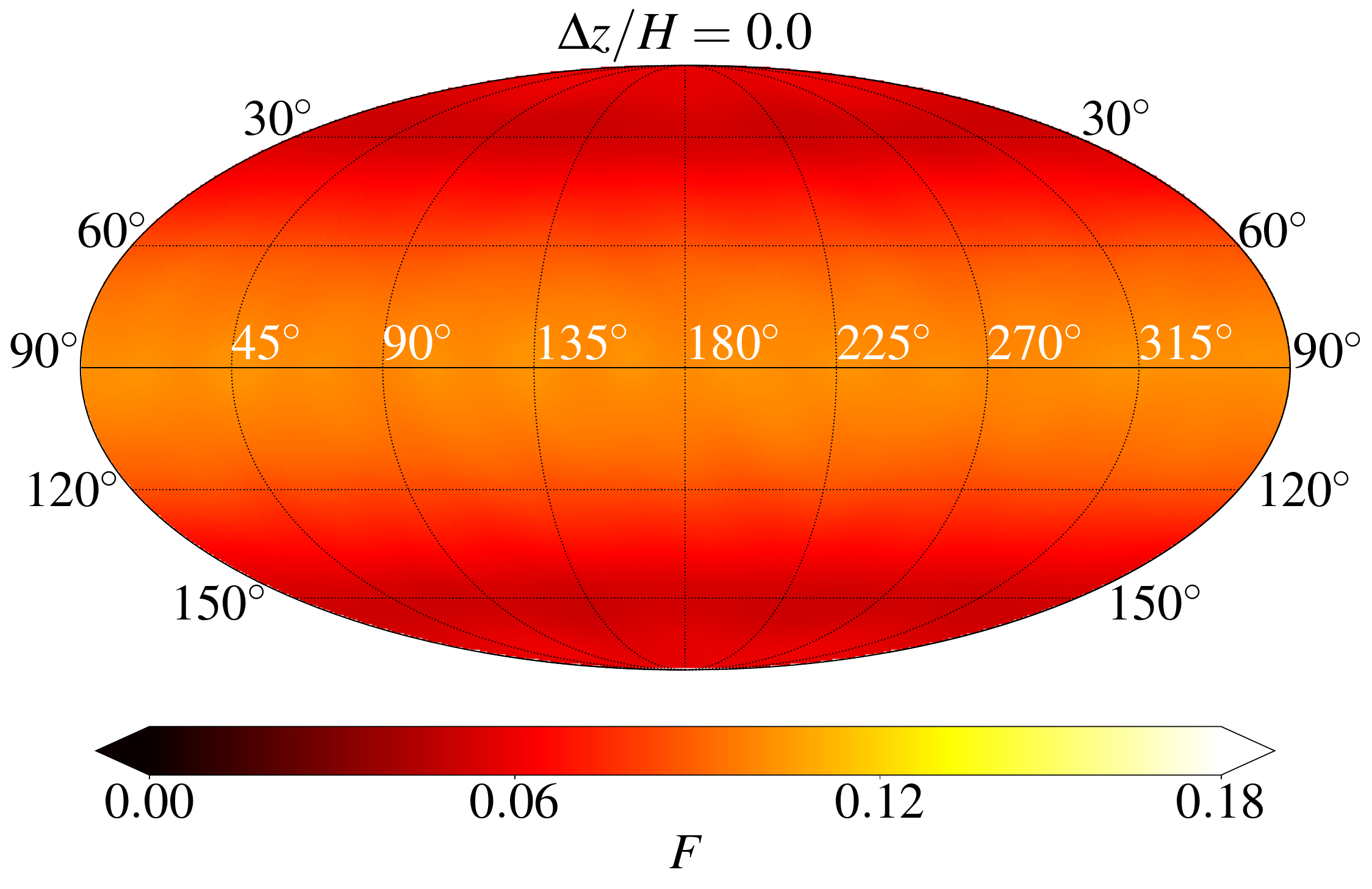}
	\end{minipage}\hfill
		\begin{minipage}[b]{.49\linewidth}
		\includegraphics[width=\textwidth]{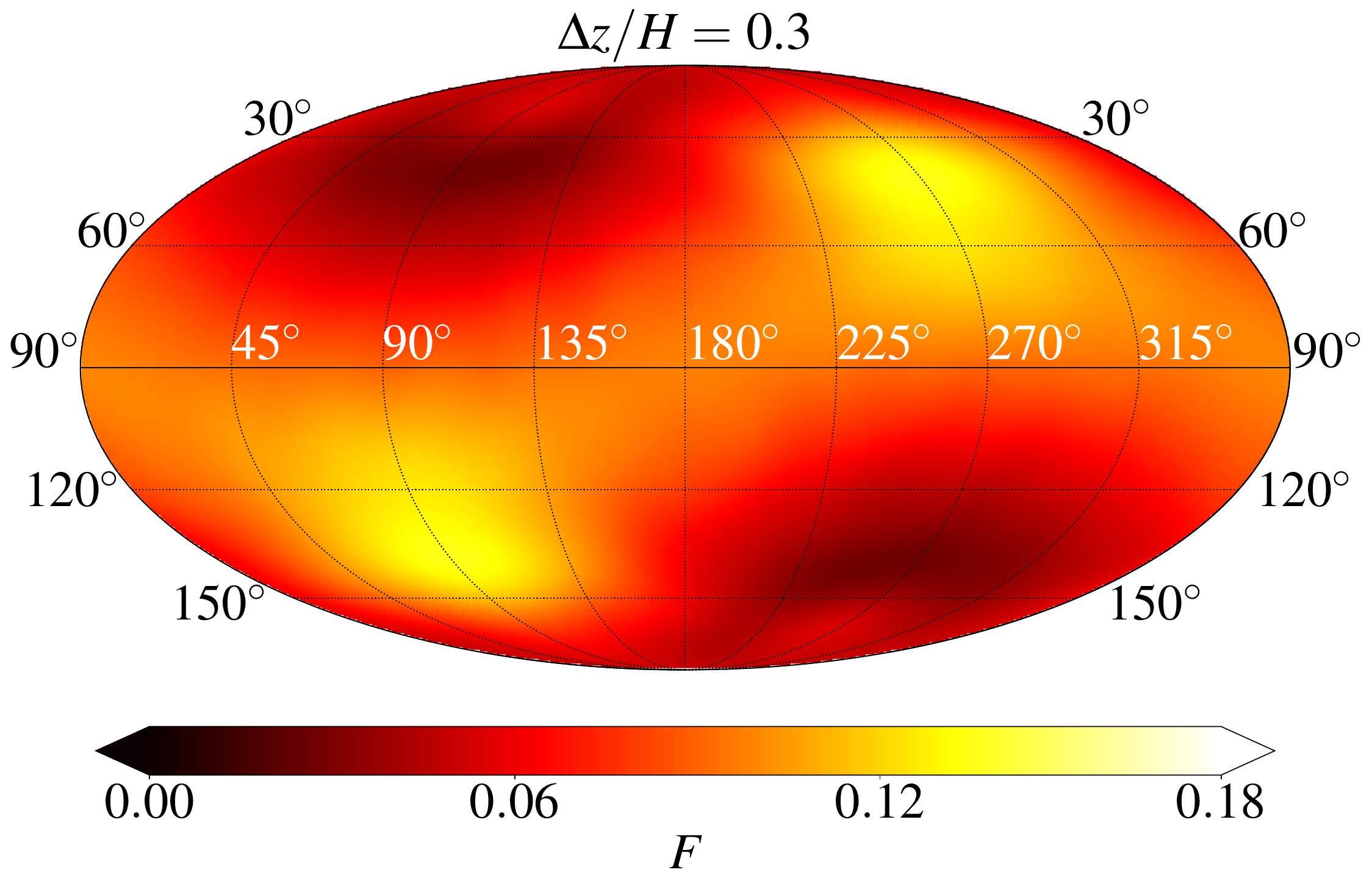}
	\end{minipage}
	\begin{minipage}[b]{.49\linewidth}
		\includegraphics[width=\textwidth]{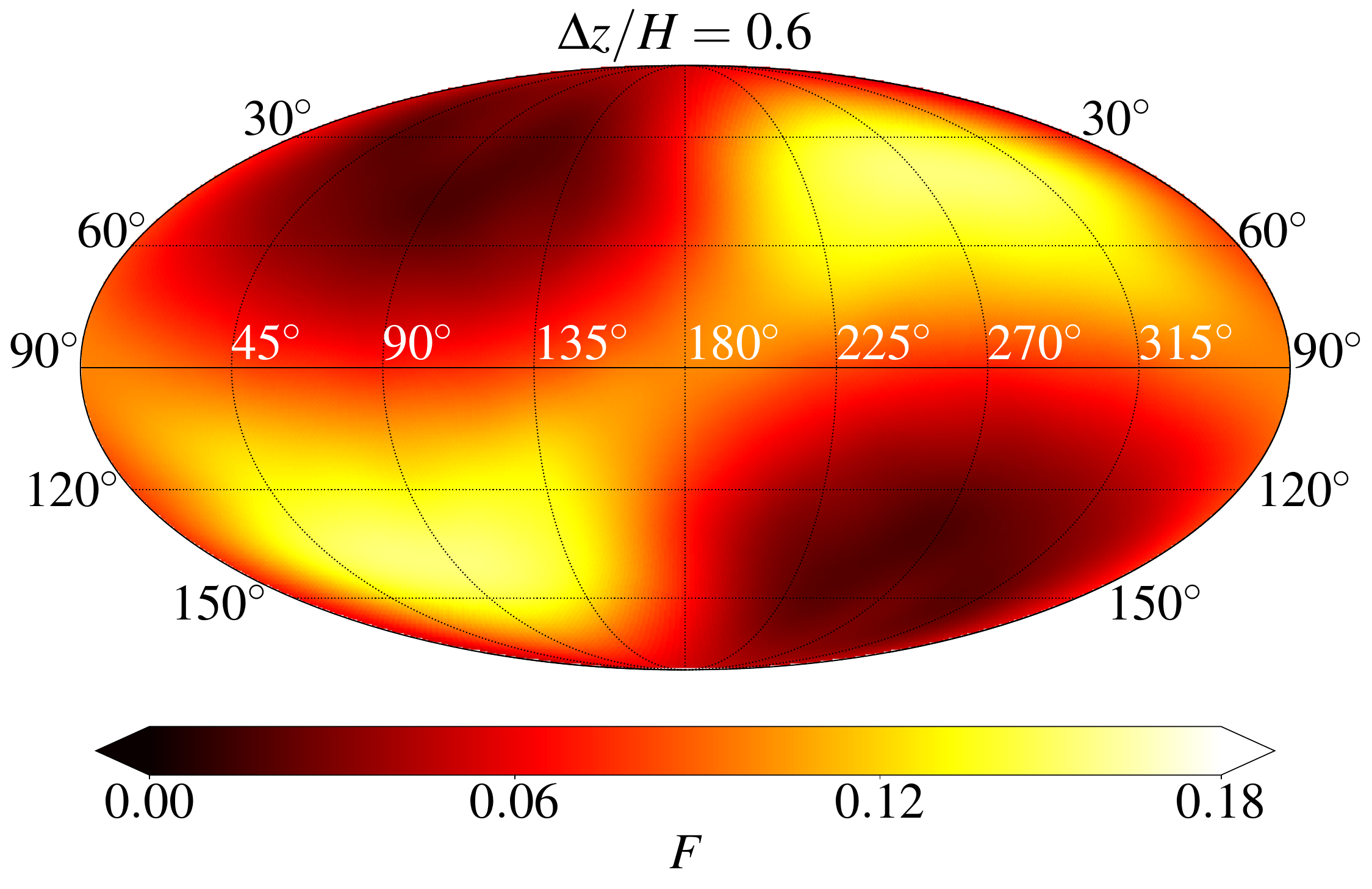}
	\end{minipage}
		\begin{minipage}[b]{.49\linewidth}
		\includegraphics[width=\textwidth]{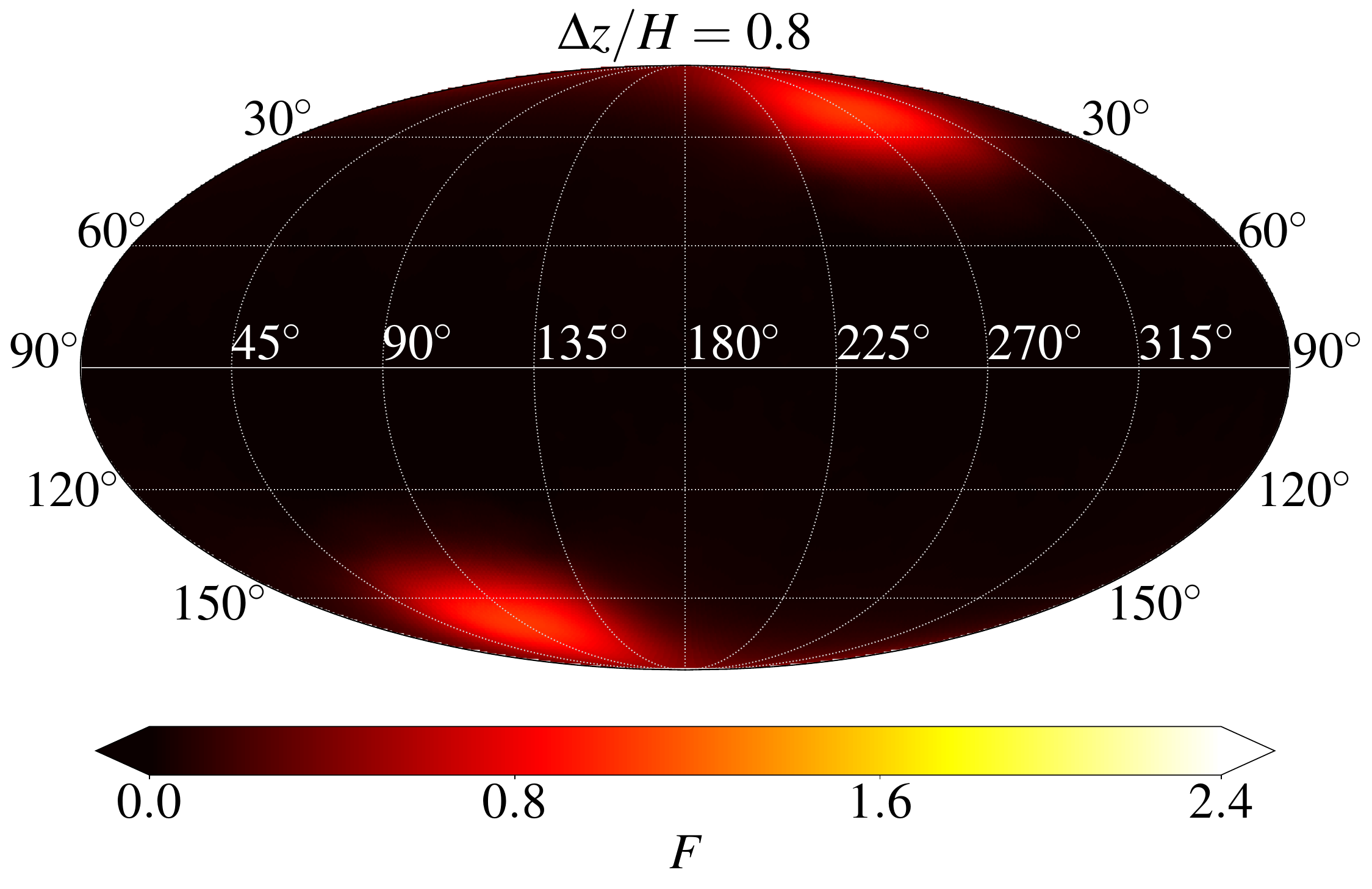}
	\end{minipage}\hfill
	\begin{minipage}[b]{.49\linewidth}
		\includegraphics[width=\textwidth]{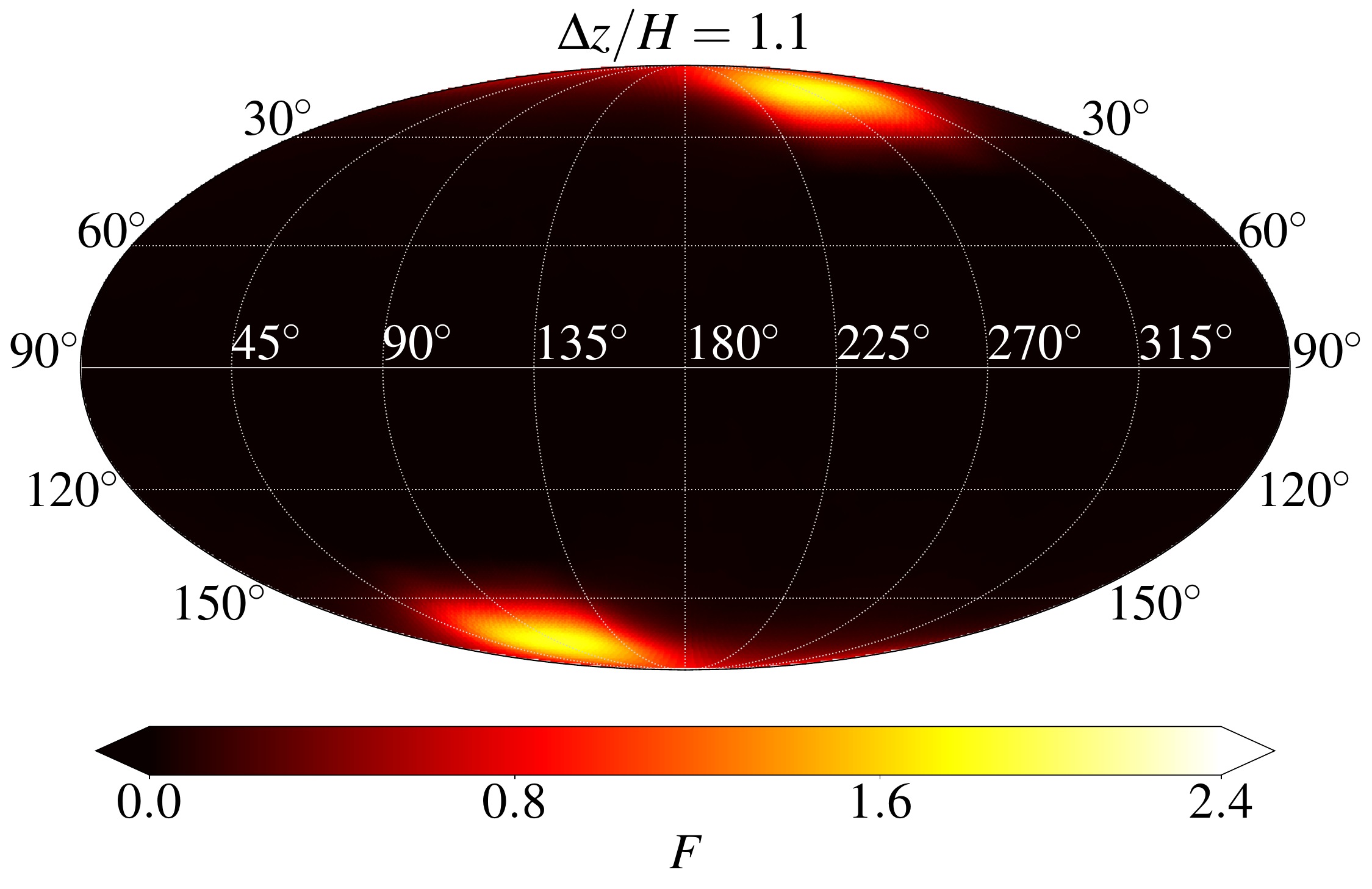}
	\end{minipage}
		\begin{minipage}[b]{.49\linewidth}
		\includegraphics[width=\textwidth]{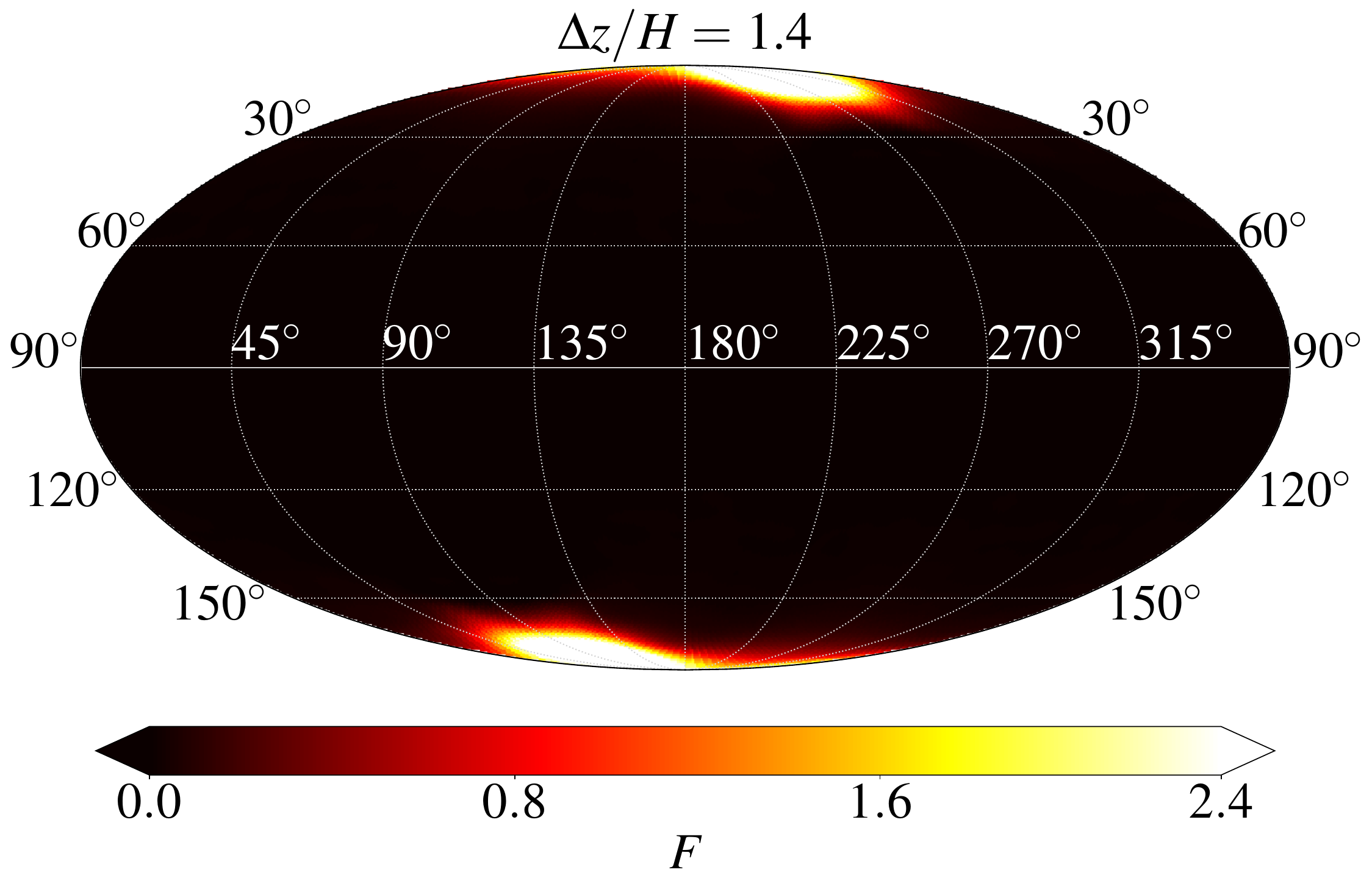}
	\end{minipage}
	\caption{Filled contours of spherical (Mollweide) projections of the normalized mass flux $F$ for the gas outflowing from the self-crossing region in terms of $\theta$ (values along the edges of projections) and $\phi$ (values along the horizontal directions) for different $\Delta z$. We use different colorbar limits for the strong (top left, top right and middle left panels) and the grazing (middle right, bottom left and bottom right panels) collision regimes, due to the large difference in the gas density. Projections are smoothed with a Gaussian symmetric beam implemented in \textsc{HEALPix}.}
	\label{fig:f}
\end{figure*}

In Figure \ref{fig:f} we show spherical projections of the normalized mass flux distributions for $\Delta z/H=0, 0.3, 0.6, 0.8, 1.1, 1.4$. For $\Delta z=0$ we can see that the outflow rate is higher near the equatorial plane, consistently with the findings by \citet{Lu_2019}.\footnote{We also notice a slight increase of $F$ at polar angles within $\sim 10^\circ$ degrees of the poles compared to the work by \citet{Lu_2019}, which is likely a consequence of the colliding with itself near this location. However, this difference is restricted to a small fraction of the outflowing gas and therefore does not affect our results.} As the offset between the streams increases, the outflow is more collimated and aligned with the direction of the incoming streams, as explained in Section \ref{sec:shock}. In Figure \ref{fig:f}, we also see a clear distinction between the two regimes we defined above. In the strong collision regime ($\Delta z \lesssim 0.7H$, first three panels), the outflow is partially deflected from the equatorial plane and spans a wide range of polar directions. In the grazing collision regime ($\Delta z \gtrsim 0.7H$, last three panels), the outflow spans only a narrow range of $\theta$ and $\phi$ with substantially higher values of the mass flux.

One purpose of calculating the normalized mass flux $F$ is related to the method of simulating the subsequent phase of accretion disc formation in TDEs, as described in \citet{Bonnerot_2020} and \citet{Bonnerot_2021} who model the outflow from the self-crossing shock through an injection of SPH particles from the intersection point. $F$ is a quantity that satisfies the condition $\int_0^\pi \int_0^{2\pi}F(\theta, \phi)\sin{\theta}\mathrm{d}\phi\mathrm{d}\theta=1$ when integrated over a unit sphere, which makes it the appropriate probability distribution to model the outflow using particle injection. For a given direction determined by $\theta$ and $\phi$ we obtain $F$ from a \textsc{HEALPix} map of $\dot{M}_{\rm out}$ (see Equation \ref{eq:F}). It is obtained directly for the values of $\Delta z$ we have simulated, while for the others we rely on linear interpolation.  For example, a \textsc{HEALPix} map for  $\Delta z=0.22\, H$ is obtained by linearly interpolating between  $\Delta z=0.2\, H$ and  $\Delta z=0.3\, H$.\footnote{We tried to obtain an analytic expression for the angular distribution of the outflow rate by fitting spherical harmonics functions to \textsc{HEALPix} maps of $\dot{M}_{\rm out}$. However, we found this procedure not suitable due to a high number of independent spherical harmonics functions needed to produce a good fit.} We have made the \textsc{HEALPix} maps obtained from our simulations and the code to extract $F$ from values of $\Delta z$, $\theta$ and $\phi$ publicly available to the community at \href{https://github.com/tajjankovic/Spin-induced-offset-stream-self-crossing-shocks-in-TDEs.git}{https://github.com/tajjankovic/Spin-induced-offset-stream-self-crossing-shocks-in-TDEs.git} for use in future research.

 \subsection{Subsequent evolution of the outflowing gas} \label{sec:bh_frame}
\noindent The outflow from the self-crossing shock can be either unbound and escape or bound and return to the black hole's vicinity, where it may form an accretion disc. The fate of the outflow can be determined by transforming the gas properties from the local co-moving frame to the black hole's reference frame, and calculating its specific energy $\epsilon$ and specific angular momentum $\ell$ as well as determining its trajectories. We use the description of the self-crossing shock presented in \citet{Lu_2019} and \citet{bonnerot_2020_book}, adopting a Newtonian description of the black hole's gravity for simplicity. 

We consider disruptions of a star with mass $M_\star$ and radius $R_\star$ by a black hole with mass $M_\mathrm{bh}$. The star is initially on a parabolic orbit, with an angular momentum vector inclined with respect to the black hole spin vector, and characterized by the impact parameter $\beta=R_\mathrm{t}/R_\mathrm{p}$, where $R_\mathrm{p}$ is the pericenter distance and $R_\mathrm{t}=R_\star (M_\mathrm{bh}/M_\star)^{1/3}$ is the tidal radius. The star has an initial specific angular momentum $\ell_\star = \sqrt{2GM_\mathrm{bh}R_\mathrm{p}}$. Disruption by the black hole's tidal forces induces an energy spread of the debris $\Delta\epsilon=GM_\mathrm{bh}R_\star/R_\mathrm{t}^2$ \citep{Stone_2013}. After the disruption, bound debris moves on a range of elliptic orbits, the most bound one having a specific energy $-\Delta \epsilon$ and eccentricity $e=1-\left ( 2/\beta \right ) \left(  M_\mathrm{bh}/M_\star\right)^{-1/3}$. We make a choice that the streams have the same energy $-\Delta \epsilon$, and we discuss the consequences of the streams having different energies in Section \ref{subsec:different_stream_properties}. Following its pericenter passage, the most bound debris now moving away from the black hole collides (for a low enough vertical offset) with the still-infalling gas at a distance given by the intersection radius

\begin{equation}\label{eq:Rint}
    R_\mathrm{int} = R_{\rm p} \frac{1+e}{1- e \cos (\Delta \phi/2)},
\end{equation}
where $\Delta \phi \approx 3\pi R_\mathrm{g}/R_\mathrm{p}$ is the apsidal precession. $R_\mathrm{int}$ determines the position in the black hole's reference frame of the local co-moving frame where the above simulations were carried out. We assume that this position is point-like, due to the condition $H\ll R_\mathrm{int}$ \citep{bonnerot2021nozzle}, and located along the $y$ axis at $z=0$. 

Defining $\mathbfit{e}_\mathrm{z}$ along the angular momentum of the in-falling stream component, Lense-Thirring precession can deflect the receding stream either in the $+\mathbfit{e}_\mathrm{z}$ or $-\mathbfit{e}_\mathrm{z}$ direction from the orbital plane. We focus here on the situation where the deflection is along the $-\mathbfit{e}_\mathrm{z}$ direction, since the other case leads to similar dynamics as long as the value of $\Delta z/H$ is the same. In the black hole's reference frame, the stream receding from the black hole has velocity $\mathbfit{v}_1^\mathrm{int}$, while $\mathbfit{v}_2^\mathrm{int}$ is the velocity of the in-falling gas. The two stream components have the same magnitude $v_\mathrm{int}=|\mathbfit{v}_1^\mathrm{int}|=|\mathbfit{v}_2^\mathrm{int}|$. We determine the speed of the colliding flows at the intersection point as  $v_\mathrm{int}= \left ( 2G M_\mathrm{bh}/R_\mathrm{int} - 2\Delta \epsilon \right )^{1/2}$  and calculate the intersection angle measured between the two velocity vectors $\cos \psi =\mathbfit{v}_1^\mathrm{int} \cdot \mathbfit{v}_2^\mathrm{int}/v_\mathrm{int}^2$, as in \citet{Dai_2015}
\begin{equation}
    \cos{\psi} = \frac{1-2e \cos(\Delta \phi/2) + e^2\cos \Delta \phi }{1-2e \cos(\Delta \phi/2)+e^2}.
\end{equation}
We use $v_\mathrm{int}$ and $\psi$ to determine the radial $v_\mathrm{r}=\mathbfit{v}_1^\mathrm{int}\cdot \mathbfit{e}_\mathrm{y}=-\mathbfit{v}_2^\mathrm{int}\cdot \mathbfit{e}_\mathrm{y}=v_\mathrm{int} \sin (\psi/2)$ and tangential $v_\mathrm{t}=\mathbfit{v}_1^\mathrm{int}\cdot \mathbfit{e}_\mathrm{x}=\mathbfit{v}_2^\mathrm{int}\cdot \mathbfit{e}_\mathrm{x}=v_\mathrm{int} \cos (\psi/2)$ velocity components. In the local co-moving frame, moving with a velocity equal to $v_\mathrm{t}$, streams collide with velocities $\pm v_\mathrm{r}$ and a shock is formed that launches an outflow (see Section \ref{subsec:angular_dep}). We fix the velocity of the outflow to $|v_\mathrm{r}|$ (same as the collision speed) as found in the simulations. Therefore, the velocity along an arbitrary direction given by the polar and azimuthal angles $\theta$ and $\phi$ in the co-moving frame (corresponding to an arbitrary pixel in the flux map of Figure \ref{fig:f}) can be determined from $\mathbfit{v}_\mathrm{out} (\theta, \phi)=v_\mathrm{r} (\sin \theta \cos \phi \mathbfit{e}_\mathrm{x} +\cos \theta \mathbfit{e}_\mathrm{y} +\sin \theta \sin \phi \mathbfit{e}_\mathrm{z})$. The velocity components of the outflow in the black hole's frame are then obtained via the Galilean transformation $\mathbfit{v}_\mathrm{out}'(\theta, \phi)=\mathbfit{v}_\mathrm{out}(\theta, \phi) + v_\mathrm{t}\mathbfit{e}_\mathrm{x}$.

\begin{figure}
	\centering
	\begin{minipage}[b]{\linewidth}
		\includegraphics[width=\textwidth]{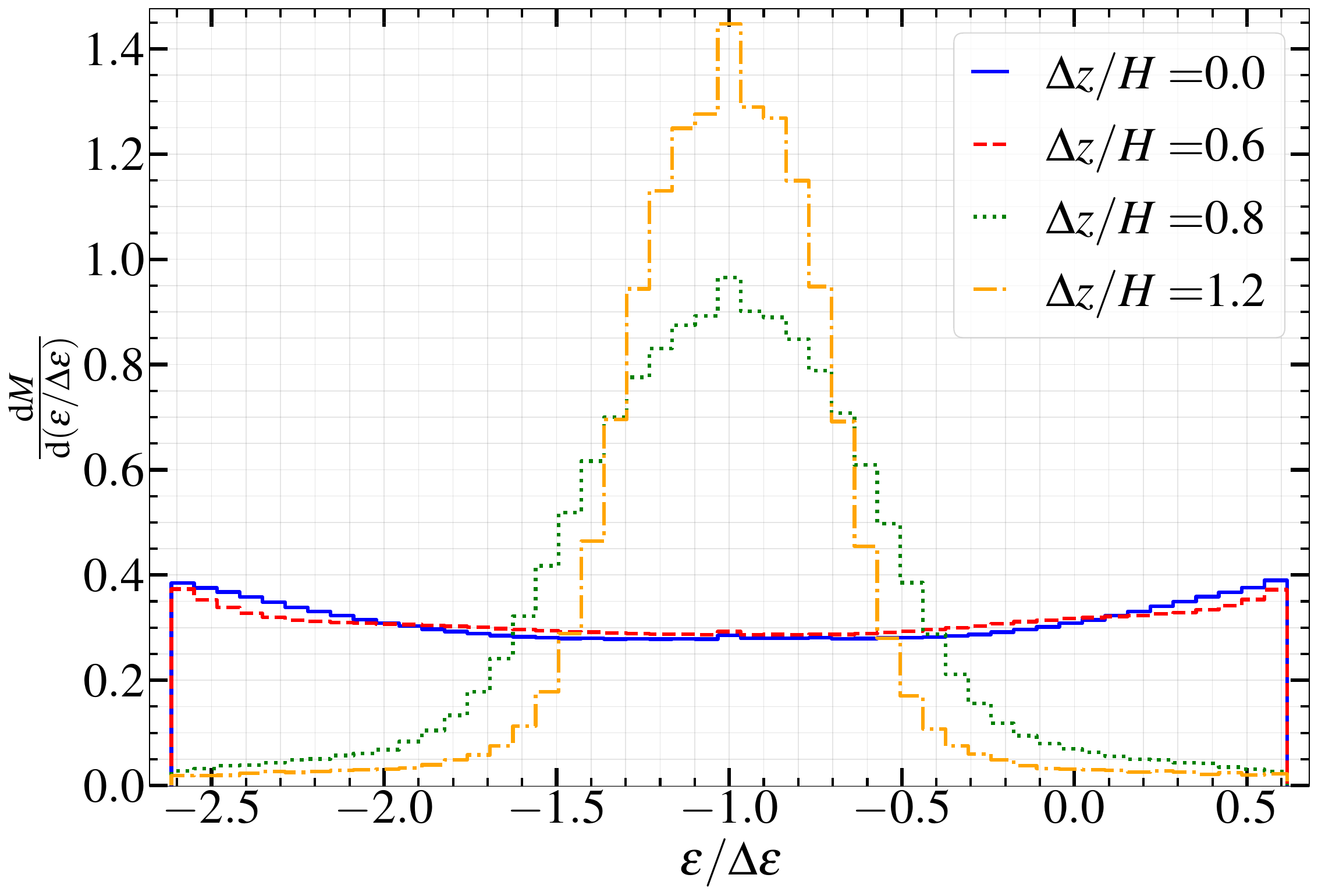}
	\end{minipage}
		\begin{minipage}[b]{\linewidth}
		\includegraphics[width=\textwidth]{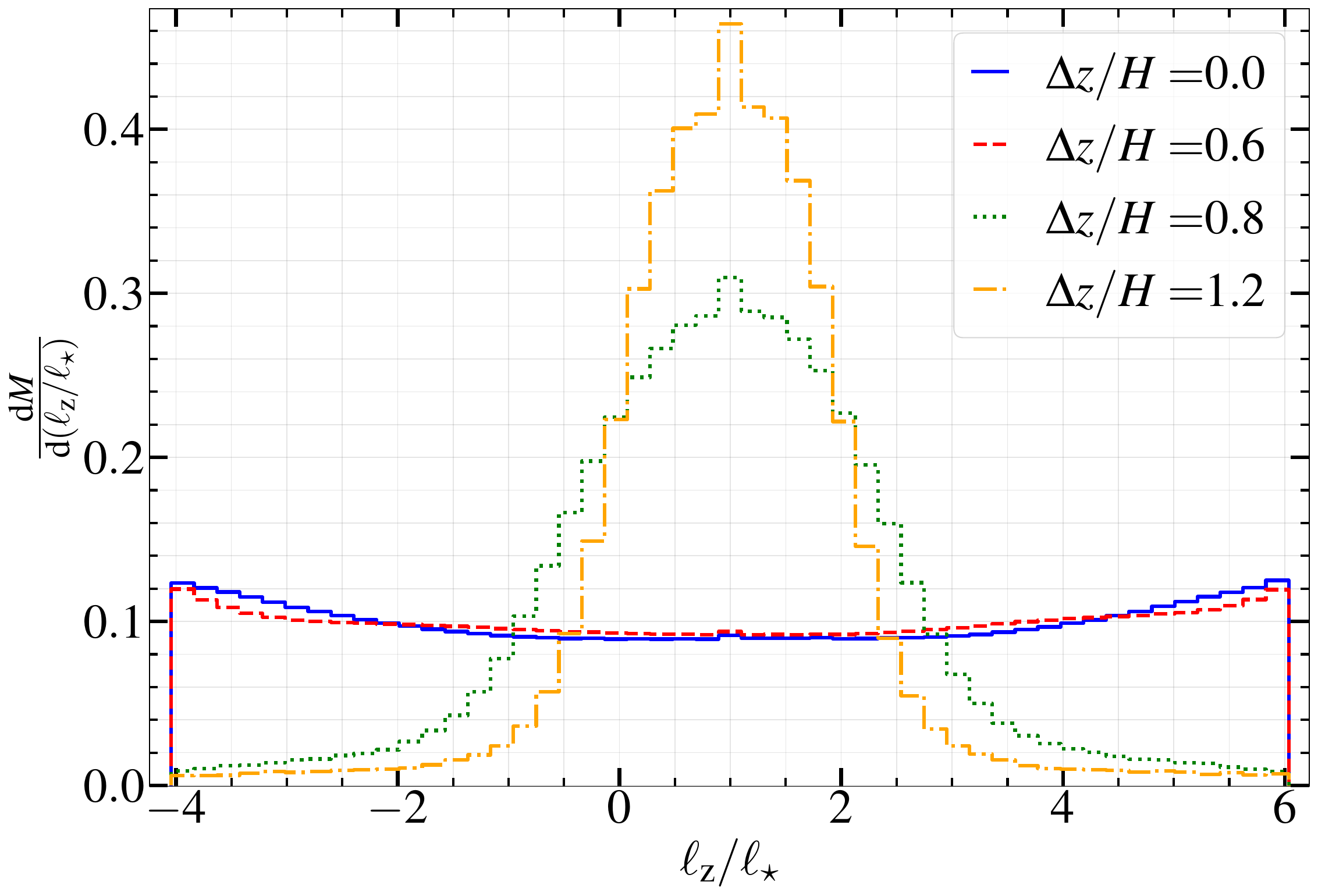}
	\end{minipage}

	\caption{Distributions of orbital energy $\epsilon$ (upper panel) and angular momentum $\ell_\mathrm{z}$ projected along the initial stellar angular momentum (lower panel) for different values of the offset $\Delta z/H$, assuming $M_\star = 0.5\, \mathrm{M_\odot}$, $R_\star=0.46\, \mathrm{R_\odot}$, $M_\mathrm{bh} = 2.5\times 10^6\, \mathrm{M_\odot}$ and $\beta=1$.  Each bin of the histograms is weighted by the total flux (mass fraction) within its indicated specific energy or angular momentum range. $\epsilon$ and $\ell_\mathrm{z}$ are normalized by the values before the collision $\Delta \epsilon$ and $\ell_\star$, respectively. The integral of distributions is normalized to 1.}
	\label{fig:dM}
\end{figure}

We calculate the specific energy $\epsilon=-GM_\mathrm{bh}/R_\mathrm{int}+v_\mathrm{out}'^2/2$ and the specific angular momentum $\boldsymbol{\ell}=  \mathbfit{R}_\mathrm{int} \times \mathbfit{v}_\mathrm{out}'$ of the outflow in the black hole's reference frame, where $\mathbfit{R}_\mathrm{int}=R_\mathrm{int} \mathbfit{e}_\mathrm{y}$ is the radial vector to the intersection point. We define a 2D grid of $\theta$ and $\phi$, and determine $\epsilon$ and $\boldsymbol{\ell}$ for every $(\theta, \phi)$ pair. We obtain the normalized flux $F$ (see Equation (\ref{eq:F})) for these angles by interpolating from \textsc{HEALPix} maps, which allows us to calculate the flux-weighted distributions of $\epsilon$ and $\ell_\mathrm{z}=\boldsymbol{\ell} \cdot \mathbfit{e}_\mathrm{z}$ shown in Figure \ref{fig:dM} for different $\Delta z/H$. We note that negative values of $\ell_\mathrm{z}/\ell_\star$ correspond to a part of the outflow that is counter-rotating (moving on retrograde orbits) with respect to the initial stellar trajectory. The stream collision induces a spread in the distributions of $\epsilon$ and $\ell_\mathrm{z}$ around the values before the collision, $-\Delta \epsilon$ and $\ell_\star$, respectively. In the strong collision regime, where $\Delta z\lesssim0.7H$, the offset distributions are very similar to the $\Delta z/H=0$ case, displaying a sharp cut-off, where the outflow is maximally accelerated or decelerated during the collision. In the grazing regime, where $\Delta z\gtrsim 0.7H$, the outflow is more collimated, and the distributions are narrower with a pronounced peak around pre-collision values. The widths of the distributions are no longer determined by a sharp cut-off, but instead gradually decrease due to the decrease in the mass flux.

At a fixed $\Delta z/H$ and $M_\mathrm{bh}$ the flux-weighted distributions of $\epsilon$ and $\ell_\mathrm{z}$  distributions have the same shape. This can be shown by calculating analytically the maximum spread in energy $\Delta \epsilon_\mathrm{max}$ and angular momentum $\Delta \ell_\mathrm{max}$ after the collision. By introducing a unit vector $\mathbfit{e}_\mathrm{r,max}$, which points radially from the collision point in the direction where the gas has the largest energy and angular momentum, it is possible to derive $\Delta \epsilon_\mathrm{max} = v_\mathrm{r}v_\mathrm{t}\mathbfit{e}_\mathrm{r,max} \cdot \mathbfit{e}_\mathrm{\phi}$ and $\Delta \ell_\mathrm{max} = R_\mathrm{int}v_\mathrm{r}\mathbfit{e}_\mathrm{r,max} \cdot \mathbfit{e}_\mathrm{\phi}$. Therefore, in a disruption of a star with $M_\star=0.5 \,\mathrm{M_\odot}$, $R_\star=0.46\, \mathrm{R_\odot}$\footnote{Stellar radius is obtained from fitting formulae for the zero-age main-sequence radii as functions of their masses presented in \citet{Tout_199610.1093/mnras/281.1.257} assuming solar metallicity.} on a $\beta=1$ orbit  by a $M_\mathrm{bh} = 2.5 \times 10^6 \,\mathrm{M_\odot}$ black hole, $\Delta \ell_\mathrm{max}/\Delta \epsilon_\mathrm{max}=R_\mathrm{int}/v_\mathrm{t}\approx  3.3 \ell_\star/\Delta \epsilon$, which is in agreement with the ratio obtained by comparing the two panels of Figure \ref{fig:dM}. The fate of the outflow can be understood by considering the spread of flux-weighted distributions of $\epsilon$ and $\ell_\mathrm{z}$, around the values before the collision, which depend on the black hole mass. In disruptions by black holes with $M_\mathrm{bh}\in [10^6,10^7]\, \mathrm{M_\odot}$, approximations  $e\approx 1$, and $0<\Delta\phi \lesssim \pi/2$ are valid. By expanding $\cos  (\Delta\phi/2)$ to the second order in Taylor series in Equation $\eqref{eq:Rint}$, we find $R_\mathrm{int}\propto M_\mathrm{bh}^{-1}$. Additionally, we use approximations $v_\mathrm{int} \approx v_\mathrm{r}$ and $ v_\mathrm{t}\approx \ell_\star/R_\mathrm{int}$ to derive $v_\mathrm{r}\propto M_\mathrm{bh} $ and $v_\mathrm{t} \propto M_\mathrm{bh}^{5/3}$, respectively.  By assuming a collision with no offset, where $\mathbfit{e}_\mathrm{r,max} \cdot \mathbfit{e}_\mathrm{\phi}=1$, we obtain $\Delta\epsilon_\mathrm{max}/\Delta\epsilon\propto M_\mathrm{bh}^{7/3}$ and $\Delta\ell_\mathrm{max}/\ell_\star\propto M_\mathrm{bh}^{-2/3}$, meaning that there is more unbound gas and less gas on retrograde orbits as $M_\mathrm{bh}$ increases.

 From the flux-weighted distributions of $\epsilon$ and $\ell_\mathrm{z}$ we calculate the distribution of the pericenter distances of the outflow. As $\Delta z/H$ increases, the distributions become narrower around the stellar pericenter $\approx R_\mathrm{p}$ due to the more stream-like outflow. However, even for $\Delta z=1.8H$ that is the most grazing collision we considered, the spread remains of order unity with $\Delta R_\mathrm{p}\approx R_\mathrm{p} \gg R_\star$, implying that the shocked stream has expanded significantly compared to its pre-collision state. This expansion along with the resulting differences in apsidal precession angles will likely affect the subsequent gas evolution, as we discuss in Section \ref{subsec:circularization}.

\begin{figure}
	\centering
		\includegraphics[width=\linewidth]{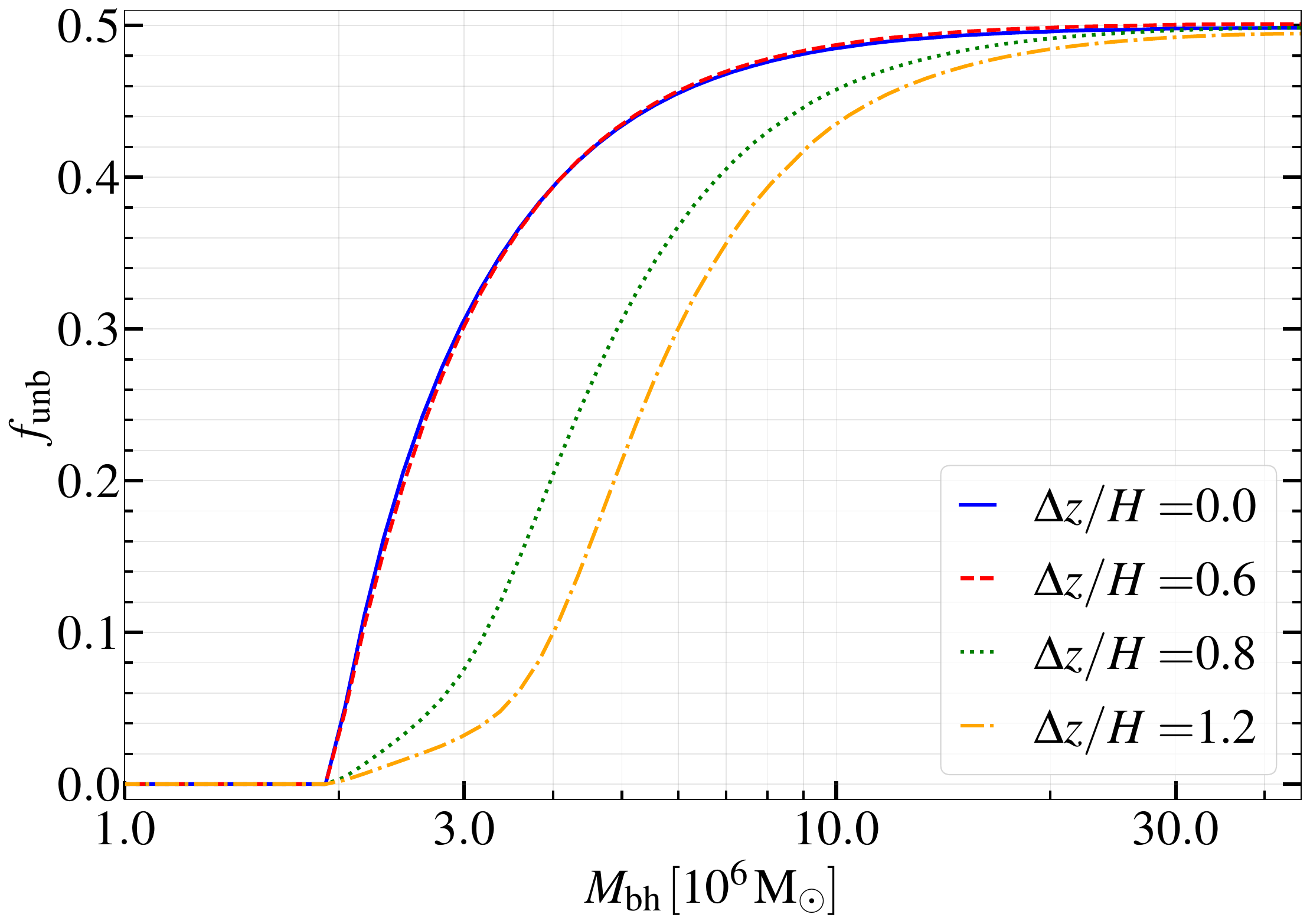}
	\caption{The mass fraction $f_\mathrm{unb}$ of the unbound gas for different stream offsets $\Delta z/H$ and black hole masses $M_\mathrm{bh}$, for $M_\star = 0.5\, \mathrm{M_\odot}$, $R_\star=0.46\, \mathrm{R_\odot}$ and $\beta=1$. }
	\label{fig:mass_frac}
\end{figure}

The unbound outflow has a positive energy, while gas with a negative energy is bound. The mass fractions $f_\mathrm{unb}$ of unbound gas for disruptions of a $M_\star=0.5\,\mathrm{M_\odot}$ star as a function of $M_\mathrm{bh}$ for $\Delta z/H=0$, 0.6, 0.8, 1.2 are shown in Figure \ref{fig:mass_frac}.\footnote{For $M_\star=0.5\, \mathrm{M_\odot}$ and $M_\mathrm{bh}\gtrsim 6.2\times 10^6\, \mathrm{M_\odot}$ the initial angular momentum $\ell_\star$ is lower than the angular momentum of the marginally bound parabolic orbit in the Schwarzschild spacetime $\ell_\mathrm{mb}=4R_\mathrm{g}c$. For clarity, we calculate $f_\mathrm{unb}$ also for higher $M_\mathrm{bh}$ even if all of the gas is swallowed by the black hole.} We see, that for $M_\mathrm{bh}\lesssim 2\times 10^6\,\mathrm{M_\odot}$ there is only bound gas, due to a smaller spread in energy. As the black hole mass increases, the fraction of unbound gas $f_\mathrm{unb}$ increases up to 50\%, which is reached for $M_\mathrm{bh}\gtrsim 3\times 10^7 \,\mathrm{M_\odot}$. The dependence on $\Delta z/H$ in the strong collision regime (blue solid and red dashed lines) is very weak due to very similar flux-weighted distributions $\epsilon$ and $\ell_\mathrm{z}$ (see Figure \ref{fig:dM}). The change in $f_\mathrm{unb}$ is most apparent in the transition from the strong to the grazing collision regime (green dotted and orange dash-dotted lines).

\begin{figure}
	\centering
 \begin{minipage}[b]{0.96\linewidth}
		\includegraphics[width=\textwidth]{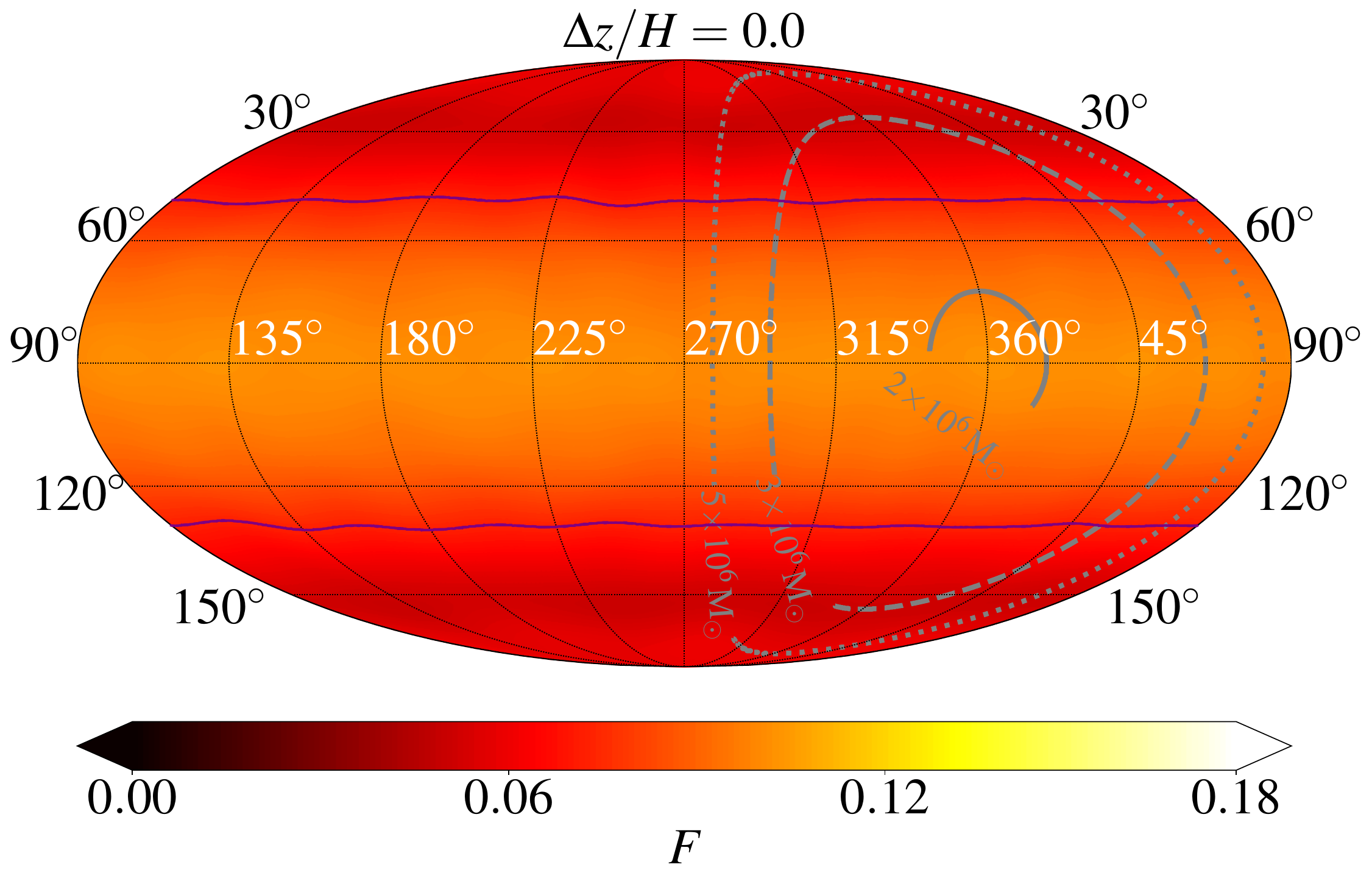}
	\end{minipage}
	\begin{minipage}[b]{0.96\linewidth}
		\includegraphics[width=\textwidth]{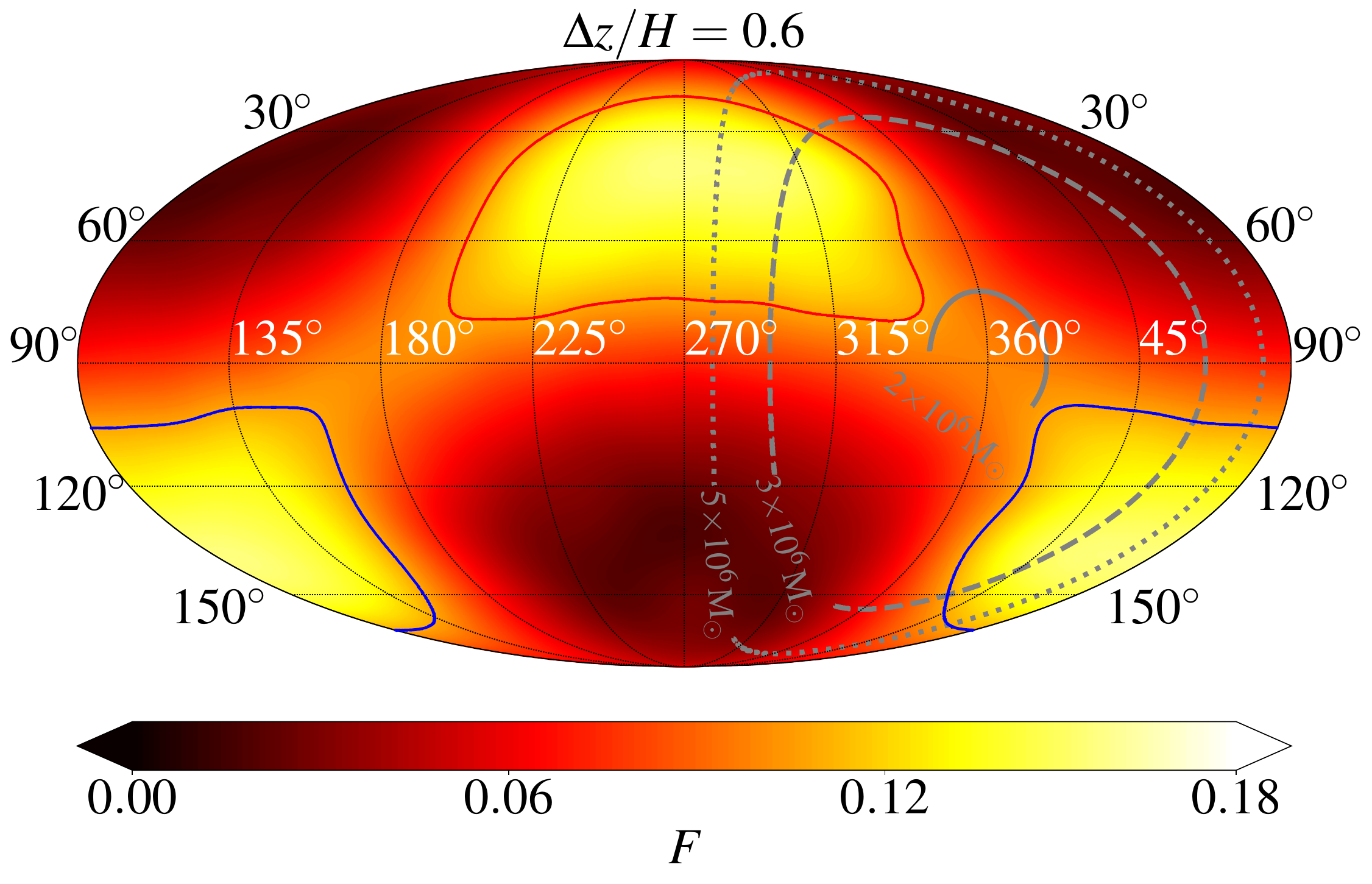}
	\end{minipage}
	\begin{minipage}[b]{0.96\linewidth}
		\includegraphics[width=\textwidth]{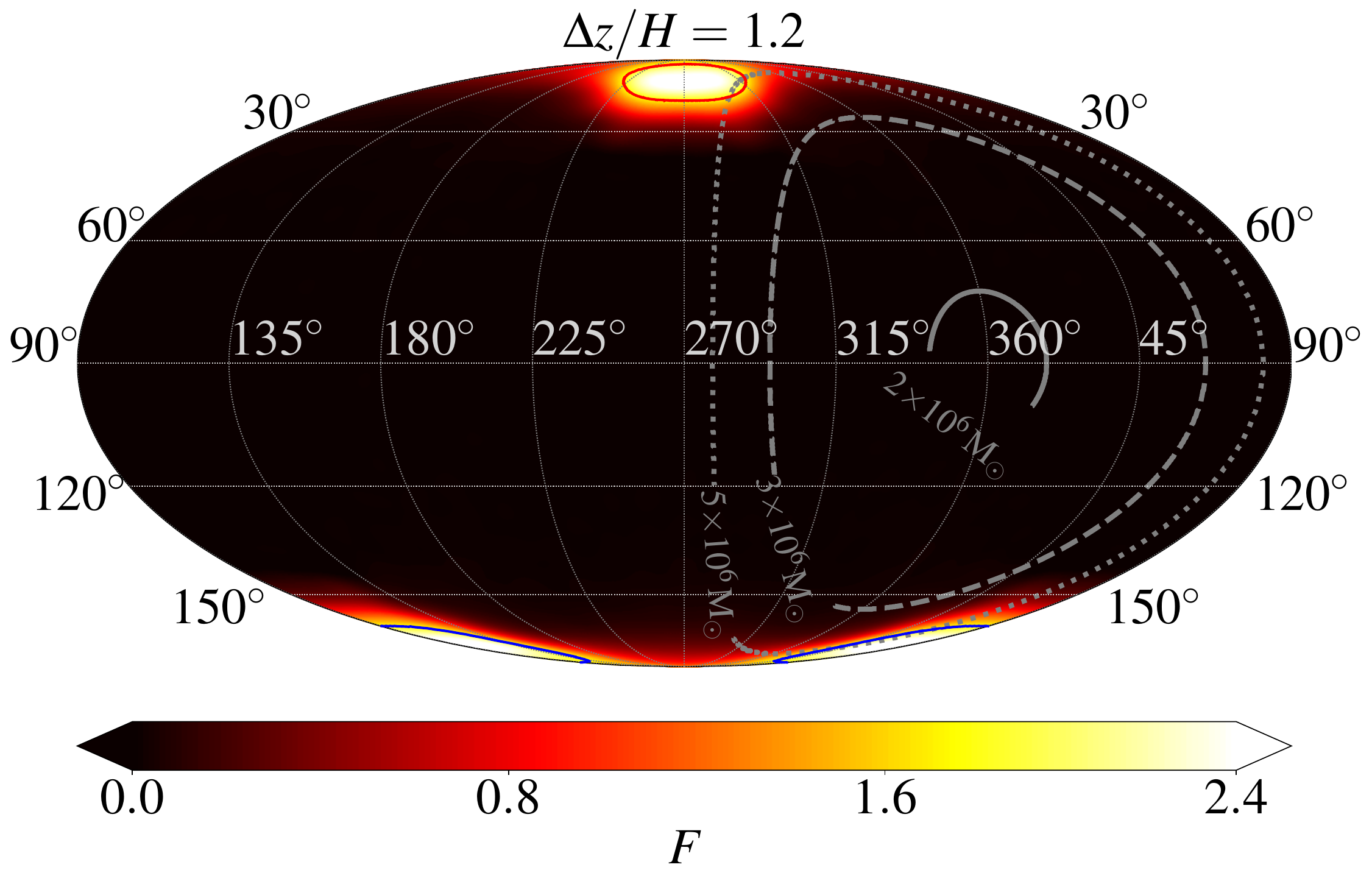}
	\end{minipage}

	\caption{Filled contours of spherical (Mollweide) projections of the normalized mass flux $F$ for the gas outflowing from the self-crossing region for $\Delta z/H=0$ (top row), 0.6 (middle row) and 1.2 (bottom row). We use different colourbar limits for the strong and the grazing collision regime, due to the large difference in the gas density. Projections are smoothed with a Gaussian symmetric beam implemented in \textsc{HEALPix}. Grey lines represent the zero-energy contours for $M_\mathrm{bh}\, =2$, 3 and $5\times 10^6\, \mathrm{M_\odot}$, which are denoted by solid, dashed and dotted line styles, respectively. We use values $M_\star =0.5\, \mathrm{M_\odot}$,  $R_\star=0.46\, \mathrm{R_\odot}$ and $\beta=1$. We note that flux projections are rotated by an angle $\phi=-90^\circ$ in comparison to Figure \ref{fig:f}. Purple, red, and blue lines denote contours corresponding to a value of $F$ equal to 70\% of its maximal value. We use these contours to select the outflow segments for which we calculate the trajectories shown in Figure \ref{fig:3dtraj}. }
	\label{fig:flux_fate}
\end{figure}

At a fixed $M_\mathrm{bh}$ there is less unbound outflow for higher $\Delta z/H$. This can be better understood from Figure \ref{fig:flux_fate} where we show spherical projections of the normalized mass flux $F$ from the self-crossing region\footnote{Figure \ref{fig:flux_fate} is similar to Figure \ref{fig:f}, except that it is rotated by an angle $\phi=-90^\circ$, meaning that the center and the boundary of Figure \ref{fig:flux_fate} are in the $-\mathbfit{e}_\mathrm{z}$ and $+\mathbfit{e}_\mathrm{z}$ direction, respectively. From Figure \ref{fig:f} one might expect that the stream components for $\Delta z/H=1.2$ would appear symmetric with respect to the horizontal axis. However, the stream moving toward $-\mathbfit{e}_\mathrm{z}$ direction is located on the boundary of the Figure \ref{fig:flux_fate} and therefore appears to be more extended due to projection effects.} and highlight the zero-energy contours --- the unbound gas is located inside the grey borders for $M_\mathrm{bh}=2\times 10^6\, \mathrm{M_\odot}$ (solid line), $3\times 10^6\, \mathrm{M_\odot}$ (dashed line) and $5\times 10^6\, \mathrm{M_\odot}$ (dotted line). In the strong collision regime (two uppermost panels),  the contour for $M_\mathrm{bh}= 3\times 10^6\, \mathrm{M_\odot}$ already encloses a large fraction of the unbound outflow. However, in the grazing collision regime (lowermost panel), the unbound region extends over the outflow only for $M_\mathrm{bh}= 5\times 10^6\, \mathrm{M_\odot}$, which is consistent with the black hole mass above which the unbound fraction becomes significant (see Figure \ref{fig:mass_frac}). 

\begin{figure*}
	\centering
		\begin{minipage}[b]{0.4\linewidth}
		\includegraphics[width=\textwidth]{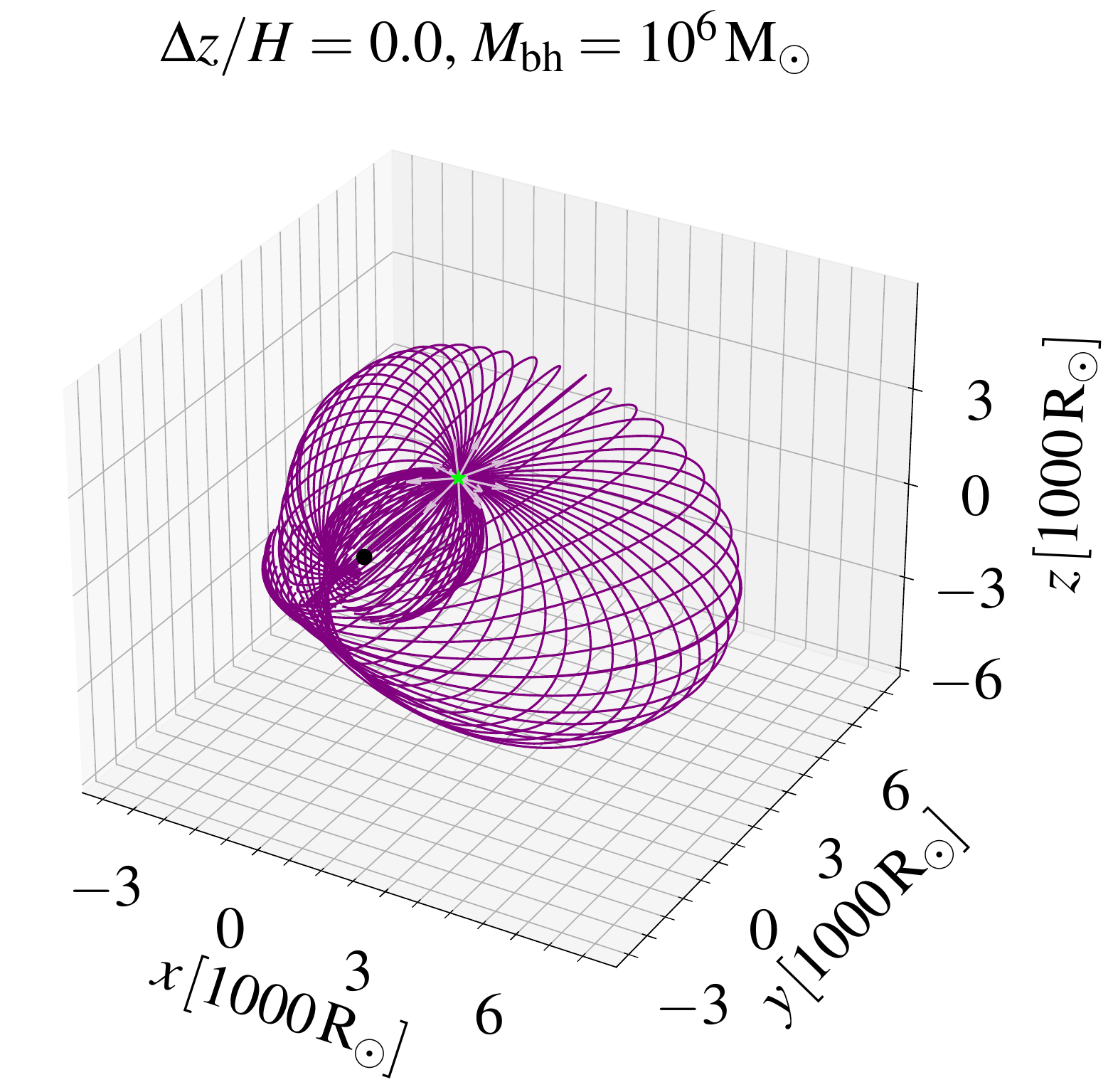}
	\end{minipage}
	\begin{minipage}[b]{0.4\linewidth}
		\includegraphics[width=\textwidth]{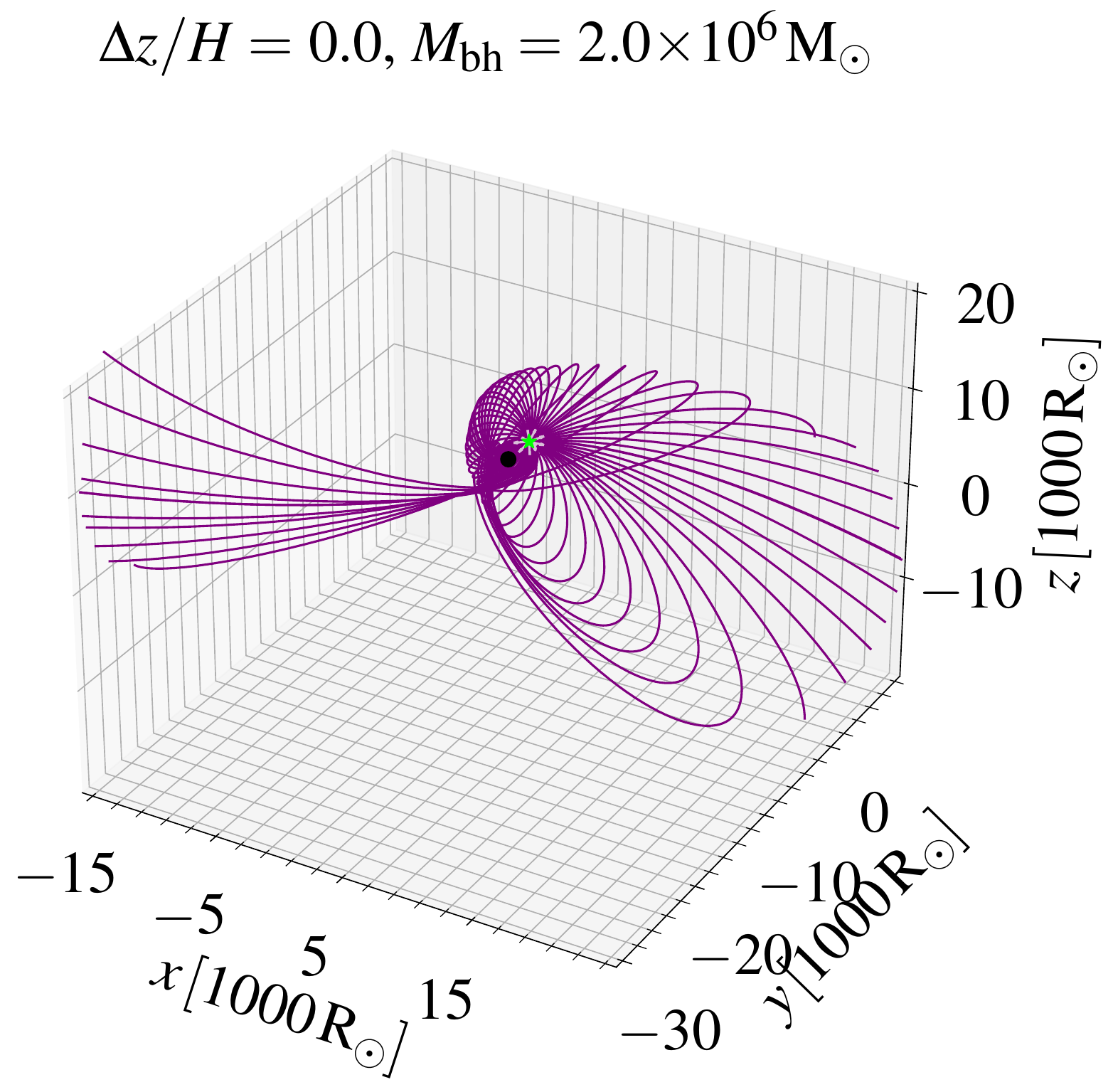}
	\end{minipage}
			\begin{minipage}[b]{0.4\linewidth}
		\includegraphics[width=\textwidth]{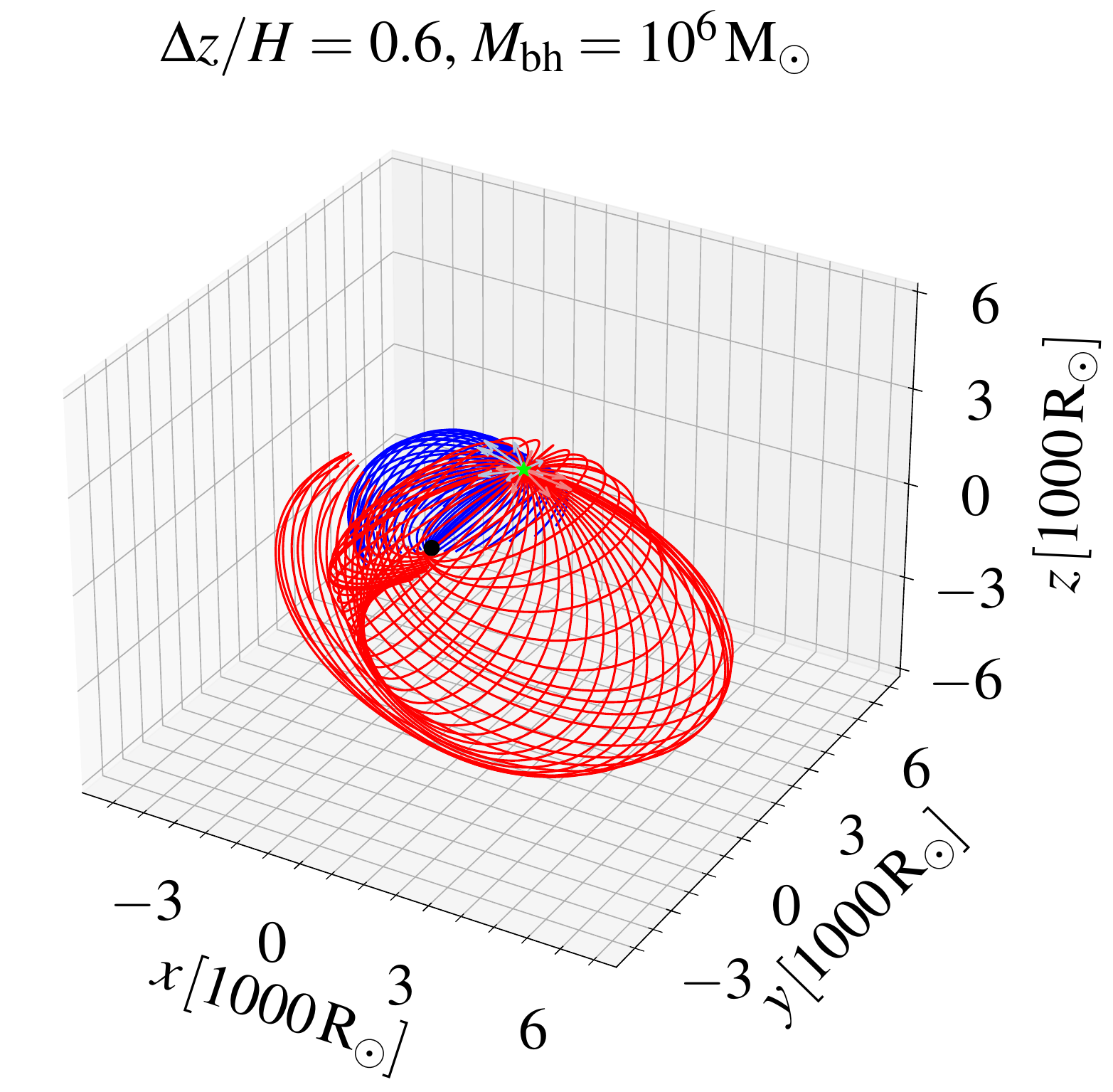}
	\end{minipage}
			\begin{minipage}[b]{0.4\linewidth}
		\includegraphics[width=\textwidth]{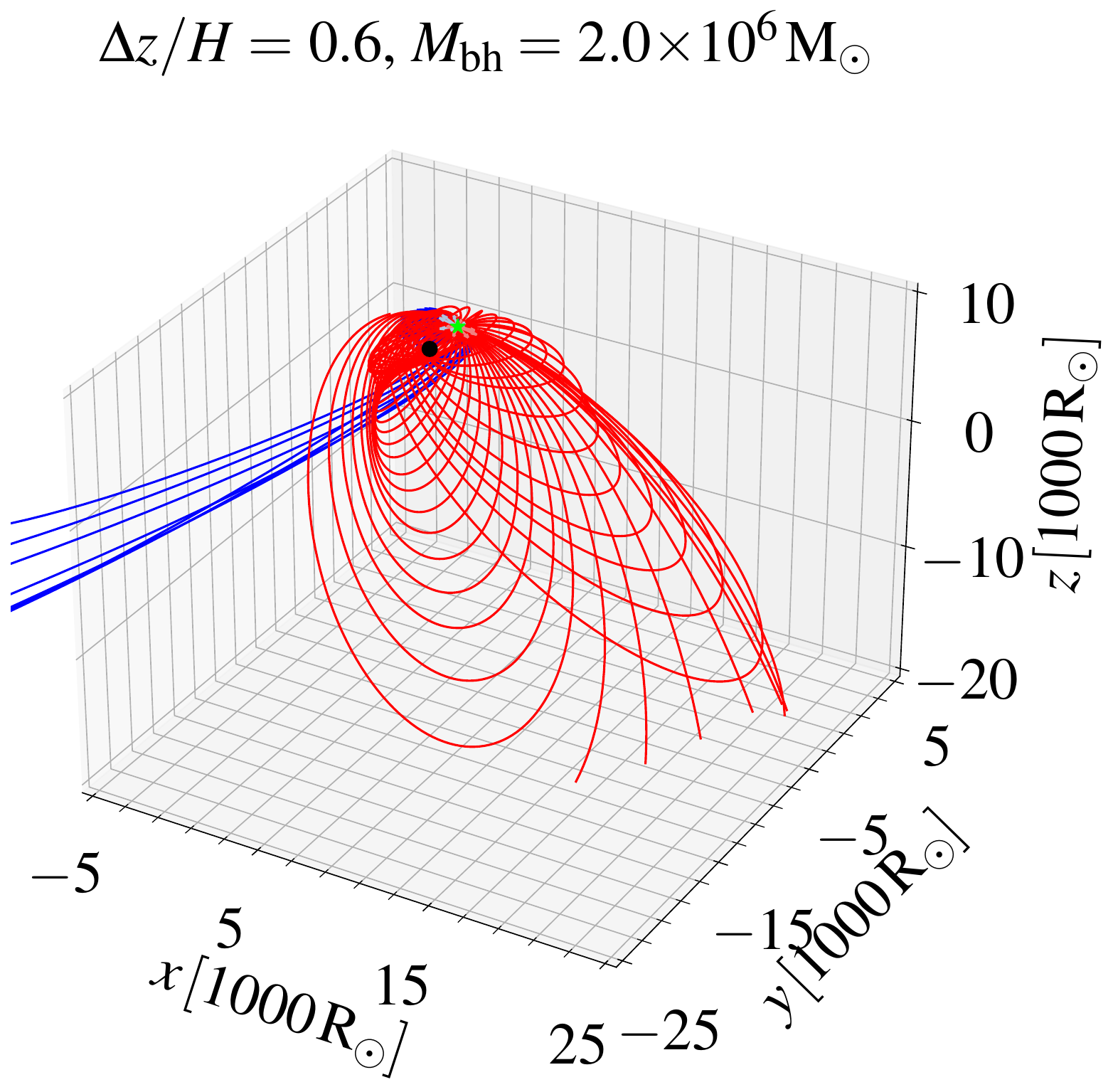}
	\end{minipage}
			\begin{minipage}[b]{0.4\linewidth}
		\includegraphics[width=\textwidth]{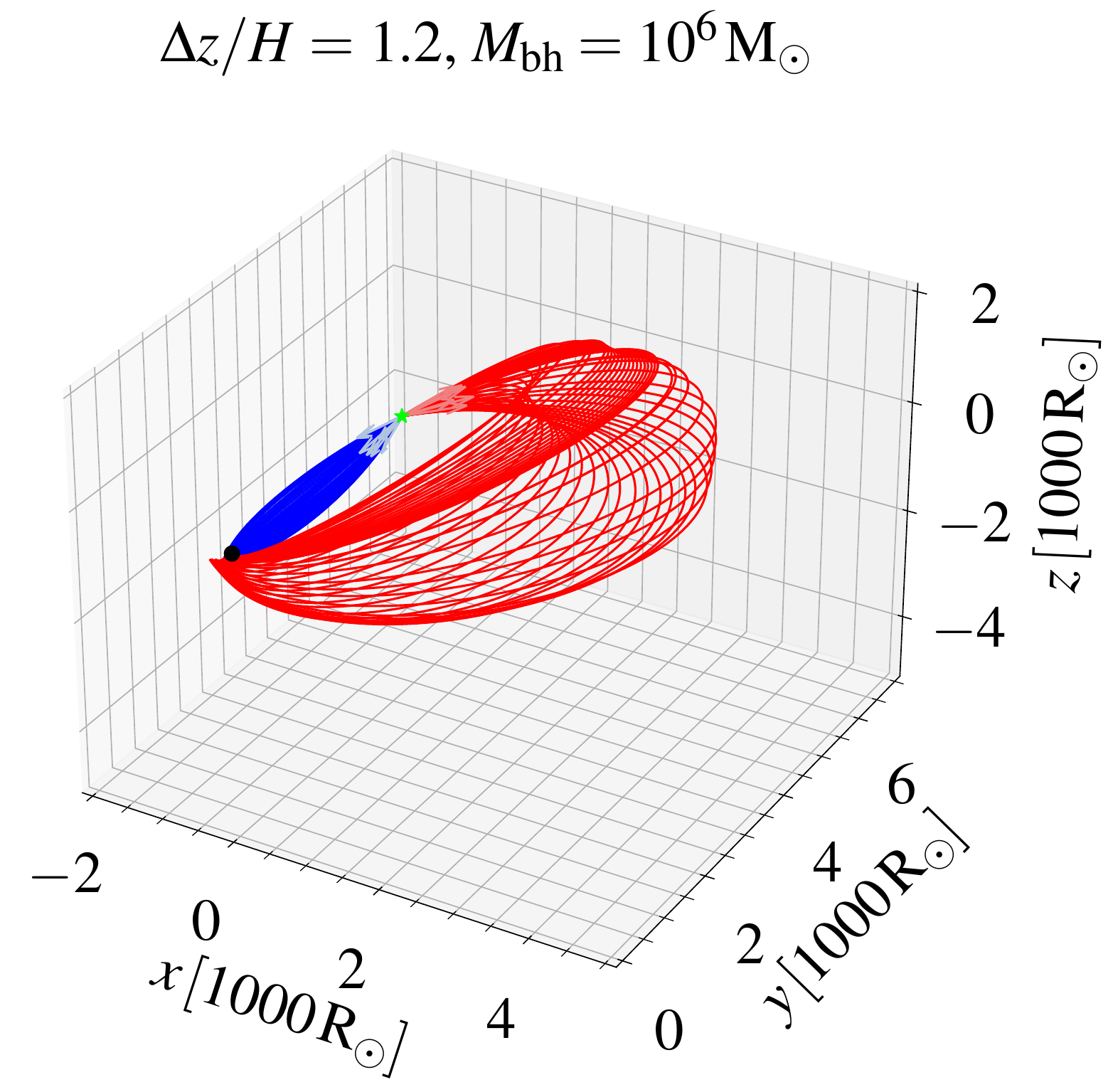}
	\end{minipage}
			\begin{minipage}[b]{0.4\linewidth}
		\includegraphics[width=\textwidth]{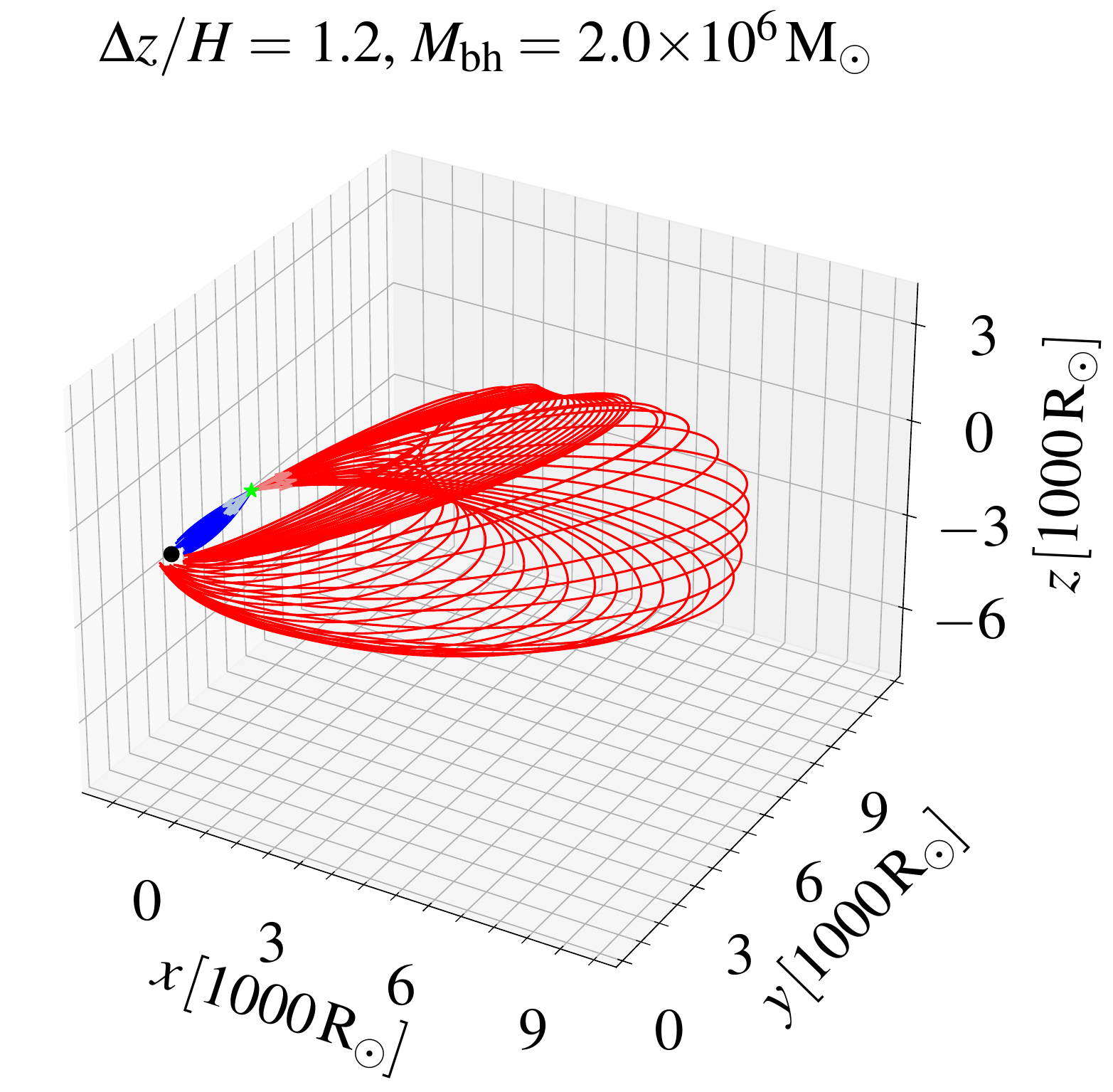}
	\end{minipage}
		
	\caption{Isometric views of 3D trajectories in the black hole's reference frame of 100 outflow segments selected from contours where $F$ is equal to 70\% of its maximal value for $\Delta z/H=0$, 0.6, 1.2 and $M_\mathrm{bh}= 10^6\, \mathrm{M_\odot}$, $2\times 10^6\, \mathrm{M_\odot}$ for $M_\star = 0.5\, \mathrm{M_\odot}$, $R_\star=0.46\, \mathrm{R_\odot}$ and $\beta=1$. We use red and blue colours for segments located on the northern and southern contours, respectively (see Figure \ref{fig:flux_fate}). For $\Delta z/H=0$ we use only purple lines to illustrate that the bulk of the outflow is contained inside a single region, while for offset collisions the outflow is contained inside two regions. Most elliptic trajectories are integrated for radial distances smaller than the pericenter distance, while hyperbolic trajectories are integrated for radial distances $r<3\times 10^4\, \mathrm{R_\odot}$ (we use the same condition for highly energetic elliptic trajectories with $|\epsilon|<0.001\Delta \epsilon$). The green ``$\star$'' symbol and the black ``$\bullet$'' symbol represent the collision point and the black hole, respectively. Arrows indicate the direction of the outflow movement from the self-crossing region.}
	\label{fig:3dtraj}
\end{figure*}

In Figure \ref{fig:3dtraj} we show an isometric view of 3D trajectories for 100 outflow segments, selected from contours where $F$ is equal to 70\% of its maximal value, in a way that the distances between neighboring segments are approximately equal. There are two such contours, located at the northern and southern hemispheres, denoted by the red and blue lines, respectively, in Figure \ref{fig:flux_fate}. For $\Delta z/H>0$ these contours delimit two distinct regions (trajectories of segments from the northern and southern contours denoted by red and blue lines, respectively, in Figure \ref{fig:3dtraj}), where the outflow is more collimated. For $\Delta z/H=0$ these contours are denoted by purple lines to emphasize, that the bulk of the outflow is contained in a single region between the two contours and launched towards the equatorial plane (trajectories of segments denoted by purple lines in Figure \ref{fig:3dtraj}). In Figure \ref{fig:3dtraj} the green ``$\star$'' and the black ``$\bullet$'' symbols represent the collision point and the black hole, respectively, while arrows indicate the direction of the outflow from the self-crossing region. We study the deflection along the $-\mathbfit{e}_\mathrm{z}$ direction and note that the results are symmetric with respect to the orbital plane of the in-falling gas if the receding gas is deflected along the $+\mathbfit{e}_\mathrm{z}$ direction.  We calculate trajectories for values of the offset $\Delta z/H=0$ (top row), 0.6 (middle row), 1.2 (bottom row) and black hole masses $M_\mathrm{bh}= 10^6\, \mathrm{M_\odot}$ (left column), $2\times 10^6\, \mathrm{M_\odot}$ (right column) in a disruption of a $M_\star = 0.5\, \mathrm{M_\odot}$ star. Most elliptic trajectories are integrated until the outflow segments reach the pericenter. For hyperbolic trajectories and highly energetic elliptic trajectories (with $|\epsilon|<0.001\Delta \epsilon$), we integrate orbits only for radial distances $r<3\times 10^4\, \mathrm{R_\odot}$.

In a disruption by a $M_\mathrm{bh}=10^6\, \mathrm{M_\odot}$ black hole (left panels of Figure \ref{fig:3dtraj}) all trajectories are elliptical, since there is no unbound gas (see Figure \ref{fig:mass_frac}). For $\Delta z/H=0$ the outflow from the self-crossing region is quasi-spherical. Only a minor degree of spherical asymmetry is present due to a higher amount of gas moving on prograde, than on retrograde orbits. As the offset between streams increases, the geometry of the outflow becomes less spherically symmetric and more stream-like, which is especially apparent for $\Delta z/H=1.2$ (see also Figure \ref{fig:flux_fate}). Increasing the black hole mass to $M_\mathrm{bh}=2\times 10^6\, \mathrm{M_\odot}$ (right panels of Figure \ref{fig:3dtraj}) the outflow is less spherically symmetric and trajectories extend to larger distances due to the larger energy spread induced by the stream collision.

\subsection{Dependence of collision properties on black hole's spin}\label{subsec:bh_spin}

\noindent We now relate the parameters used to simulate offset collisions to physical parameters, particularly the black hole spin. Due to Lense-Thirring precession, several revolutions, or windings of the stream around the black hole may be necessary for a collision to be successful. We denote the number of such windings by $N_\mathrm{w}$, where $N_\mathrm{w}= 0$ corresponds to a prompt collision, i.e. after the first pericenter passage of the stream following a stellar disruption. In the following, we  use an analytical treatment similar to that of \citet{Batra_2022}, to evaluate the properties of offset collisions. 

During the pericenter passage, the angular momentum vector of the debris $\mathbfit{L}$, inclined by an angle $i\in[0,\pi]$ with respect to the black hole's spin vector $\mathbfit{S}$, changes by $\Delta \mathbfit{L}= \Delta\boldsymbol{\Phi}\times \mathbfit{L}$, where 

\begin{equation}\label{eq:dphi}
    \Delta\boldsymbol{\Phi}=\boldsymbol{ \Omega}_\mathrm{p}\Delta t \approx \sqrt{2}\pi a \left( \frac{R_\mathrm{g}}{R_\mathrm{p}}\right)^{3/2} \boldsymbol{\hat{s}}
\end{equation}
is the precession angle, $\boldsymbol{\hat{s}}=\mathbfit{S}/|\mathbfit{S}|$ is the unit vector aligned with the direction of $\mathbfit{S}$ and $a$ is the dimensionless spin parameter. In Equation (\ref{eq:dphi}),
$\boldsymbol{\Omega}_\mathrm{p} = 2a R_\mathrm{g}^2 c/(R_\mathrm{p}^3 ) \boldsymbol{\hat{s}}$ is the precession frequency \citep{Nixon_2012} and $\Delta t \approx \pi \sqrt{R_\mathrm{p}^3/(2GM_\mathrm{bh})}$ is the time the debris spends at the pericenter.

The in-falling and receding streams, therefore, follow different orbital planes, which intersect along a line $L_\mathrm{LT}$ at an angle $\gamma$ from the major axis of the initial orbit. We define this angle as increasing in the direction opposite from the stream's orbital motion and restrict its range to $\gamma + \Delta \phi/2 \in [0,2\pi]$ for clarity. Note that $L_\mathrm{LT}$ coincides with the projection $\mathbfit{S}_\mathrm{p}$ of the black hole spin vector along the orbital plane of the in-falling stream since $\Delta \mathbfit{L} \propto \mathbfit{S}\times \mathbfit{L}$. Due to Lense-Thirring precession, $\mathbfit{L}$ gets shifted by an angle \citep{stone19}
\begin{equation} \label{eq:domega}
\Delta \Omega = |\Delta \boldsymbol{\Phi} | \sin i.
\end{equation}
We calculate $\Delta z$ from the projected distance between the intersection point and $L_\mathrm{LT}$ 
\begin{equation} \label{eq:dz}
    \Delta z = R_\mathrm{int} \sin \left ( \gamma +(N_\mathrm{w}+1/2) \Delta \phi \right ) \Delta \Omega.
\end{equation}
By convention, $\Delta z>0$ corresponds to deflections along the direction of $\mathbfit{L}$ and the opposite direction for $\Delta z<0$. During each pericenter passage the orbital plane of the stream changes, which shifts $\mathbfit{S}_\mathrm{p}$, leading to a rotation of $L_\mathrm{LT}$ by an angle $\Delta \gamma = - N_\mathrm{w}  |\Delta\boldsymbol{\Phi} | \cot i$. This formula can be understood from the fact that if $i = \pi/2$, $\mathbfit{S}$ coincides with $\mathbfit{S}_\mathrm{p}$ and $L_\mathrm{LT}$ does not change. If $i\approx 0$, $L_\mathrm{LT}$ changes the most due to orbital plane precession because $\mathbfit{S}$ is the furthest from being initially contained in the orbital plane. Taking into account this effect leads to a more accurate expression for the vertical offset
\begin{equation} \label{eq:dz_LT}
    \Delta z' = R_\mathrm{int} \sin \left ( \gamma +\Delta \gamma+(N_\mathrm{w}+1/2) \Delta \phi \right ) \Delta \Omega.
\end{equation}

As long as its evolution is dominated by the black hole's tidal force, the stream is confined between two orbital planes, inclined by an angle $\alpha$, that intersect along a line, which we refer to as $L_\mathrm{H}$. During its first approach, the in-falling stream is in hydrostatic equilibrium until the tidal force takes over, which we assume happens at the semi-minor axis of the most bound debris $b_\mathrm{orb}=\ell_\star \left( 2\Delta\epsilon  \right )^{-1/2}$. This implies (see \citealt{Bonnerot_2022}) that $L_\mathrm{H}$ is along the major axis of the initial orbit, and that $\alpha=H_0/b_\mathrm{orb}$,  where $H_0$ is the stream width. We calculate the width of colliding streams $H_\mathrm{int}$ from the projected distance between the intersection point and $L_\mathrm{H}$, which yields
\begin{equation} \label{eq:Hint}
    H_\mathrm{int}=\left |R_\mathrm{int} \sin \left ((N_\mathrm{w}+1/2) \Delta \phi \right ) \right | \alpha.
\end{equation}
The width of the in-falling stream decreases until it reaches a minimal value near pericenter, where a nozzle shock occurs that makes the stream bounce back. By imposing an absolute value in Equation (\ref{eq:Hint}) we assume that this bounce is perfect, i.e. that it instantaneously reverses the sign of the velocity component orthogonal to the orbital plane. This assumption is likely oversimplified (see \citealt{bonnerot2021nozzle}) but we nevertheless adopt it here for simplicity, keeping in mind that it may affect our results, especially for a large number of windings.

\begin{figure*}
	\centering
		\begin{minipage}[b]{0.32\linewidth}
		\includegraphics[width=\textwidth]{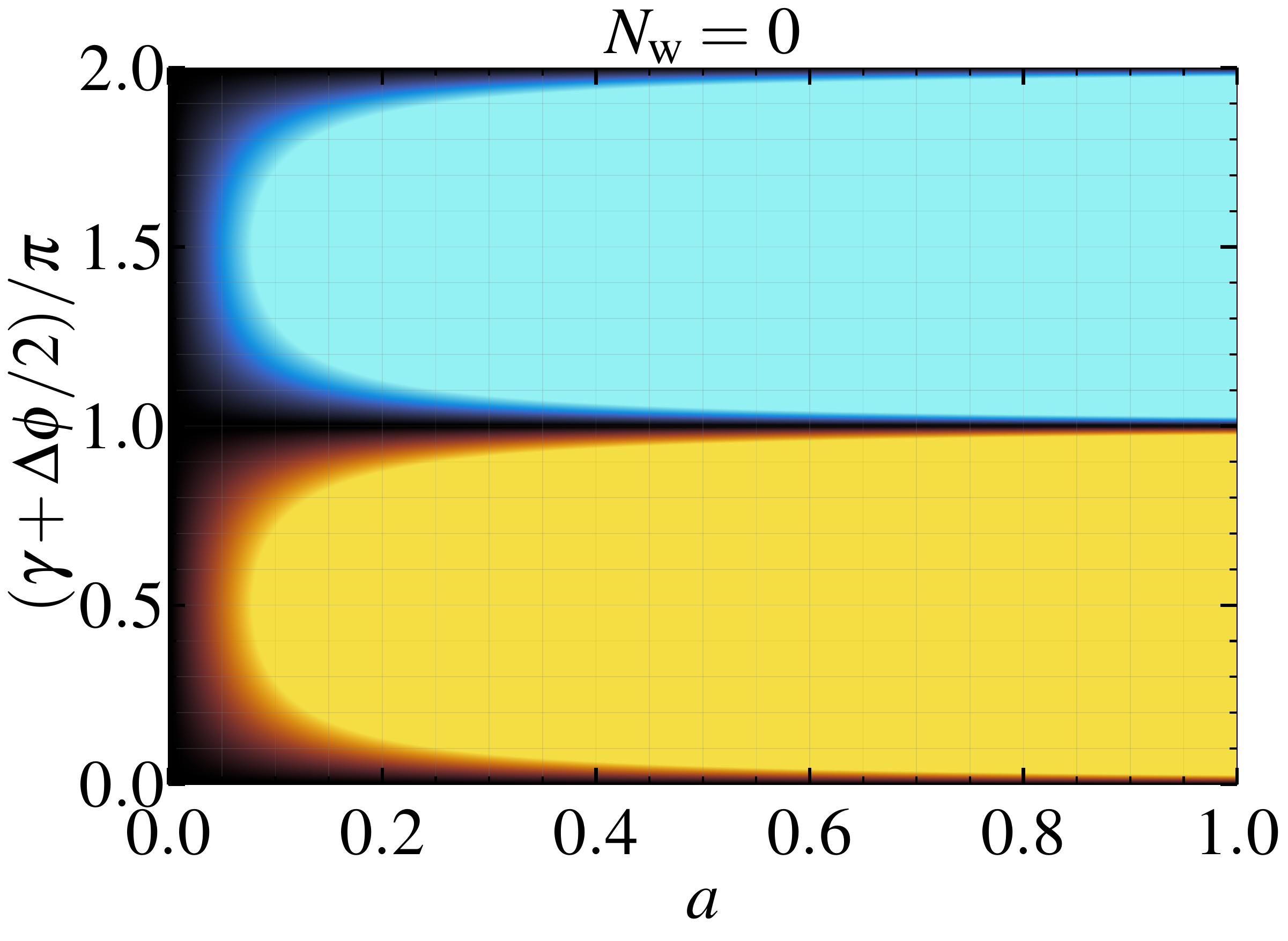}
	\end{minipage}
	\begin{minipage}[b]{0.32\linewidth}
		\includegraphics[width=\textwidth]{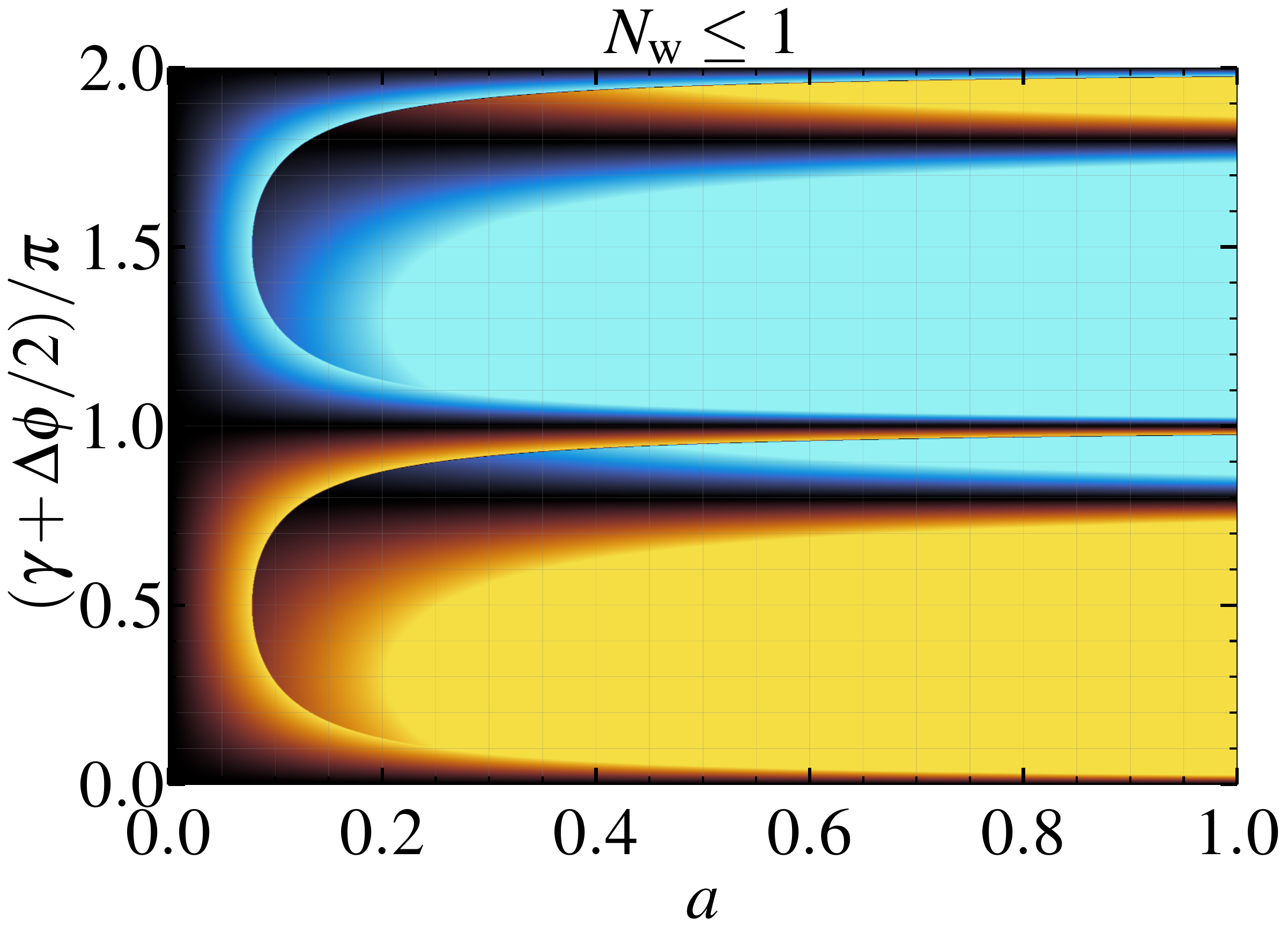}
	\end{minipage}
			\begin{minipage}[b]{0.32\linewidth}
		\includegraphics[width=\textwidth]{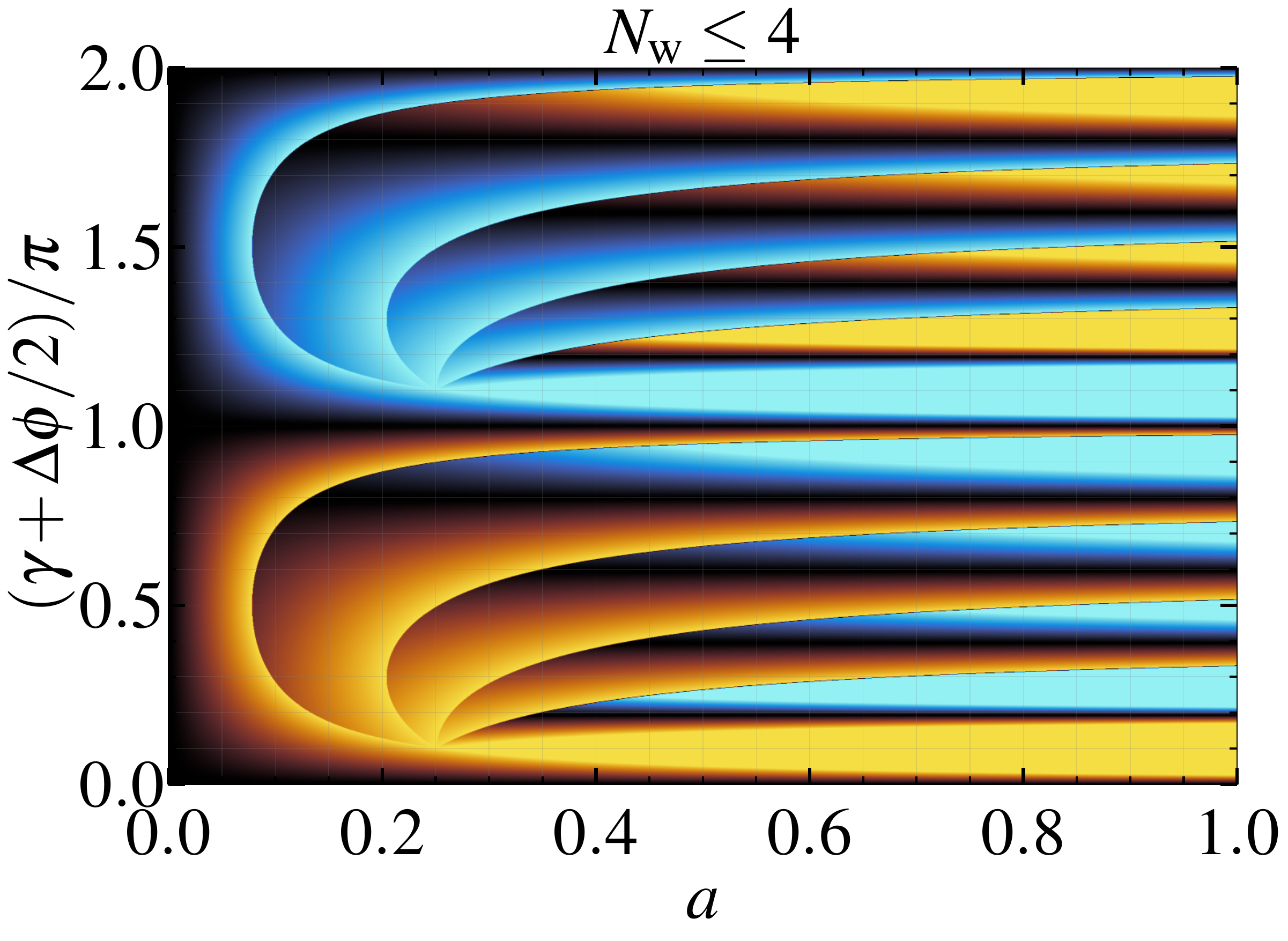}
	\end{minipage}
			\begin{minipage}[b]{0.49\linewidth}
		\includegraphics[width=\textwidth]{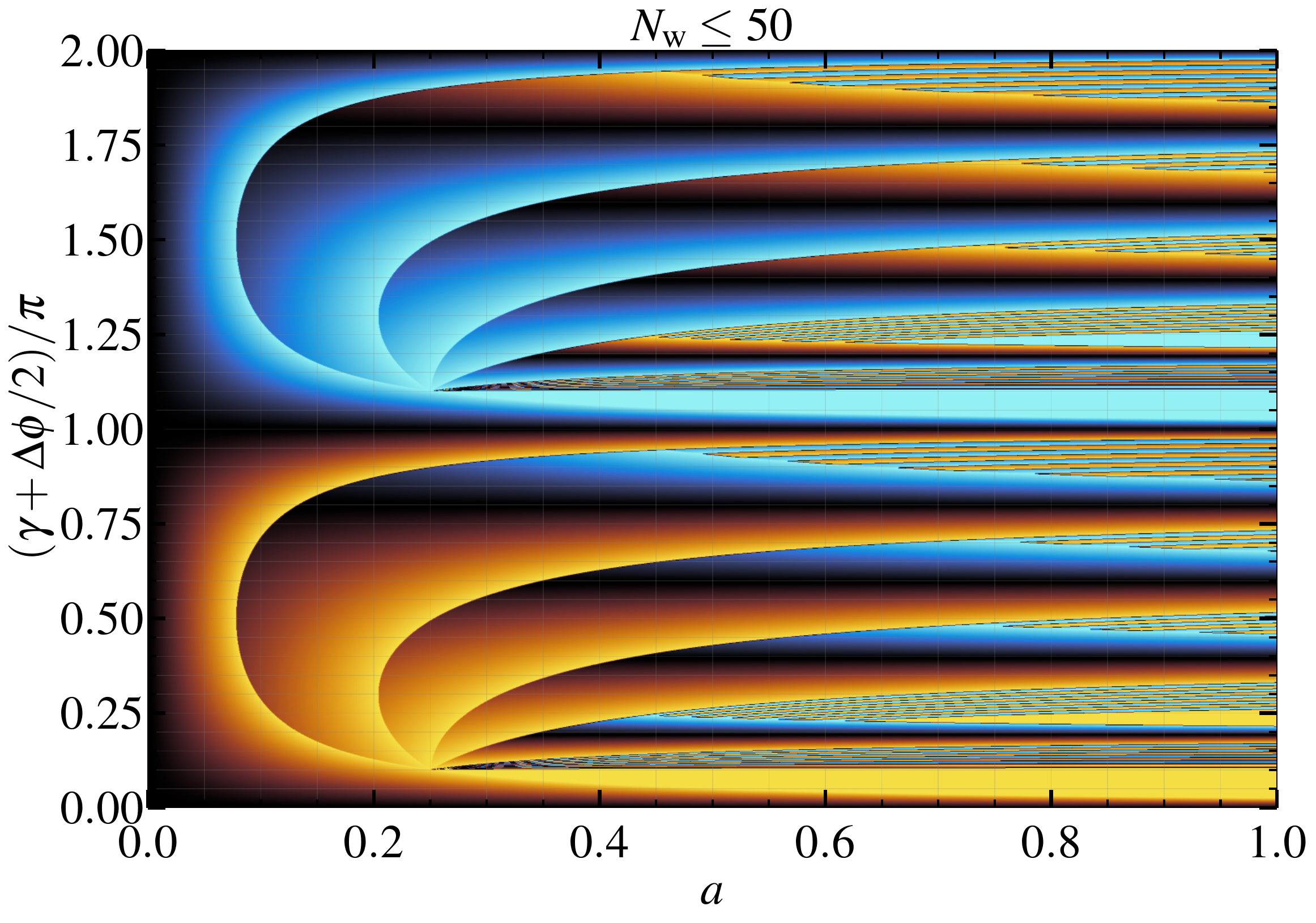}
	\end{minipage}
			\begin{minipage}[b]{0.49\linewidth}
		\includegraphics[width=\textwidth]{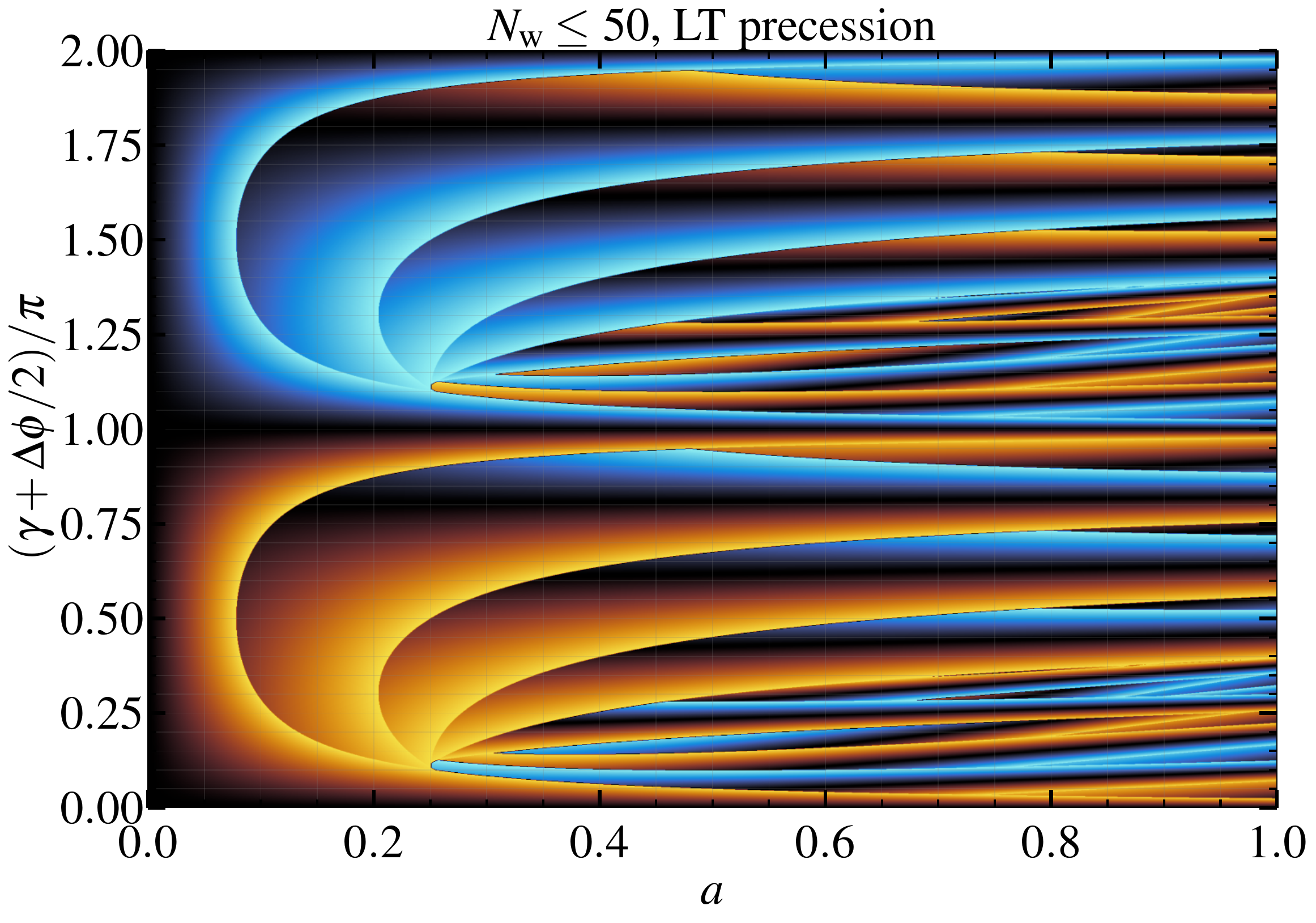}
	\end{minipage}
					\includegraphics[width=\textwidth]{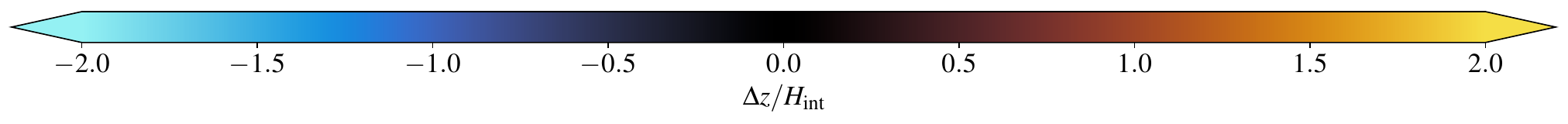}

	\caption{$\Delta z/H_\mathrm{int}$ as a function of the black hole spin's magnitude and orientation specified by $a$ and $\gamma$ for  $M_\star =0.5\, \mathrm{M_\odot}$,  $R_\star=0.46\, \mathrm{R_\odot}$, $M_\mathrm{bh}=2.5\times 10^6\, \mathrm{M_\odot}$, $\beta=1$, $i=30^\circ$, and $H_\mathrm{0}=5 \mathrm{R_\odot}$. We calculate $\Delta z$ from Equation (\ref{eq:dz}) (first four panels) and Equation (\ref{eq:dz_LT}) (the last panel) for a number of windings $N_\mathrm{w} = 0$ (top left panel) and $N_\mathrm{w} \leq 1$, 4, 50.}
	\label{fig:dz_hint}
\end{figure*}

The streams collide after $N_\mathrm{w}$ orbital windings if the condition  $|\Delta z| < 2 H_\mathrm{int}$ is met\footnote{We take the absolute value of $\Delta z$ because we compare the width of streams to the distance between their centers of mass.} \citep{Batra_2022}. In Figure \ref{fig:dz_hint}, we show $\Delta z/H_\mathrm{int}$ as a function of $a$ and $\gamma$ for $N_\mathrm{w} = 0$ (top left panel) and $N_\mathrm{w} \leq 1$ (top middle panel), 4 (top right panel) and 50 (bottom panels). We calculate $\Delta z$ from Equation (\ref{eq:dz}) (first four panels) and Equation (\ref{eq:dz_LT}) (the last panel) for $M_\star =0.5\, \mathrm{M_\odot}$,  $R_\star=0.46\, \mathrm{R_\odot}$, $M_\mathrm{bh}=2.5\times 10^6\, \mathrm{M_\odot}$, $\beta=1$, $i=30^\circ$ and $H_\mathrm{0}=5\, \mathrm{R_\odot}$ (for these parameters $\Delta \phi =36^\circ$). In the case of prompt collisions ($N_\mathrm{w} = 0$), the vertical offset is antisymmetric along the horizontal line at $(\gamma + \Delta \phi/2)/\pi = 1$, with $\Delta z/H_\mathrm{int}<0$ and $\Delta z/H_\mathrm{int}>0$ for $(\gamma + \Delta \phi/2)/\pi > 1$ and $(\gamma + \Delta \phi/2)/\pi < 1$, respectively. For values of $(\gamma + \Delta \phi/2)/\pi = 0$, 1 and 2, the projected distance between the intersection point and $L_\mathrm{LT}$ equals zero, which results in a self-crossing collision with $\Delta z=0$. At a fixed $\gamma$ in offset collisions, $\Delta z$ increases with $a$ until a prompt collision is avoided, meaning that it will occur for a larger number of windings $N_\mathrm{w} > 0$. The lowest value of $a\approx 0.1$, for which the collision can be avoided, is reached for $(\gamma + \Delta \phi/2)/\pi = 1/2$ and $3/2$, that is when the projected distance from the intersection point to $L_\mathrm{LT}$ (see Equation (\ref{eq:dz})) is the largest.

When $N_\mathrm{w}$ increases, successful collisions become possible in new regions of the parameter space where they were avoided for lower $N_\mathrm{w}$. If $N_\mathrm{w}\Delta \phi< \pi$, these regions exhibit a similar trend as in the case of prompt collisions (see top middle and top right panels), but are progressively shifted towards lower values of $\gamma$ with increasing $N_\mathrm{w}$. This can be understood by considering the change in $\gamma$ of regions where $\Delta z= 0$. For these parts of the parameter space, the projected distance from the intersection point to $L_\mathrm{LT}$ does not change with $N_\mathrm{w}$ (the argument of the sine function in Equation (\ref{eq:dz}) is constant), meaning that the regions shift towards lower values of $\gamma$ by a factor of $-N_\mathrm{w}\Delta \phi$ when $N_\mathrm{w}$ increases. These new regions are also shifted towards higher values of $a$. This can be inferred by considering the shift of the left-most boundary of a new region, where $\Delta z/H_\mathrm{int} = 2$ for a value of $\Delta z$ that is maximized, and therefore independent of $N_\mathrm{w}$. As long as $H_\mathrm{int}$ increases with $N_\mathrm{w}$, which is the case for $(N_\mathrm{w}+1/2)\Delta \phi/\pi< 1/2$ (implying $N_\mathrm{w}<3$ from Equation (\ref{eq:Hint})), a successful collision can happen for a larger $a$, thus shifting the boundary to higher spin values when $N_\mathrm{w}$ is increased. We also notice that the boundaries between regions with a different $N_\mathrm{w}$ all intersect at two locations: $(\gamma + \Delta \phi/2)/\pi = \Delta \phi/(2\pi) \approx \{0.1, 1.1\}$ and $a \approx 0.25$. This is because for $\gamma=0$, $\Delta z/H_\mathrm{int}$ does not depend on $N_\mathrm{w}$, and $\Delta z/H_\mathrm{int}=2$ is therefore reached for the same value of $a$, independent of $N_\mathrm{w}$. We find that collisions are possible for the majority of the parameter space for $N_\mathrm{w}\leq 4$. The rest of the parameter space corresponds to $N_\mathrm{w}\Delta \phi > \pi$ and is covered cyclically from highest to lowest $\gamma$, creating thin horizontal bands of alternate colours with $|\Delta z|/H_\mathrm{int} \approx 2$ (bottom left panel of Figure \ref{fig:dz_hint}).

When taking into account the rotation of $L_\mathrm{LT}$  (see Equation (\ref{eq:dz_LT})), regions of constant $\Delta z/H_\mathrm{int}$ are shifted towards higher values of $\gamma$ by $|\Delta \gamma|$ and the shift is greater as $a$ increases (see the right bottom panel in Figure \ref{fig:dz_hint}). In addition, for $N_\mathrm{w} > 4$ and $a \gtrsim 0.5$, the projected distance from $L_\mathrm{LT}$ to the intersection point is significantly affected, resulting in a wider range of $\Delta z/H_\mathrm{int}$ in these regions that replace the thin horizontal bands described above. However, the rest of the parameter space is only marginally affected, implying that the streams mostly collide with a similar offset as in the case when the rotation of $L_\mathrm{LT}$ is not taken into account (bottom left panel of Figure \ref{fig:dz_hint}).

The dependence on $i$, $H_0$ and $M_\mathrm{bh}$ can be understood by taking the ratio of Equations (\ref{eq:dz}) and (\ref{eq:Hint}), which gives
\begin{flalign}
   \nonumber \frac{|\Delta z|}{H_\mathrm{int}} &= \xi \pi a |\sin i| \left ( \frac{R_\mathrm{g}}{R_\mathrm{p}} \right )^{3/2}  \frac{\  \ell_\star  }{\sqrt{\Delta \epsilon }H_0}&& 
  \label{eq:dz_hint}\\
		&\approx \xi a |\sin i|\left ( \frac{M_\mathrm{bh}}{10^6\,\mathrm{M_\odot}}\right)^{3/2}\left ( \frac{H_\mathrm{0}}{5\, \mathrm{R_\odot}}\right)^{-1} &&
	\end{flalign}
where we assume $R_\star=0.46\, \mathrm{R_\odot}$, $\beta=1$,  and introduce $\xi$ as the ratio of the two sine functions. By assuming $\xi=1$, which is motivated by the fact that $\xi \approx 0$ only for specific values of $\gamma$ and $N_\mathrm{w}$, we determine that for most values of $\gamma$, collisions become more likely with decreasing $M_\mathrm{bh}$, increasing $H_0$, and values of $i$ closer to 0 or $\pi$. In Figure \ref{fig:dz_hint}, these changes would be reflected by an overall shift towards larger values of $a$ such that more of the parameter space corresponds to $|\Delta z| < 2 H_\mathrm{int}$. Additionally, the vertical space between regions with $\Delta z=0$ would be reduced when decreasing $M_\mathrm{bh}$ due to lower apsidal precession.

\begin{figure}
	\centering

					\includegraphics[width=\linewidth]{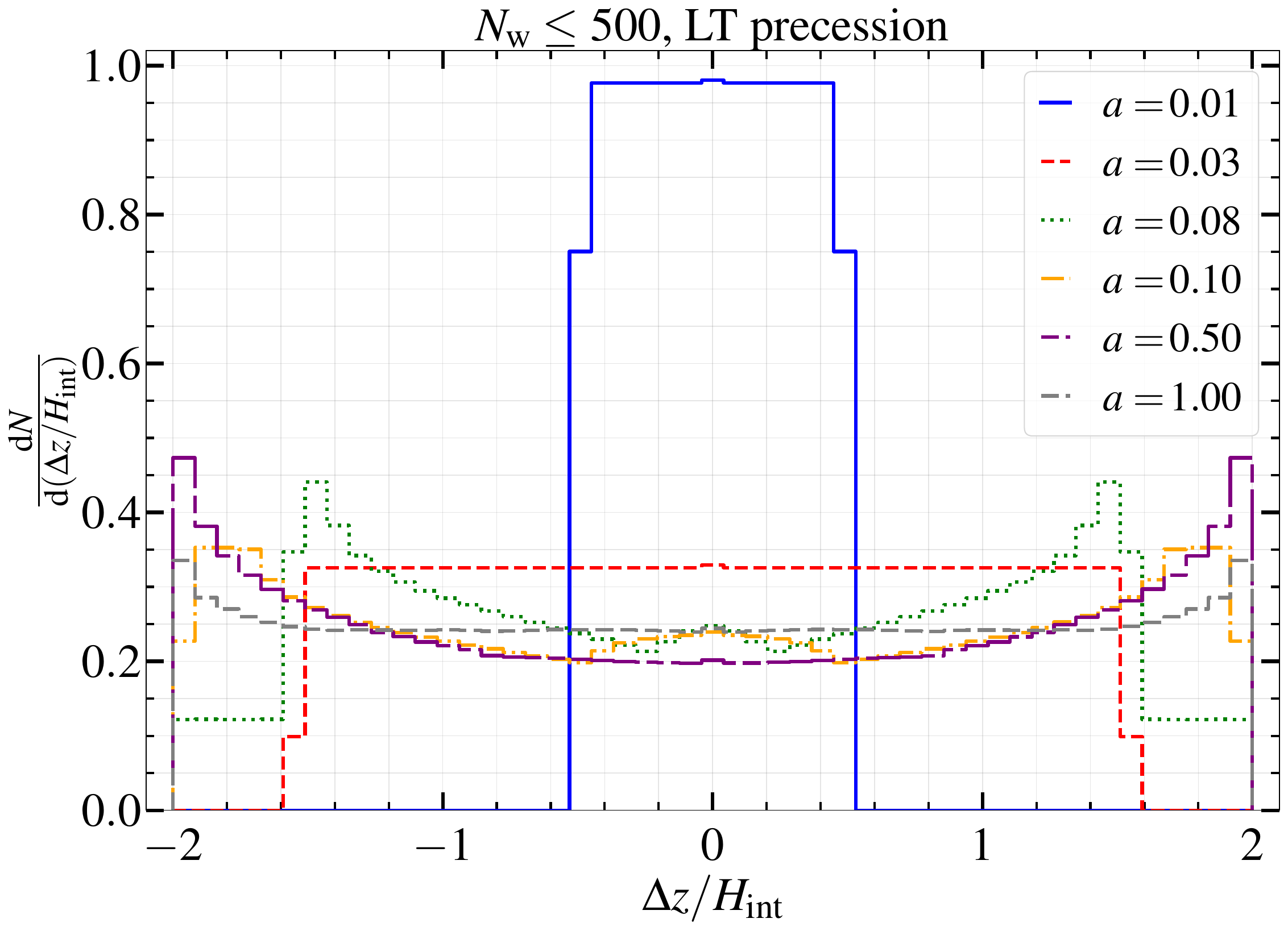}

	\caption{{Distributions of $\Delta z/H_\mathrm{int}$ for different values of the black hole spin's magnitude $a$ for   $M_\star =0.5\, \mathrm{M_\odot}$,  $R_\star=0.46\, \mathrm{R_\odot}$, $M_\mathrm{bh}=2.5\times 10^6\, \mathrm{M_\odot}$, $\beta=1$, and $H_\mathrm{0}=5 \mathrm{R_\odot}$. We calculate $\Delta z$ from Equation (\ref{eq:dz_LT}) for a number of windings $N_\mathrm{w} \leq 500$. The integral of distributions is normalized to 1.}}
	\label{fig:dz_histogram}
\end{figure}

{In Figure \ref{fig:dz_histogram} we show probability distributions of $\Delta z/H_\mathrm{int}$ for different $a$. Assuming a random orientation between the black hole's spin vector and the star's angular momentum vector, the distributions of angles $i \in [0,\pi]$ and $\gamma \in [0,2\pi]$ are taken to be uniform in $\cos i$ and $\gamma$, respectively. When $a\lesssim 0.03$ and collisions are typically prompt ($N_\mathrm{w} = 0$), we find that streams generally collide with low offsets, meaning that most collisions are strong. However, when $a\gtrsim 0.03$, we discover that there can be both strong and grazing collisions for all values of $a$ since streams collide at a higher number of windings $N_\mathrm{w}$. For $a\lesssim 0.5$ the grazing collisions ($\left|\Delta z\right|/H_\mathrm{int}\approx 2$) become more likely as $a$ increases, but this trend reverses for $a\gtrsim 0.5$, consistently with the prevalence of low vertical offsets visible at such high spin values in Figure \ref{fig:dz_hint} (lower-right panel).}

\section{Discussion} \label{sec4}

\subsection{Effects of the different properties of the colliding streams}\label{subsec:different_stream_properties}

\noindent Our choice of the stream collision configuration assumes that the streams have a uniform density profile and that they are identical, meaning that they have the same $H$, $\dot{M}$, and velocities. We now explore possible deviations from this assumption and the consequences on the outflow properties found in Section \ref{sec3}. Despite not leading to significant net expansion, the study by \citet{bonnerot2021nozzle} suggests that the nozzle shock can modify the width of the colliding stream components by a factor of a few. As a consequence, a part of the receding gas could avoid a direct collision and the denser in-falling stream would be less affected, resulting in a decrease in energy dissipation. Additionally, the streams may reach the intersection point with different $\dot{M}$ (or velocities) due to the slight difference in orbital energy between them (e.g. \citealt{Jankovic_2023https://doi.org/10.48550/arxiv.2302.00607}). In this case, the kinetic energy associated with the radial motion would be less efficiently dissipated resulting in a more asymmetric outflow, where the outflow component from the stream with higher $\dot{M}$ is more collimated and expands at a faster rate. In addition, \citet{Rossi_2021} find that due to different $\dot{M}$ the intersection point does not remain fixed but instead evolves with time.

It is expected that the density distribution inside the colliding streams is not uniform. Instead, the density is the highest at the stream's center of mass and decreases toward the stream's edge \citep{Jiang_2016, bonnerot2021nozzle}. As a consequence, the mass of the colliding stream parts would be different in the collisions of streams with a uniform and 
 a non-uniform density distribution for the same $\Delta z/H$ (even if streams have the same $\dot{M}$, $H$ and velocity), which would affect energy dissipation. This effect is likely most prominent for $\Delta z \lesssim 2H$, where the mass of directly colliding gas would be significantly lower for a non-uniform density profile, implying that a lower vertical offset may be required to reach the same collision strength as in the uniform case. For $\Delta z/H \ll 1$ we expect that this effect does not significantly change the outcome of collisions because mass of colliding gas is less affected by the density distribution. 
 
 In Section \ref{sec:bh_frame} we evaluate the outflow properties only in the case of the most bound debris. In typical situations ($M_\mathrm{bh}\approx 10^6\,\mathrm{M_\odot}$, $M_\star\approx1\,\mathrm{M_\odot}$, $\beta\approx1$), apsidal precession is high enough for the intersection radius to be significantly lower than the stream apocenter distance with $R_\mathrm{int}\lesssim a_\mathrm{min}$, where $a_\mathrm{min}=R_\star (M_\mathrm{bh}/M_\star)^{2/3}/2$ is the semi-major axis of the most bound debris. Because the trajectories at this location are not strongly affected by changing the orbital energy, we do not expect the collisions to be qualitatively modified.

\subsection{Circularization of the outflow}\label{subsec:circularization}

\noindent The debris from TDEs can form an accretion disc if its energy is efficiently dissipated. While the self-crossing shock provides an initial source of dissipation that can affect the gas trajectories, it is usually not enough on its own to cause significant circularization. As demonstrated by \citet{Bonnerot_2020} and \citet{Bonnerot_2021}, even when the self-crossing shock is strong and results in a large-scale outflow, additional secondary shocks closer to the black hole are necessary to promptly complete disc formation. The efficiency of this process appears to be enhanced by the presence of outflowing gas on retrograde orbits, as it leads to strong head-on collisions near the pericenter. 

For the offset self-crossing shock we focus on here, we expect a qualitatively similar evolution towards disc formation only in the strong collision regime, where the fraction of retrograde gas is significant. The evolution towards disc formation may be slower in the grazing collision regime because there is less gas on retrograde orbits. However, as explained in Section \ref{sec:bh_frame}, even the most grazing collision we considered with $\Delta z = 1.8H$ leads to significant stream expansion associated with an order unity spread in pericenter $\Delta R_\mathrm{p} \approx 10R_\mathrm{g}\approx R_\mathrm{p}$. Consequently, some of the gas in the expanded stream has very short pericenter distances, leading to strong apsidal precession and collisions close to the black hole, where the energy dissipation is enhanced. In addition, the collision at the next winding will be strong despite the vertical offset induced by spin because of the increased stream width. This indicates that a significant delay in the accretion disc formation will occur only if the streams completely miss each other. In this case, several revolutions may take place before an offset collision occurs (see Section \ref{subsec:bh_spin} and \citet{Batra_2022}), after which we expect an accretion disc to promptly assemble. 

\subsection{Observational features}\label{subsec:observ_features}

For $\Delta z/H \approx 0$ the outflow from the self-crossing region is close to spherical as seen in Figure \ref{fig:f}. In this case, the photons produced by the self-crossing shock suffer from large adiabatic losses, making the associated radiation likely undetectable \citep{Bonnerot_2020}. As $\Delta z/H$ increases, the outflow becomes more collimated and aligned with the direction of the incoming streams. Along other directions where the gas density is lower, radiation may be able to more efficiently emerge from the system without suffering as many adiabatic losses. 

\begin{figure}
	\centering

					\includegraphics[width=\linewidth]{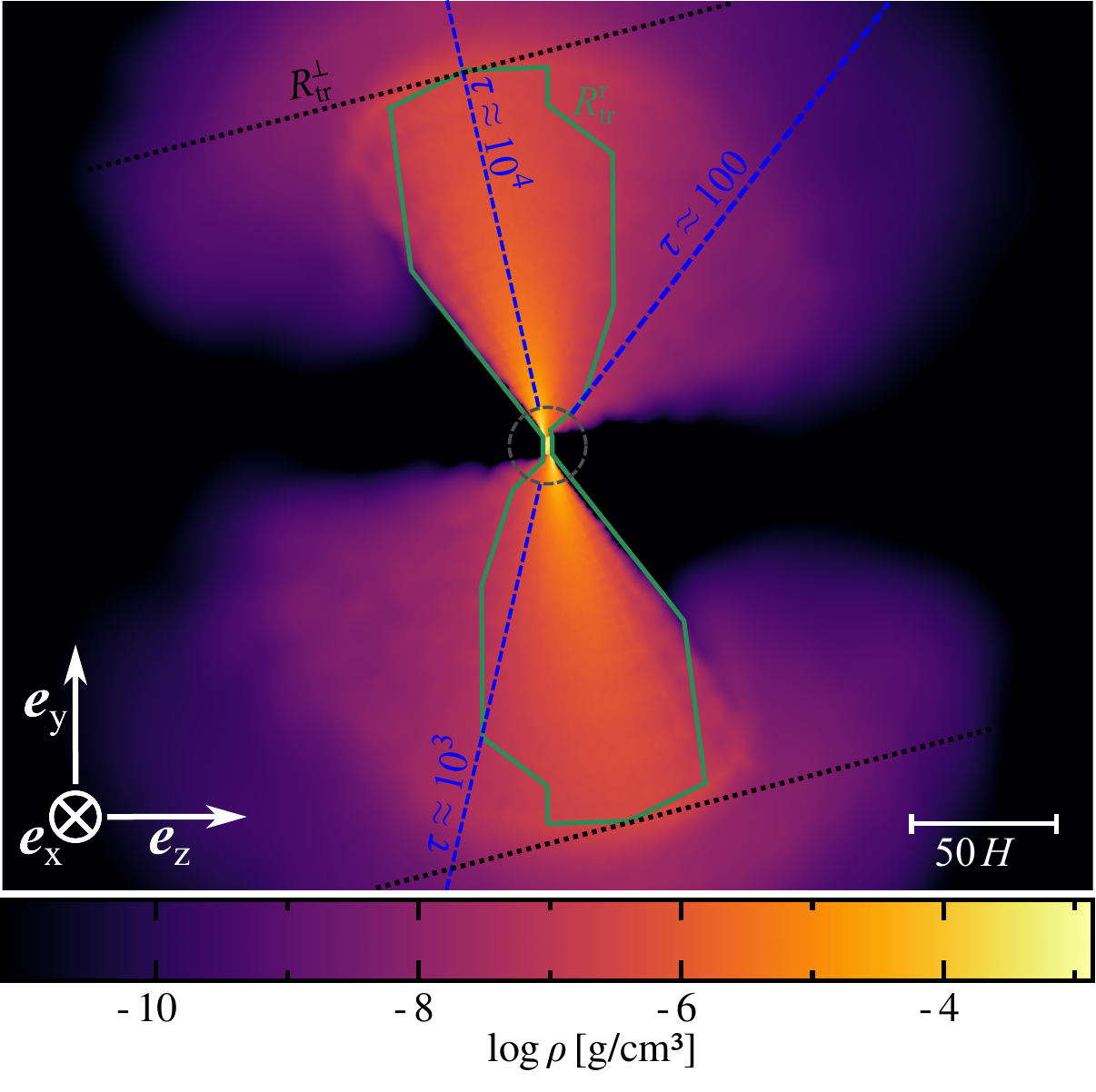}

	\caption{A density slice of the outflow (in the local reference frame) in the $zy$ plane at $x=0$ at time $t=150$ for $\Delta z=1.2H$. Length units are expressed in units of the stream widths $H=10\, R_\odot$. {The green solid line and the black dotted lines represent the locations of the trapping surfaces defined in Equations (\ref{eq:Rtr}) and (\ref{eq:Rtr_perp}), respectively}. We show values of the optical depth $\tau$, integrated from $R=15H$ (grey dashed circle) to infinity, along several directions marked by blue dashed lines.}
	\label{fig:Rtr}
\end{figure}

{To quantify this effect we evaluate and compare the radial trapping radius $R_\mathrm{tr}^\mathrm{r}$ and the transverse trapping radius $R_\mathrm{tr}^\mathrm{\perp}$, above which the photons decouple from the gas by diffusing away in the radial and transverse directions, respectively. $R_\mathrm{tr}^\mathrm{r}$ is} calculated by equating the dynamical timescale of the gas $t_\mathrm{dyn}\approx R_\mathrm{}/v(R_\mathrm{})$ to the diffusion timescale of photons $t_\mathrm{diff}\approx \tau_\mathrm{r} R_\mathrm{}/c$, where $R$ is the radial distance from the point of self-crossing, $v(R_\mathrm{})$ is the velocity of the gas and $\tau_\mathrm{r}$ is the optical depth, {integrated along the radial direction} from $R$ to infinity. {$R_\mathrm{tr}^\perp$ is defined as the radial distance $R$ from the point of self-crossing where $t_\mathrm{dyn}$ and the diffusion timescale along the transverse direction $t_\mathrm{diff}^\perp = \tau_\perp  R_\perp / c$ are equal, assuming a conical outflow. $\tau_\perp$ is the transverse optical depth, integrated along the transverse distance $R_\perp= R\theta_0$ from the line where the density peaks to infinity, where $\theta_0$ is obtained from simulations as half of the opening angle of a conical outflow. Therefore, $R_\mathrm{tr}^\mathrm{r}$ can be expressed from}
\begin{equation} \label{eq:Rtr}
c/v\left(R_\mathrm{tr}^\mathrm{r}\right) = \int_{R_\mathrm{tr}^\mathrm{r}}^{+\infty} \rho \kappa_\mathrm{s} \mathrm{d} R,
\end{equation}
{and $R_\mathrm{tr}^\perp$ from}
\begin{equation} \label{eq:Rtr_perp}
c/v\left(R_\mathrm{tr}^\perp\right)= \theta_0\int_{0}^{+\infty} \rho\kappa_\mathrm{s} \mathrm{d} R_\perp
\end{equation}
where $\kappa_\mathrm{s}=0.34\, \mathrm{cm^2 g^{-1}}$ is the electron scattering opacity and $\rho$ is the gas density.  {$R_\mathrm{tr}^\mathrm{r}$} is evaluated for different values of $\theta$ and $\phi$, by computing the integral of Equation (\ref{eq:Rtr}) from outward in until its value reaches the local ratio of the speed of light to the gas velocity. {$R_\mathrm{tr}^\perp$ is evaluated for different gas sections along the direction of the maximum density inside the outflow, by computing the integral of Equation (\ref{eq:Rtr_perp}) from the center of mass to the edge of the outflow (obtained from simulations) until its value reaches the local ratio of the speed of light to the gas velocity modified by $\theta_0$.} In Figure \ref{fig:Rtr} we show the density of gas contained in a slice in the $zy$ plane at $x=0$ for $\Delta z/H=1.2$, where we overplot the {radial and the transverse trapping surfaces with green solid and black dotted lines, respectively.} {The density is taken from the simulation with $\Delta z/H=1.2$, and its value in physical units is obtained by assuming $H=10\,\mathrm{R_\odot}$, $v_\mathrm{r}=0.01c$ and $\dot{M}=0.7\,\mathrm{M_\odot\,yr^{-1}}$.} Dashed blue lines indicate values of $\tau$ obtained by integration from $R=15H$ (dashed grey circle) to infinity along several directions. {We see that the radial trapping surface extends to distances $R_\mathrm{tr}^\mathrm{r} \gg H$ in most directions, except near the equatorial plane where the density is reduced. Additionally, the transverse trapping surface is found to be larger than the radial one, implying that photons diffuse only radially. Nonetheless, we expect that transverse diffusion becomes important with $R_\mathrm{tr}^\perp<R_\mathrm{tr}^\mathrm{r}$ below a critical outflow rate, whose value is $\dot{M}_\mathrm{crit}\approx 0.4\, \mathrm{M_\odot/yr}$ for $\Delta z/H=1.2$\footnote{{Above this critical outflow rate, both trapping radii are limited to the finite extent of the outflow, where the density drops sharply. For $\dot{M}<\dot{M}_\mathrm{crit}$, these trapping radii recede inside the outflow, with the transverse one being lower in the grazing collision regime. By assuming a conical outflow and a linearly decreasing transverse density profile, we obtain an analytic estimate $R_\mathrm{tr}^\perp/R_\mathrm{tr}^\mathrm{r}\approx 1.5\theta_0^2$ from Equations (\ref{eq:Rtr}) and (\ref{eq:Rtr_perp}).}}. However, even in the grazing collision regime with $\dot{M}<\dot{M}_\mathrm{crit}$, $R_\mathrm{tr}^\perp$ is only a few times lower than the (radial) trapping radius of a spherical outflow launched in collisions with $\Delta z/H = 0$. This indicates that the geometry of the outflow does not significantly affect the adiabatic losses of photons produced by the self-crossing shock.}

{In the following, we focus on cases where $\dot{M}>\dot{M}_\mathrm{crit}$ and estimate the emerging bolometric luminosity $L_\mathrm{bol}$ assuming only diffusion along the radial direction. At the trapping surface, the fluxes of energy given by diffusion and advection are by definition similar. Consequently, we estimate $L_\mathrm{bol}$ by integrating the advection flux $\mathbfit{F}=\rho_\mathrm{} u_\mathrm{} \mathbfit{v}_{}$ across the trapping surface, where $\rho_\mathrm{}$, $u_\mathrm{}$, $\mathbfit{v}_{\rm }$ are the gas density, specific internal energy, and velocity at the trapping surface obtained from simulations, respectively. For $H=10\,\mathrm{R_\odot}$, $v_\mathrm{r}=0.01c$ and $\dot{M}=0.7\,\mathrm{M_\odot\,yr^{-1}}$, we find that $L_\mathrm{bol}\approx 10^{41}\,\mathrm{erg/s}$ in the strong collision regime and then decreases by a factor $\sim 3$ in the grazing collision regime. This reduction is mostly caused by the change in the shock heating rate $\dot{E}$ (see Figure \ref{fig:dotE}) rather than a variation of adiabatic losses, consistently with our above prediction of them being unimportant. A more careful determination of the emerging radiation requires radiation-hydrodynamics simulations, which we defer to future research.}

In the later stages of a TDE, when the accretion disc has formed, X-ray radiation from the disc can be absorbed and then reprocessed to the optical band by the outflow \citep{Roth_2016, Lu_2019}. For more offset collisions, we expect the solid angle covered by the outflow to decrease due to a deviation from sphericity (see Figure \ref{fig:3dtraj}), such that X-ray photons may only promptly escape along specific viewing angles while being efficiently reprocessed along others. This effect could explain the variety of optical to X-ray luminosity ratios found observationally (e.g. \citealt{van_Velzen_2021}, {\citealt{sazonov_2021MNRAS.508.3820S}}, {\citealt{Hammerstein_2023ApJ...942....9H}}). {Additionally, we find that the apocenter distances of the outflow are typically between $10^{14}-10^{15}\,\mathrm{cm}$ which is consistent with the blackbody radii inferred from optical emission of TDEs (e.g. \citealt{van_Velzen_2021}).}

{Another potentially observable effect arises from polarimetric properties of TDEs, which can be determined by means of spectropolarimetry \citep{Leloudas_2022NatAs...6.1193L, Liodakis_2023Sci...380..656L}. In TDEs, the light undergoes polarization mainly due to the Thomson scattering, and the observed polarization is determined at the final scattering surface. Since TDEs are spatially unresolved, the observed polarization is determined by the photon electric vectors on the projected photosphere, which represents the apparent surface as seen from the observer's perspective. If the projected photosphere is circularly symmetric, randomly oriented electric vectors cancel each other out, resulting in a zero net polarization. However, if the projected photosphere deviates from circular symmetry, the cancellation of electric vectors becomes incomplete, leading to a non-zero net polarization \citep{Patra_2022MNRAS.515..138P}. This effect may lead to different degrees of polarizations for stream collisions with different offsets. In the strong collision regime, the outflow is more spherical leading to low polarizations, while in the grazing collision regime, the outflow is highly aspherical resulting in higher polarizations.}

\section{Summary} \label{sec5}

\noindent After a star is disrupted, the part of the elongated stream of debris that passed pericenter collides with the still-infalling gas due to the relativistic apsidal precession, leading to a self-crossing shock. For rotating black holes, Lense-Thirring precession causes a misalignment between the colliding streams. We study the impact of this effect on the outflow from the self-crossing shock by locally simulating the collision between two streams with widths $H$ offset by a distance $\Delta z$. We analyze the gas evolution during the collision, the geometry of the outflow, and its subsequent dynamics, as well as consequences on accretion disc formation and observational signatures.  Our main conclusions are as follows.

\begin{enumerate}[label=(\roman*),align=left]
\item Based on the amount of dissipated kinetic energy of streams during the collision and the outflow geometry, we identify two regimes of collisions. In the strong collision regime ($\Delta z \lesssim 0.7H$), the collision results in higher dissipated energy, and the outflow remains close to spherical. In the grazing collision regime ($\Delta z \gtrsim 0.7H$), less energy gets dissipated and the outflow is collimated along the directions of the incoming streams.
\item We have made the outflow properties found from simulations available online at \href{https://github.com/tajjankovic/Spin-induced-offset-stream-self-crossing-shocks-in-TDEs.git}{https://github.com/tajjankovic/Spin-induced-offset-stream-self-crossing-shocks-in-TDEs.git}. The provided code returns the mass flux normalized by the total outflow rate (see Equation (\ref{eq:F})) given a direction of motion and a vertical offset $\Delta z$ specified by the user.
\item The fractions of unbound gas and gas on retrograde orbits decrease with $\Delta z/H$ due to a lower dissipated energy. Increasing the black hole mass also causes more of the outflow to be unbound, but less of it to move on retrograde orbits. 
\item When $a\gtrsim 0.1$, we analytically predict that there can be both strong and grazing collisions for all values of $a$ depending on the projection of the black hole's spin vector on the in-falling stream's orbital plane.
\item Even in the most extreme grazing encounter, the outflow components are wider than streams before the collision, which likely causes strong collisions later on, with some of the gas colliding closer to the black hole, increasing the energy dissipation. As a consequence, a significant delay in accretion disc formation would likely occur only if the collision is entirely prevented.
\item {The observed signal emerging from the self-crossing region is less luminous for offset collisions than aligned ones, since the energy injected by the self-crossing shock is lower. Additionally, the deviation from outflow sphericity may increase the degree of the observed polarization and allow X-ray radiation generated by the accretion disc to emerge along specific directions without being absorbed. }
\end{enumerate}

We provided the first systematic study of the outflow properties from the self-crossing shock produced by colliding streams that are offset due to the black hole's rotation. In the future, we aim to quantify the radiative output of these interactions and explore the impact they may have on the ensuing accretion disc formation, as these processes hold promise to provide constraints on the black hole spin from observed TDEs.

\section*{Acknowledgements}
\addcontentsline{toc}{section}{Acknowledgements}

\noindent {T. J. and A. G. acknowledge the financial support from the Slovenian Research Agency (research core funding P1-0031, infrastructure program I0-0033, and project grants No. J1-8136, J1-2460, N1-0344)}. This project has also received funding from the European Union’s Horizon 2020 Framework Programme under the Marie Sklodowska-Curie grant agreement no. 836751. We acknowledge the use of \textsc{phantom} for detailed hydrodynamical simulations and \textsc{SPLASH} for the visualization of the output \citep{price}.  We gratefully acknowledge the HPC RIVR consortium (\href{www.hpc-rivr.si}{www.hpc-rivr.si}) and EuroHPC JU (\href{eurohpc-ju.europa.eu}{eurohpc-ju.europa.eu}) for funding this research by providing computing resources of the HPC system Vega at the Institute of Information Science (\href{www.izum.si}{www.izum.si}). Some of the results in this paper have been derived using the \textsc{healpy} and \textsc{HEALPix} package \citep{2005ApJ...622..759G, Zonca2019}. We thank GWverse (\href{https://gwverse.tecnico.ulisboa.pt/}{gwverse.tecnico.ulisboa.pt/}), which funded T.J.'s visit to the Niels Bohr Academy within the eCOST Action CA16104 as the Short Term Scientific Mission, and made possible the start of this study. We thank M. Pessah, J. Stone, and W. Lu for the useful discussions.

\section*{Data availability}
\addcontentsline{toc}{section}{Data availability}

The data underlying this article are available on
Github at \href{https://github.com/tajjankovic/Spin-induced-offset-stream-self-crossing-shocks-in-TDEs.git}{https://github.com/tajjankovic/Spin-induced-offset-stream-self-crossing-shocks-in-TDEs.git}.

\noindent 
\bibliographystyle{mnras}
\bibliography{cite}

\begin{thebibliography}{}
\makeatletter
\relax
\def\mn@urlcharsother{\let\do\@makeother \do\$\do\&\do\#\do\^\do\_\do\%\do\~}
\def\mn@doi{\begingroup\mn@urlcharsother \@ifnextchar [ {\mn@doi@}
  {\mn@doi@[]}}
\def\mn@doi@[#1]#2{\def\@tempa{#1}\ifx\@tempa\@empty \href
  {http://dx.doi.org/#2} {doi:#2}\else \href {http://dx.doi.org/#2} {#1}\fi
  \endgroup}
\def\mn@eprint#1#2{\mn@eprint@#1:#2::\@nil}
\def\mn@eprint@arXiv#1{\href {http://arxiv.org/abs/#1} {{\tt arXiv:#1}}}
\def\mn@eprint@dblp#1{\href {http://dblp.uni-trier.de/rec/bibtex/#1.xml}
  {dblp:#1}}
\def\mn@eprint@#1:#2:#3:#4\@nil{\def\@tempa {#1}\def\@tempb {#2}\def\@tempc
  {#3}\ifx \@tempc \@empty \let \@tempc \@tempb \let \@tempb \@tempa \fi \ifx
  \@tempb \@empty \def\@tempb {arXiv}\fi \@ifundefined
  {mn@eprint@\@tempb}{\@tempb:\@tempc}{\expandafter \expandafter \csname
  mn@eprint@\@tempb\endcsname \expandafter{\@tempc}}}

\bibitem[\protect\citeauthoryear{Alexander, van Velzen, Horesh  \&
  Zauderer}{Alexander et~al.}{2020}]{Alexander_2020}
Alexander K.~D.,  van Velzen S.,  Horesh A.,   Zauderer B.~A.,  2020, \mn@doi
  [Space Science Reviews] {10.1007/s11214-020-00702-w}, 216

\bibitem[\protect\citeauthoryear{{Batra}, {Lu}, {Bonnerot}  \&
  {Phinney}}{{Batra} et~al.}{2023}]{Batra_2022}
{Batra} G.,  {Lu} W.,  {Bonnerot} C.,   {Phinney} E.~S.,  2023, \mn@doi
  [\mnras] {10.1093/mnras/stad318}, \href
  {https://ui.adsabs.harvard.edu/abs/2023MNRAS.520.5192B} {520, 5192}

\bibitem[\protect\citeauthoryear{Bonnerot \& Lu}{Bonnerot \&
  Lu}{2020}]{Bonnerot_2020}
Bonnerot C.,  Lu W.,  2020, \mn@doi [Monthly Notices of the Royal Astronomical
  Society] {10.1093/mnras/staa1246}

\bibitem[\protect\citeauthoryear{{Bonnerot} \& {Lu}}{{Bonnerot} \&
  {Lu}}{2022}]{bonnerot2021nozzle}
{Bonnerot} C.,  {Lu} W.,  2022, \mn@doi [\mnras] {10.1093/mnras/stac146}, \href
  {https://ui.adsabs.harvard.edu/abs/2022MNRAS.511.2147B} {511, 2147}

\bibitem[\protect\citeauthoryear{{Bonnerot} \& {Stone}}{{Bonnerot} \&
  {Stone}}{2021}]{bonnerot_2020_book}
{Bonnerot} C.,  {Stone} N.~C.,  2021, \mn@doi [\ssr]
  {10.1007/s11214-020-00789-1}, \href
  {https://ui.adsabs.harvard.edu/abs/2021SSRv..217...16B} {217, 16}

\bibitem[\protect\citeauthoryear{{Bonnerot}, {Rossi}, {Lodato}  \&
  {Price}}{{Bonnerot} et~al.}{2016}]{Bonnerot_2016eccentric_disks}
{Bonnerot} C.,  {Rossi} E.~M.,  {Lodato} G.,   {Price} D.~J.,  2016, \mn@doi
  [\mnras] {10.1093/mnras/stv2411}, \href
  {https://ui.adsabs.harvard.edu/abs/2016MNRAS.455.2253B} {455, 2253}

\bibitem[\protect\citeauthoryear{Bonnerot, Lu  \& Hopkins}{Bonnerot
  et~al.}{2021}]{Bonnerot_2021}
Bonnerot C.,  Lu W.,   Hopkins P.~F.,  2021, \mn@doi [Monthly Notices of the
  Royal Astronomical Society] {10.1093/mnras/stab398}, 504, 4885–4905

\bibitem[\protect\citeauthoryear{Bonnerot, Pessah  \& Lu}{Bonnerot
  et~al.}{2022}]{Bonnerot_2022}
Bonnerot C.,  Pessah M.~E.,   Lu W.,  2022, \mn@doi [The Astrophysical Journal
  Letters] {10.3847/2041-8213/ac6950}, 931, L6

\bibitem[\protect\citeauthoryear{Bricman \& Gomboc}{Bricman \&
  Gomboc}{2020}]{Bricman_2020}
Bricman K.,  Gomboc A.,  2020, \mn@doi [The Astrophysical Journal]
  {10.3847/1538-4357/ab6989}, 890, 73

\bibitem[\protect\citeauthoryear{Clerici \& Gomboc}{Clerici \&
  Gomboc}{2020}]{Clerici_Gomboc_2020}
Clerici A.,  Gomboc A.,  2020, \mn@doi [Astronomy \& Astrophysics]
  {10.1051/0004-6361/202037641}, 642

\bibitem[\protect\citeauthoryear{Dai, McKinney  \& Miller}{Dai
  et~al.}{2015}]{Dai_2015}
Dai L.,  McKinney J.~C.,   Miller M.~C.,  2015, \mn@doi [The Astrophysical
  Journal] {10.1088/2041-8205/812/2/l39}, 812, L39

\bibitem[\protect\citeauthoryear{{G{\'o}rski}, {Hivon}, {Banday}, {Wandelt},
  {Hansen}, {Reinecke}  \& {Bartelmann}}{{G{\'o}rski}
  et~al.}{2005}]{2005ApJ...622..759G}
{G{\'o}rski} K.~M.,  {Hivon} E.,  {Banday} A.~J.,  {Wandelt} B.~D.,  {Hansen}
  F.~K.,  {Reinecke} M.,   {Bartelmann} M.,  2005, \mn@doi [\apj]
  {10.1086/427976}, \href {http://adsabs.harvard.edu/abs/2005ApJ...622..759G}
  {622, 759}

\bibitem[\protect\citeauthoryear{Guillochon \& Ramirez-Ruiz}{Guillochon \&
  Ramirez-Ruiz}{2013}]{Guillochon_2013}
Guillochon J.,  Ramirez-Ruiz E.,  2013, \mn@doi [The Astrophysical Journal]
  {10.1088/0004-637x/767/1/25}, 767, 25

\bibitem[\protect\citeauthoryear{Guillochon \& Ramirez-Ruiz}{Guillochon \&
  Ramirez-Ruiz}{2015}]{Guillochon_2015}
Guillochon J.,  Ramirez-Ruiz E.,  2015, \mn@doi [The Astrophysical Journal]
  {10.1088/0004-637x/809/2/166}, 809, 166

\bibitem[\protect\citeauthoryear{{Hammerstein} et~al.,}{{Hammerstein}
  et~al.}{2023}]{Hammerstein_2023ApJ...942....9H}
{Hammerstein} E.,  et~al., 2023, \mn@doi [\apj] {10.3847/1538-4357/aca283},
  \href {https://ui.adsabs.harvard.edu/abs/2023ApJ...942....9H} {942, 9}

\bibitem[\protect\citeauthoryear{Hayasaki, Stone  \& Loeb}{Hayasaki
  et~al.}{2013}]{Hayasaki_2013}
Hayasaki K.,  Stone N.,   Loeb A.,  2013, \mn@doi [Monthly Notices of the Royal
  Astronomical Society] {10.1093/mnras/stt871}, 434, 909–924

\bibitem[\protect\citeauthoryear{{Huang}, {Davis}  \& {Jiang}}{{Huang}
  et~al.}{2023}]{huang_2023arXiv230317443H}
{Huang} X.,  {Davis} S.~W.,   {Jiang} Y.-f.,  2023, \mn@doi [arXiv e-prints]
  {10.48550/arXiv.2303.17443}, \href
  {https://ui.adsabs.harvard.edu/abs/2023arXiv230317443H} {p. arXiv:2303.17443}

\bibitem[\protect\citeauthoryear{Jankovič \& Gomboc}{Jankovič \&
  Gomboc}{2023}]{Jankovic_2023https://doi.org/10.48550/arxiv.2302.00607}
Jankovič T.,  Gomboc A.,  2023, \mn@doi [The Astrophysical Journal]
  {10.3847/1538-4357/acb8b0}, 946, 25

\bibitem[\protect\citeauthoryear{Jiang, Guillochon  \& Loeb}{Jiang
  et~al.}{2016}]{Jiang_2016}
Jiang Y.-F.,  Guillochon J.,   Loeb A.,  2016, \mn@doi [The Astrophysical
  Journal] {10.3847/0004-637x/830/2/125}, 830, 125

\bibitem[\protect\citeauthoryear{Komossa}{Komossa}{2015}]{komossa}
Komossa S.,  2015, \mn@doi [Journal of High Energy Astrophysics]
  {10.1016/j.jheap.2015.04.006}, 7, 148

\bibitem[\protect\citeauthoryear{{Leloudas} et~al.,}{{Leloudas}
  et~al.}{2022}]{Leloudas_2022NatAs...6.1193L}
{Leloudas} G.,  et~al., 2022, \mn@doi [Nature Astronomy]
  {10.1038/s41550-022-01767-z}, \href
  {https://ui.adsabs.harvard.edu/abs/2022NatAs...6.1193L} {6, 1193}

\bibitem[\protect\citeauthoryear{{Liodakis} et~al.,}{{Liodakis}
  et~al.}{2023}]{Liodakis_2023Sci...380..656L}
{Liodakis} I.,  et~al., 2023, \mn@doi [Science] {10.1126/science.abj9570},
  \href {https://ui.adsabs.harvard.edu/abs/2023Sci...380..656L} {380, 656}

\bibitem[\protect\citeauthoryear{Liptai, Price, Mandel  \& Lodato}{Liptai
  et~al.}{2019}]{liptai2019disc}
Liptai D.,  Price D.~J.,  Mandel I.,   Lodato G.,  2019, Disc formation from
  tidal disruption of stars on eccentric orbits by {K}err black holes using
  general relativistic smoothed particle hydrodynamics (\mn@eprint {arXiv}
  {1910.10154})

\bibitem[\protect\citeauthoryear{{Lodato}, {King}  \& {Pringle}}{{Lodato}
  et~al.}{2009}]{Lodato_2009MNRAS.392..332L}
{Lodato} G.,  {King} A.~R.,   {Pringle} J.~E.,  2009, \mn@doi [\mnras]
  {10.1111/j.1365-2966.2008.14049.x}, \href
  {https://ui.adsabs.harvard.edu/abs/2009MNRAS.392..332L} {392, 332}

\bibitem[\protect\citeauthoryear{Lu \& Bonnerot}{Lu \&
  Bonnerot}{2019}]{Lu_2019}
Lu W.,  Bonnerot C.,  2019, \mn@doi [Monthly Notices of the Royal Astronomical
  Society] {10.1093/mnras/stz3405}, 492, 686–707

\bibitem[\protect\citeauthoryear{{Nixon} \& {King}}{{Nixon} \&
  {King}}{2012}]{Nixon_2012}
{Nixon} C.~J.,  {King} A.~R.,  2012, \mn@doi [\mnras]
  {10.1111/j.1365-2966.2011.20377.x}, \href
  {https://ui.adsabs.harvard.edu/abs/2012MNRAS.421.1201N} {421, 1201}

\bibitem[\protect\citeauthoryear{{Patra}, {Lu}, {Brink}, {Yang}, {Filippenko}
  \& {Vasylyev}}{{Patra} et~al.}{2022}]{Patra_2022MNRAS.515..138P}
{Patra} K.~C.,  {Lu} W.,  {Brink} T.~G.,  {Yang} Y.,  {Filippenko} A.~V.,
  {Vasylyev} S.~S.,  2022, \mn@doi [\mnras] {10.1093/mnras/stac1727}, \href
  {https://ui.adsabs.harvard.edu/abs/2022MNRAS.515..138P} {515, 138}

\bibitem[\protect\citeauthoryear{Petrushevska et~al.,}{Petrushevska
  et~al.}{2023}]{Petrushevska_2023}
Petrushevska T.,  et~al., 2023, \mn@doi [Astronomy \& Astrophysics]
  {10.1051/0004-6361/202244623}, 669, A140

\bibitem[\protect\citeauthoryear{Price \& et. al.}{Price \& et.
  al.}{2018}]{price}
Price D.~J.,  et. al. 2018, \mn@doi [Publications of the Astronomical Society
  of Australia] {10.1017/pasa.2018.25}, 35

\bibitem[\protect\citeauthoryear{Rees}{Rees}{1988}]{rees}
Rees M.~J.,  1988, \mn@doi [Nature] {10.1038/333523a0}, 333, 523

\bibitem[\protect\citeauthoryear{Rossi, Servin  \& Kesden}{Rossi
  et~al.}{2021}]{Rossi_2021}
Rossi J.,  Servin J.,   Kesden M.,  2021, \mn@doi [Physical Review D]
  {10.1103/physrevd.104.103019}, 104

\bibitem[\protect\citeauthoryear{Roth, Kasen, Guillochon  \& Ramirez-Ruiz}{Roth
  et~al.}{2016}]{Roth_2016}
Roth N.,  Kasen D.,  Guillochon J.,   Ramirez-Ruiz E.,  2016, \mn@doi [The
  Astrophysical Journal] {10.3847/0004-637X/827/1/3}, 827, 3

\bibitem[\protect\citeauthoryear{S{\k{a}}dowski, Tejeda, Gafton, Rosswog  \&
  Abarca}{S{\k{a}}dowski et~al.}{2016}]{Sadowski_2016}
S{\k{a}}dowski A.,  Tejeda E.,  Gafton E.,  Rosswog S.,   Abarca D.,  2016,
  \mn@doi [Monthly Notices of the Royal Astronomical Society]
  {10.1093/mnras/stw589}, 458, 4250

\bibitem[\protect\citeauthoryear{Saxton, Komossa, Auchettl  \& Jonker}{Saxton
  et~al.}{2021}]{Saxton_2021}
Saxton R.,  Komossa S.,  Auchettl K.,   Jonker P.~G.,  2021, \mn@doi [Space
  Science Reviews] {10.1007/s11214-020-00759-7}, 217

\bibitem[\protect\citeauthoryear{{Sazonov} et~al.,}{{Sazonov}
  et~al.}{2021}]{sazonov_2021MNRAS.508.3820S}
{Sazonov} S.,  et~al., 2021, \mn@doi [\mnras] {10.1093/mnras/stab2843}, \href
  {https://ui.adsabs.harvard.edu/abs/2021MNRAS.508.3820S} {508, 3820}

\bibitem[\protect\citeauthoryear{Shiokawa, Krolik, Cheng, Piran  \&
  Noble}{Shiokawa et~al.}{2015}]{Shiokawa_2015}
Shiokawa H.,  Krolik J.~H.,  Cheng R.~M.,  Piran T.,   Noble S.~C.,  2015,
  \mn@doi [The Astrophysical Journal] {10.1088/0004-637x/804/2/85}, 804, 85

\bibitem[\protect\citeauthoryear{Stone, Sari  \& Loeb}{Stone
  et~al.}{2013}]{Stone_2013}
Stone N.,  Sari R.,   Loeb A.,  2013, \mn@doi [Monthly Notices of the Royal
  Astronomical Society] {10.1093/mnras/stt1270}, 435, 1809

\bibitem[\protect\citeauthoryear{Stone, Kesden, Cheng  \& van Velzen}{Stone
  et~al.}{2019}]{stone19}
Stone N.~C.,  Kesden M.,  Cheng R.~M.,   van Velzen S.,  2019, \mn@doi [General
  Relativity and Gravitation] {10.1007/s10714-019-2510-9}, 51, 30

\bibitem[\protect\citeauthoryear{Tout, Pols, Eggleton  \& Han}{Tout
  et~al.}{1996}]{Tout_199610.1093/mnras/281.1.257}
Tout C.~A.,  Pols O.~R.,  Eggleton P.~P.,   Han Z.,  1996, \mn@doi [Monthly
  Notices of the Royal Astronomical Society] {10.1093/mnras/281.1.257}, 281,
  257

\bibitem[\protect\citeauthoryear{{Yao} et~al.,}{{Yao}
  et~al.}{2023}]{Yao_2023arXiv230306523Y}
{Yao} Y.,  et~al., 2023, \mn@doi [arXiv e-prints] {10.48550/arXiv.2303.06523},
  \href {https://ui.adsabs.harvard.edu/abs/2023arXiv230306523Y} {p.
  arXiv:2303.06523}

\bibitem[\protect\citeauthoryear{Zonca, Singer, Lenz, Reinecke, Rosset, Hivon
  \& Gorski}{Zonca et~al.}{2019}]{Zonca2019}
Zonca A.,  Singer L.,  Lenz D.,  Reinecke M.,  Rosset C.,  Hivon E.,   Gorski
  K.,  2019, \mn@doi [Journal of Open Source Software] {10.21105/joss.01298},
  4, 1298

\bibitem[\protect\citeauthoryear{van Velzen \& et al.}{van Velzen \&
  et~al.}{2019}]{van_Velzen_2019}
van Velzen S.,  et al. 2019, \mn@doi [The Astrophysical Journal]
  {10.3847/1538-4357/aafe0c}, 872, 198

\bibitem[\protect\citeauthoryear{van Velzen et~al.,}{van Velzen
  et~al.}{2021}]{van_Velzen_2021}
van Velzen S.,  et~al., 2021, \mn@doi [The Astrophysical Journal]
  {10.3847/1538-4357/abc258}, 908, 4

\makeatother
\end{thebibliography}

\end{document}